\documentclass{emulateapj}

\usepackage{amsmath}
\usepackage{graphicx}
\usepackage{float}
\usepackage{amssymb}
\usepackage{color}
\usepackage{subfigure}
\usepackage{sidecap}

\shorttitle{Supernova explosions triggered by a quark-hadron phase transition}
\shortauthors{Fischer et al.}

\begin{document}

\title{Core-collapse supernova explosions triggered by a
quark-hadron phase transition during the early post-bounce phase}

\author{T.~Fischer$^{1,2,3}$,
I.~Sagert$^4$,
G.~Pagliara$^5$,
M.~Hempel$^3$,
J.~Schaffner-Bielich$^5$,
T.~Rauscher$^3$,
F.-K.~Thielemann$^3$,
R.~K{\"a}ppeli$^3$,
G.~Mart{\'i}nez-Pinedo$^1$
and
M.~Liebend{\"o}rfer$^3$}

\affil{
\begin{tabular}{l}
1 GSI, Helmholtzzentrum f\"ur Schwerioneneforschung GmbH,
Planckstra{\ss}e~1, 64291 Darmstadt, Germany\\
2 Technische Universit{\"a}t Darmstadt, Schlossgartenstra{\ss}e 9,
64289 Darmstadt, Germany\\
3  Department of Physics, University of Basel, Klingelbergstra{\ss}e~82,
4056 Basel, Switzerland \\
4 Department of Physics \& Astronomy, Michigan State University,
East Lansing, MI 48824, U.S.A \\
5 Institut f\"ur Theoretische Physik, Ruprecht-Karls-Universit\"at,
Philosophenweg 16, 69129 Heidelberg, Germany
\end{tabular}
}

\date{\today}

\begin{abstract}
We explore explosions of massive stars, which are triggered via
the quark-hadron phase transition during the early post bounce
phase of core-collapse supernovae.
We construct a quark equation of state, based on the bag model
for strange quark matter.
The transition between the hadronic and the quark phases is
constructed applying Gibbs conditions.
The resulting quark-hadron hybrid equations of state are used in
core-collapse supernova simulations, based on general relativistic
radiation hydrodynamics and three flavor Boltzmann neutrino transport
in spherical symmetry.
The formation of a mixed phase reduces the adiabatic index,
which induces the gravitational collapse of the central protoneutron
star.
The collapse halts in the pure quark phase, where the adiabatic
index increases.
A strong accretion shock forms, which propagates towards the
protoneutron star surface.
Due to the density decrease of several orders of magnitude,
the accretion shock turns into a dynamic shock with matter outflow.
This moment defines the onset of the explosion in supernova
models that allow for a quark-hadron phase transition,
where otherwise no explosions could be obtained.
The shock propagation across the neutrinospheres releases
a burst of neutrinos.
This serves as a strong observable identification for the
structural reconfiguration of the stellar core.
The ejected matter expands on a short timescale
and remains neutron-rich.
These conditions might be suitable for the production
of heavy elements via the $r$-process.
The neutron-rich material is followed by proton-rich
neutrino-driven ejecta in the later cooling phase of the
protoneutron star where the $\nu p$-process might occur.
\end{abstract}


\maketitle

%..................................................................................................................
\section{Introduction}

% - intro. sentense
The explosion mechanism of massive stars is an active
subject of research in theoretical astrophysics.
In this article, we explore a scenario where the explosion is
triggered due to the phase transition from hadronic
matter to quark matter.
We present specific and distinguished characteristics for
possible explosion observables.

% - core-collapse supernova phenomenology
Stars with an initial main sequence mass above 10~M$_\odot$
produce extended Fe-cores, as the end product of nuclear burning.
During the following evolution, the photodisintegration of heavy
nuclei as well as electron captures reduce the pressure of the
Fe-core.
The initial contraction of the core proceeds into a collapse,
during which density and temperature rise.
At nuclear densities the repulsive nuclear interaction causes a
significant stiffening of the equation of state (EoS), which in turn
leads to the formation of a shock wave.
As it propagates outward it continuously looses energy due to the
dissociation of in-falling heavy nuclei from the progenitor.
The central object formed at core bounce is a hot and lepton-rich
protoneutron star (PNS).
The shock propagation across the neutrinospheres, i.e. the neutrino
energy and flavor dependent spheres of last scattering,
releases a burst of electron neutrinos - known as the deleptonization
burst - emitted from electron capture at free protons.
The electron neutrino luminosity rises up to several $10^{53}$~erg/s
(depending on the progenitor) on a timescale of 5--20~ms after bounce.
This enormous energy loss, in combination with the dissociation of the
in-falling heavy nuclei, turns the dynamic shock into a standing accretion
shock (SAS) already at about 5~ms after bounce
\citep[see e.g.][]{
HillebrandtMueller:1981, MayleWilson:1987, Bruenn:1989, MyraBludman:1989}.
The deleptonization near the neutrinospheres results in a low
proton-to-baryon ratio (given by the electron fraction) of $Y_e=0.1$--$0.2$.
The PNS interior, which did not experience shock heating, stays slightly
less neutron-rich where $Y_e\simeq0.3$ at bounce
(depending on the equation of state and the electron capture rates).

% - the SN problem
In an attempt to explain core-collapse supernova explosions,
neutrino heating was suggested as a mechanism for the revival of the
SAS by \citet{BetheWilson:1985}, leading to so called neutrino-driven
explosions.
In spherical symmetry, this mechanism has been shown to
produce explosions only for the low mass $8.8$~M$_\odot$
O-Ne-Mg-core  from Nomoto~(1983,1984,1987) by
\citet{Kitaura:etal:2006} and \citet{Fischer:etal:2010b}.
Multi-dimensional phenomena, such as rotation, convection and the
development of fluid instabilities, have been shown to increase
the neutrino heating efficiency \citep[see e.g.][]{Miller:etal:1993,
Herant:etal:1994, Burrows:etal:1995, JankaMueller:1996}.
Such models help to aid the understanding of aspherical explosions
\citep[see also][and references therein]{Bruenn:etal:2009,MarekJanka:2009}.
Besides neutrino heating \citep[][]{BetheWilson:1985},
alternative explosive scenarios have been explored,
such as the dumping of magnetic energy by
\citet{LeBlankWilson:1970},
\citet{Moiseenko:etal:2007}
and
\citet{Takiwaki:etal:2009}
as well as by acoustic energy by \citet{Burrows:etal:2006}.
The later two scenarios require multi-dimensional models.

% - quark matter in core-collapse SNe
An additional scenario is obtained by looking at the conditions achieved
at the PNS interior during the post bounce phase of core-collapse supernovae.
Due to the mass accretion on the order of several $0.1$~M$_\odot$/s,
the central density rises above several times nuclear matter density
and the temperature above several tens of~MeV
(depending on the progenitor model).
With increasing degeneracy, the central electron fraction
reduces below $Y_e\lesssim 0.25$ on timescales on the order of several
100~ms post bounce.
The timescale for the PNS contraction depends on the mass accretion
rate which in turn is in direct correlation with the progenitor model.
The conditions obtained in the PNS interiors raise the question about
the state of matter and possibly the appearance of exotic matter
such as hyperons and quarks.

% - QCD general remarks
Assuming a purely hadronic composition, matter is composed of color
charged quarks which cannot propagate freely and are confined into
color neutral baryons and mesons.
At high temperatures or baryon densities, hadronic matter is expected to
change its state to a phase of deconfined quarks and gluons,
where chiral symmetry is restored.
Thereby, deconfinement and asymptotic freedom allow color charged
quarks and gluons to move as free and almost non-interacting particles.
Due to chiral symmetry restoration, the quarks obtain their current mass
values, which for the up and  down quarks are in the range of few~MeV,
while the strange quark current mass is about $m_s \sim 100$~MeV.

In Quantum Chromodynamics (QCD), which is believed to be the fundamental
theory behind the strong force, the interaction of quarks is characterized
by an effective coupling constant $\alpha_s$, which monotonically approaches
zero at large momentum transfer.
For small values of $\alpha_s$, a perturbative treatment of the strong
interaction can be applied.
However, at the energy scale of quark deconfinement, the interaction
strength becomes too large and a perturbative treatment is no longer
applicable, requiring alternative approaches.
One possible description is the phenomenological quark bag model
\citep[see e.g.][]{Detar83}, based on the spatial confinement of quarks.
In Nambu-Jona-Lasinio models, quarks are implemented as quantum
fields with point-like interactions in a chirally symmetric Lagrangian
\citep[see][and references therein]{Buballa05}.
More advanced methods to describe quark matter are based on e.g. the
Dyson-Schwinger equation \citep[see][]{Fischer06} or ab initio
calculations of QCD on the lattice.
Ab initio calculations show a smooth crossover from a hadronic phase
to a deconfined quark gluon plasma phase, at high temperatures and
low baryochemical potentials \citep[see e.g.][]{Fodor04,Aoki06}.
On the other hand, field theoretical and phenomenological calculations
point to a first order phase transition from hadronic to 
quark matter at large chemical potentials and finite temperatures
\citep[see][]{Berges99,Klevansky:1992,Pisarski84}.
The many open questions concerning critical temperatures and
densities, the order of the phase transition and whether both
chiral symmetry restoration and deconfinement happen simultaneously
or not, make QCD phase transitions an active topic of experimental
research at the Relativistic Heavy Ion Collider (RHIC) in  Brookhaven,
the Large Hadron Collider (LHC) at CERN and
the Facility for Antiproton and Ion Research (FAIR) at GSI,  Darmstadt
and at NICA in Dubna (Russia).
The experiments at the LHC and RHIC are focused on the production
of matter at high temperatures and low baryochemical potentials.
On the other hand, FAIR and NICA will be designed to study nuclear
matter at large densities and temperatures, which are conditions
that correspond to core-collapse supernova interiors.
They will enable a crosscheck of results from terrestrial experiments
and astrophysical observations.
However, the conditions in supernovae, neutron-rich matter and the
possible production of strange quarks by weak interactions,
differ from those in heavy-ion collision experiments.
As will be discussed below, a quark-hadron phase transition
at large baryon number densities in heavy-ion collisions can be
compatible with low critical densities for the appearance of quark
matter in supernovae and neutron stars.
A detailed discussion of this issue will be given in \S~2.

\begin{table*}[htp]
\centering
\caption{List of neutrino reactions considered, where
$\nu=\{\nu_e,\bar{\nu}_e,\nu_{\mu/\tau},\bar{\nu}_{\mu/\tau}\}$
and $N=\{n,p\}$.}
\begin{tabular}{ c c c}
\hline
\hline
Label & Neutrino reaction & Reference\\
\hline
1
&
$e^- + p \rightleftarrows n +\nu_e$
&
\citet{Bruenn:1985} \\
2
&
$e^+ + n \rightleftarrows p +\bar{\nu}_e$
&
\citet{Bruenn:1985} \\
3
&
$e^- + \left<A,Z\right> \rightleftarrows \left<A,Z-1\right> +\nu_e$
&
\citet{Bruenn:1985} \\
4
&
$\nu + N \rightleftarrows \nu' + N$
&
\citet{Bruenn:1985}, \citet{MezzacappaBruenn:1993b} \\
5
&
$\nu + \left<A,Z\right> \rightleftarrows \nu' + \left<A,Z\right>$
&
\citet{Bruenn:1985}, \citet{MezzacappaBruenn:1993a} \\
6
&
$\nu + e^\pm \rightleftarrows \nu' + e^\pm$
&
\citet{Bruenn:1985}, \citet{MezzacappaBruenn:1993c} \\
7
&
$e^- + e^+ \rightleftarrows \nu + \bar{\nu}$
&
\citet{Bruenn:1985}, \citet{MezzacappaMesser:1999} \\
8
&
$N + N \rightleftarrows N + N + \nu + \bar{\nu}$
&
\citet{ThompsonBurrows:2001}\\
9
&
$\nu_e + \bar{\nu}_e \rightleftarrows \nu_{\mu/\tau} + \bar{\nu}_{\mu/\tau}$
&
\citet{Buras:etal:2003}, \citet{Fischer:etal:2009} \\
\hline
\end{tabular}
\label{table-nu-reactions}
\end{table*}
%

% quark matter in SNe - hist. remarks
The first attempt to investigate the quark-hadron phase transition in
simulations of massive stars goes back to \citet{TakaharaSato:1988}.
They tried to find a connection between the appearance of quark matter
and the apparent but statistically insignificant multi-peaked neutrino
signal from SN1987a \citep[see][]{Hirata:etal:1988,Bionta:etal:1987},
using general relativistic hydrodynamics in spherical symmetry
and a polytropic EoS.
More sophisticated input physics was applied in the investigation
by \citet{Gentile:etal:1993}.
Their model was based on general relativistic hydrodynamics and
a parametrized description of the deleptonization during the Fe-core
collapse as well as an improved EoS for both hadronic and quark
matter.
The transition between these two phases was constructed via
a co-existence region, i.e. a quark-hadron mixed phase.
They found a significant softening of the EoS in the mixed phase
as well as the formation of a second shock wave as a direct
consequence of the quark-hadron phase transition.
The secondary shock followed and finally merged with the Fe-core
bounce shock, only milliseconds after its appearance.
However, both studies were not able to predict any possible features in
the neutrino signal emitted due to the lack of neutrino transport.
Simulations of the QCD phase transition of very massive ($100$
M$_\odot$) progenitors have been reported recently by
\citet{Nakazato:etal:2008a} and \citet{Nakazato:etal:2010a}.
They applied the quark bag model for the quark matter EoS with
a large bag constant and hence a high critical density for the
onset of deconfinement.
The transition from hadronic matter to quark matter was modeled
via an extended mixed phase applying the Gibbs conditions.
They confirmed a significant softening of the EoS in the
mixed phase due to the reduced adiabatic index.
A characteristic neutrino signature was not found due to the
immediate formation of a black hole during the phase transition.
However, they observed a significant shortening of the post bounce accretion
time until black hole formation, due to the softening of the EoS at
conditions where quark matter was found to occur.

The present article follows a different approach.
It is the continuation of \citet{Sagert:etal:2009},
where first results were discussed.
We construct a quark-hadron hybrid EoS, i.e. using the microscopic EoS
from \citet{Shen:etal:1998} for hadrons and the bag model based EoS
for strange quark matter.
The resulting EoS is applied to core-collapse supernova simulations
of low and intermediate mass Fe-core progenitors.
Our model is based on general relativistic radiation hydrodynamics and
three flavor Boltzmann neutrino transport in spherical symmetry.
For these progenitors, spherically symmetric simulations do not result in
explosions via the delayed neutrino heating mechanism
\citep[see e.g.][]{RamppJanka:2000, Liebendoerfer:etal:2001b, Janka:2001,
Thompson:etal:2003}.
Here, we explore the possibility that a quark-hadron phase transition
during the early post bounce evolution can result in an explosion and
discuss the observational consequences.
The temperatures on the order of tens of MeV and the large
neutron excess of matter in supernova interiors, enables a
phase transition at low critical densities close to nuclear
saturation density (depending on the bag model parameters).
In this respect, the physical conditions in core-collapse supernova
interiors differ from matter in heavy-ion collision experiments,
where temperatures on the order of few 100~MeV are obtained
but matter stays more or less isospin symmetric with an electron
fraction of $Y_e\simeq0.5$.
The expected critical density is on the order of several
times nuclear saturation density in heavy-ion collision experiments.
Furthermore, the supernova timescales (1--100~ms) and the timescale
in heavy-ion collisions ($\sim10^{-23}$~seconds) are substantially different.
These differences will be further discussed in \S2.
In the core-collapse supernova simulations under investigation,
the phase transition takes place within the first $500$~ms post bounce
(depending on the critical conditions and the progenitor model).
We observe an accelerated PNS contraction due to the softening
of the EoS in the mixed phase, in comparison to simulations using
purely hadronic EoSs.
The contraction proceeds into an adiabatic collapse when the maximum
stable mass, depending on the hybrid EoS, is reached.
During the collapse, density and temperature increase and a pure
quark phase develops where the EoS is significantly stiffer than the
EoS in the mixed phase.
The collapse halts and a strong second shock front forms.
This shock wave accelerates at the PNS surface where the density
drops over several orders of magnitude, resulting in explosions
with matter at outflow velocities on the order of $10^5$~km/s.
Furthermore, the shock propagation across the neutrinospheres
which are located in the hadronic phase, releases an additional
neutrino burst dominated by $\bar{\nu}_e$
and $(\nu_{\mu/\tau},\bar{\nu}_{\mu/\tau})$.
We analyze the emitted neutrino signal from this explosion
scenario for several examples of massive progenitors.
The possible detection of the QCD induced neutrino burst has been
investigated by \citet{Dasgupta:etal:2010} for a Galactic event,
based on results from \citet{Sagert:etal:2009}.

% - organisation of the manuscript
The manuscript is organized as follows.
In \S~2 we briefly discuss our core-collapse supernova model
including the standard hadronic EoS.
Furthermore, we explain the construction of the hybrid EoS in detail.
We explore different choices of parameters and discuss the 
corresponding phase diagrams and hybrid star mass-radius relations.
\S~3 is devoted to the illustration of the results obtained
for the 10.8~M$_\odot$ progenitor from \citet{Woosley:etal:2002},
that will be used as the reference model.
The results are discussed in \S~4, where we compare the reference
model with simulations performed with different progenitor masses
and different hybrid EoSs.
Furthermore, we explore the nucleosynthesis relevant
conditions and draw conclusions about the possibility of reaching
high magnetic fields on the order of $10^{13}-10^{15}$~G
during the PNS evolution with quark matter. 
We close the manuscript with the summary in \S~5.
%

%.............................................................................
\section{Neutrino radiation hydrodynamics}

\begin{figure*}[htp]
\centering
\subfigure[15~M$_\odot$]{
\includegraphics[width=0.42\textwidth]{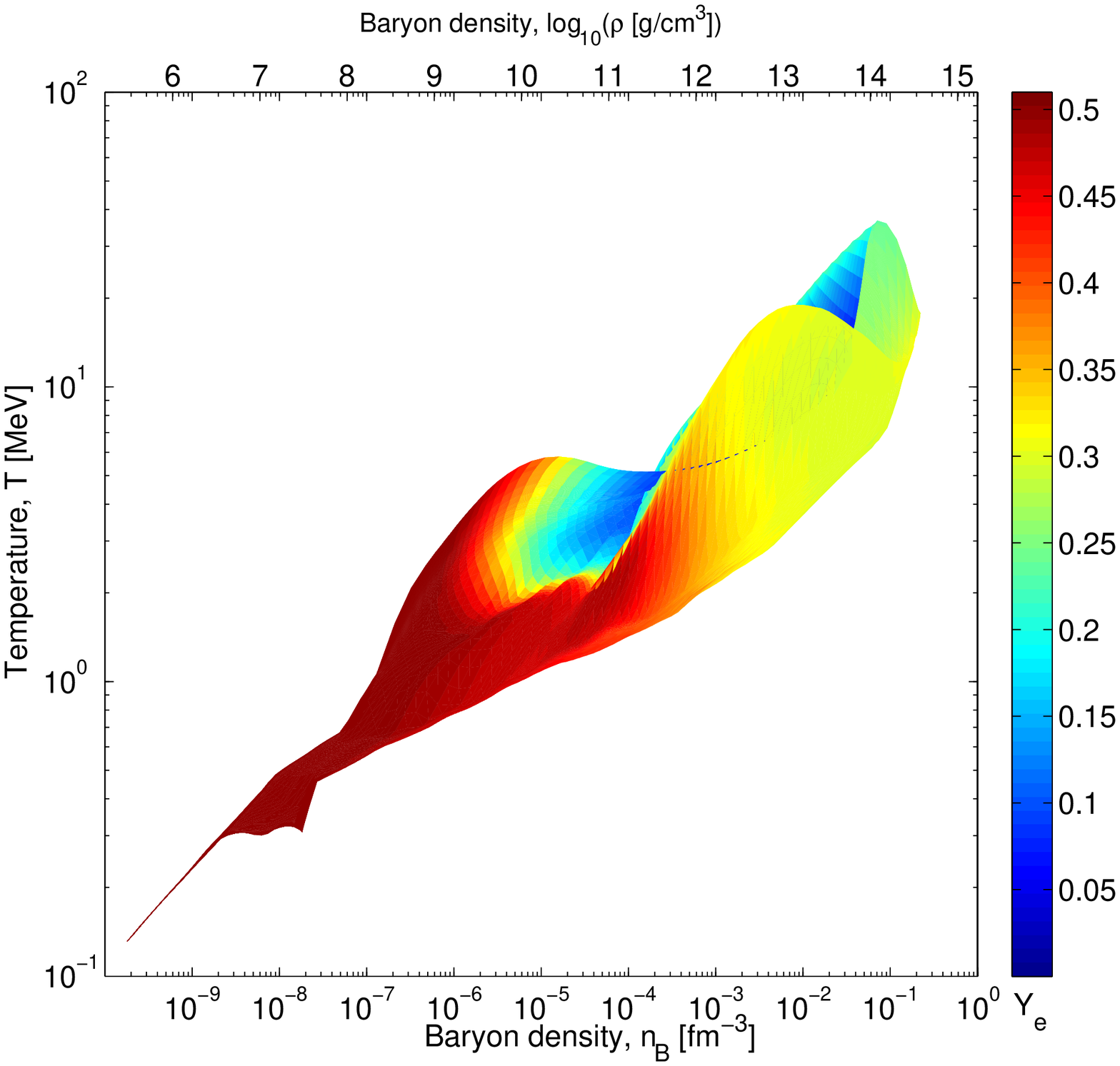}}
\hspace{10mm}
\subfigure[$40$~M$_\odot$]{
\includegraphics[width=0.42\textwidth]{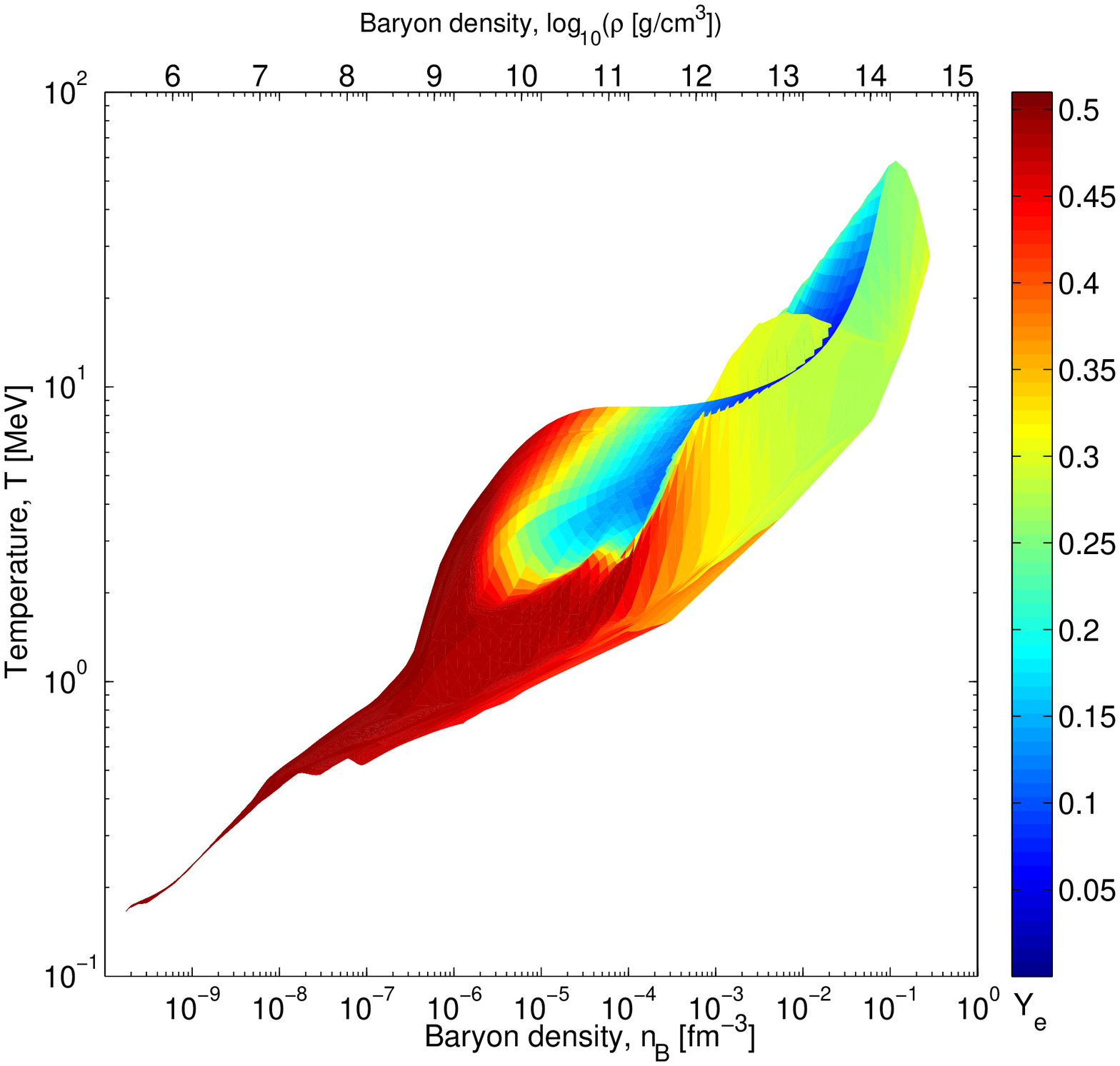}}
\caption{Phase space covered in core-collapse supernova simulations
at the example of two progenitor models from \citet{WoosleyWeaver:1995},
both using the pure hadronic EoS from \citet{Shen:etal:1998}.
The white background corresponds to densities and
temperatures which are not obtained, where the colored
domain belongs to densities and temperatures which are
obtained. Color-scaled is the electron fraction which ranges from
neutron matter (blue) $Y_e=0$ to symmetric matter (red)
$Y_e\simeq 0.5$.}
\label{fig-eos-domain}
\end{figure*}

The spherically symmetric radiation hydrodynamics model
AGILE-BOLTZTRAN is based on three flavor Boltzmann neutrino
transport.
It solves the evolution of the neutrino distribution functions
$f_\nu$ (phase-space functions) for each neutrino flavor
$\nu=\{\nu_e,\bar{\nu}_e,\nu_{\mu/\tau},\bar{\nu}_{\mu/\tau}\}$,
\begin{equation}
\frac{df_\nu}{dt} = \left. \frac{df_\nu}{dt} \right\vert_\text{collision}
\label{eq-Boltzmann}
\end{equation}
depending on the phase-space derivative (l.h.s of Eq.(\ref{eq-Boltzmann}))
and neutrino-matter interactions (r.h.s of Eq.(\ref{eq-Boltzmann})).
The neutrino reactions for the various interactions considered
are listed in Table~\ref{table-nu-reactions}, including the references.
In spherical symmetry, the distribution functions depend on the
phase-space coordinates system time $t$, baryon mass $a$,
the cosine of the neutrino propagation angle $\mu=\cos\theta$ and
the neutrino energy $E$, $f_\nu(t,a,\mu,E)$, where the latter two
are the momentum space representations.
The original Newtonian variant of AGILE-BOLTZTRAN
\citep[see][]{MezzacappaBruenn:1993a, MezzacappaBruenn:1993b,
MezzacappaBruenn:1993c}
has been extended by
\citet{Liebendoerfer:etal:2001a}
to solve the general relativistic equations, based on non-stationary
spacetime represented by the following line element 
\begin{equation*}
ds^2 = -\alpha(t,a)^2dt^2 + \left(\frac{r(t,a)'}{\Gamma(t,a)}\right)^2da^2
+ r(t,a)^2 d\Omega,
\end{equation*}
where $d\Omega = d\theta^2+\sin^2\theta d\phi^2$
is the line element of a 2-sphere of radius $r$ in spherical coordinates
$(\theta, \phi)$, and $r' = \partial r/\partial a$.
Note that the metric functions $\alpha$ and $\Gamma$ depend
on the coordinates system time $t$ and baryon mass $a$.
The evolution equations for energy and momentum are obtained from
the covariant derivative of the stress-energy tensor,
$\nabla_i T^{ik}=0$, with the following choice of the stress-energy tensor
\begin{eqnarray*}
T^{tt} &=& \rho(1+e+J), \,\,\,\,
T^{aa} = p + \rho K, \,\,\,\, 
T^{at} = T^{ta} = \rho H, \\
T^{\theta\theta} &=& T^{\phi\phi} = p + \frac{1}{2}\rho(J-K),
\end{eqnarray*}
where $\rho$,  $p$ and $e$ are the rest mass density, matter pressure
and the internal energy density.
$J$, $H$ and $K$ are the neutrino moments, which are given by the
momentum integrals of the neutrino distribution functions
\citep[for details, see][]{Liebendoerfer:etal:2004}.
Special attention has been devoted to accurately conserve
lepton number, energy and momentum, as described in details in
\citet{Liebendoerfer:etal:2004}.
General relativistic effects have been investigated with respect to the
PNS collapse and black hole formation for several massive progenitor
stars of 40 and 50~M$_\odot$ in \citet{Fischer:etal:2009}.

The numerical implementation of the above mentioned physical
equations, including the finite differencing mass conserving scheme,
has been introduced in \citet{Liebendoerfer:etal:2002} and
\citet{Liebendoerfer:etal:2004},
where recent improvements of the finite differencing scheme for
the evolution of the enclosed baryon mass have been introduced
in \citet{Fischer:etal:2010b}.
With these improvements, stable solutions for the evolution equations
can be obtained for long simulation times on the order of seconds.

In the following subsections, we will introduce the EoS where
the sophisticated baryon EoS from \citet{Shen:etal:1998} has 
been extended with a quark-hadron mixed and a pure quark phase
at high densities and temperatures.

\subsection{The EoS in core-collapse supernova simulations}

The EoS has to handle a large variety of physical conditions
obtained in core-collapse supernova simulations.
The covered domain is illustrated in Fig.~\ref{fig-eos-domain}
for the examples of the 15 and 40~M$_\odot$ progenitor models
from \citet{WoosleyWeaver:1995} for the first second post bounce,
both using the pure hadronic EoS from \citet{Shen:etal:1998}.
Temperatures and baryon densities reach from below
0.1~MeV and 10~g/cm$^3$ ($6\times10^{-15}$~fm$^{-3}$)
up to
100~MeV and $10^{15}$ g/cm$^3$ (0.6~fm$^{-3}$).
The electron fraction is represented on a color-scale,
where low values of $Y_e\simeq 0.1$ are obtained close to
the neutrinospheres at sub-saturation densities.
For the low density domain, which is entirely given by the progenitor,
matter is isospin symmetric with an electron fraction of $Y_e\simeq0.5$.
Furthermore, the figures show that only a narrow middle band
of the ($T,\rho$)-diagram is covered in core-collapse supernova
simulations of massive stars, e.g. low temperatures and high densities
(which corresponds to neutron star matter)
as well as vice versa are not obtained.
This is because the entropy per baryon
does not change over many orders of magnitude
during the early (up to 1~second) post bounce evolution
which is illustrated in Fig.~\ref{fig-eos-domain}.

\subsubsection{Hadronic matter}

For temperatures roughly below 0.5~MeV, the rates for nuclear
reactions are important and the assumption of nuclear statistical
equilibrium (NSE) cannot be applied.
The baryon contributions to the EoS are given by
nuclear abundances which must be evolved in time.
The dynamical evolution of these abundances are calculated using
the nuclear reaction network developed by \citet{Thielemann:etal:2004}
(and references therein) which employs tabulated reaction rates.
Due to current computational limitations, we employ the free nucleons,
and in addition $^3$He and $^4$He as well as the
13 symmetric nuclei starting from $^{12}$C to $^{56}$Ni.
Furthermore, we use the following 3 asymmetric nuclei
$^{53}$Fe, $^{54}$Fe and $^{56}$Fe.
The details of the implementation of the network into AGILE-BOLTZTRAN
as well as the motivation for our choice of nuclei and the use in
core-collapse supernova models, is given in \citet{Fischer:etal:2010b}.
It is of special importance to mimic the energy generation from
nuclear burning processes in explosion models, where the
explosion shock travels through the low temperature and density
envelop of the progenitor and the simplification of the ideal gas of
Si-nuclei cannot be used.

For higher temperatures, matter is assumed to be in NSE and the
production and destruction of nuclei are in thermal and chemical
equilibrium with respect to strong interactions.
The abundances depend only on the thermodynamic state
given by the temperature $T$, the baryon density $\rho$ (alternatively
$n_B$) and the electron fraction $Y_e$.
For this regime, we use the hadronic EoS from \citet{Shen:etal:1998}.
It is based on the relativistic mean field (RMF) approach for
homogeneous matter and the Thomas-Fermi-approximation for 
a single average heavy nucleus.
It has an incompressibility modulus of 281~MeV and
an asymmetry energy of 36.9~MeV.
For this stiff EoS, cold neutron stars with maximum mass
of 2.2~M$_\odot$ are obtained
(see the black solid line in Fig.~\ref{plot_mr_as03_wf}).

On top of the baryons, contributions from electrons and positrons as well
as photons and coulomb corrections (only for non-NSE) are added,
based on \citet{TimmesArnett:1999}.

\subsubsection{Quark matter in the bag model}

Bag models are phenomenological models which were originally introduced
to describe quark confinement.
Due to their simple handling and ability to reproduce hadron properties
\citep[see e.g.][]{Chodos74b, Degrand75, Detar83},
they are also often applied for bulk quark matter and
phase transitions in compact star interiors.
The first description of a bag model goes back to \citet{Bogolyubov68}
and was improved a few years later by \citet{Chodos74b},
known today as the MIT bag model.
The main idea of quark bag models is that the true vacuum of QCD is a
medium which refuses  the penetration of quarks and confines them within a 
sphere (or a bag) to form a color neutral hadron.
The QCD vacuum exerts a pressure on the hadron, represented 
phenomenologically by the bag constant $B$, which is opposed by the
motion of the interior quarks.
Within the bag, quarks are assumed to move in asymptotic freedom
with vanishing or current masses.
Their total pressure $p^Q$, energy density $\epsilon^Q$, entropy density
$s^Q$, and baryon number density  $n_B^Q$ can be written as follows
\begin{eqnarray}
p^Q = \sum_{i} p_{i} -B, & \: \: & \epsilon^Q =
\sum_{i} \epsilon_{i} + B, 
\label{e_p} \\
s^Q = \sum_{i} s_{i}, & \: \: &  n_B^Q =
\frac{1}{3} \sum_{i} n_{i},
\label{s_n}
\end{eqnarray}
where the sums run over all present quark flavors $i$.
In the simple bag model quarks are treated as non-interacting fermions
and their contributions in the above sums can be calculated by solving the
corresponding Fermi integrals for given temperature $T$,
quark chemical potential $\mu_i$, and mass $m_i$ as follows
\begin{eqnarray}
p_{i}(m_i,T,\mu_i)&=& \frac{1}{3} \frac{g_i}{2 \pi^2} \int_0^\infty
k^2 dk\,\,k\,\frac{\partial E_i (k)}{\partial k} 
\nonumber \\
&\times& \left[f (k,\mu_i) + f (k,-\mu_i) \right]
\label{fermi_pressure}
\end{eqnarray}
\begin{eqnarray}
\epsilon_{i}(m_i,T,\mu_i)&=& \frac{g_i}{2 \pi^2} \int_0^\infty E_i (k)
k^2 dk
\nonumber \\
&\times& \left[f (k,\mu_i) + f (k,-\mu_i) \right]
\label{fermi_energy}
\end{eqnarray}
\begin{eqnarray}
s_i (m_i,T,\mu_i) &=& \frac{g_i}{2\pi^2} \int_0^\infty k^2 dk 
\left[
- f (k,\mu_i) \mathrm{ln} f (k,\mu_i) \right.
\nonumber \\
&-& ( 1 - f (k,\mu_i)) \mathrm{ln}(1-f (k,\mu_i)) 
\nonumber\\ 
&-&
f (k,-\mu_i) \mathrm{ln} f (k,-\mu_i)
\nonumber \\
&-& \left. ( 1 - f (k,-\mu_i)) \mathrm{ln}(1-f (k,-\mu_i)) 
\right] 
\label{entropy_f}
\end{eqnarray}
\begin{eqnarray}
n_i(m_i,T,\mu_i) &=& \frac{g_i}{2 \pi^2} \int_0^\infty k^2 dk
\nonumber \\
&\times& \left[f (k,\mu_i) - f (k,-\mu_i) \right].
\label{n_f1}
\end{eqnarray}
$f(k,\pm \mu_i)$ are the Fermi distribution functions with chemical
potentials for particles ($+ \mu_i$) and antiparticles ($- \mu_i$),
\begin{eqnarray}
 f(k, \pm \mu_i) = \frac{1}{e^{(E_i(k) \mp \mu_i)/T} +1},
\end{eqnarray}
where $k$ is the momentum and $E_i (k) = \sqrt{m_i^2 + k^2}$ is the
quark Fermi energy.
The number of degrees of freedom  for each flavor $g_i$ consists of
2 spin states and 3 colors.

Various approaches have been introduced to extend and improve the
simple bag model \citep[see e.g.][]{Detar83},
including first order corrections
for the strong coupling constant $\alpha_s$
\citep[see e.g.][]{FarhiJaffe:1984}.
Since analytical expressions for the thermodynamic potentials at finite
temperature and including $\alpha_s$ corrections can only be obtained
for massless quarks, we calculate the pressure of massive quarks for
the flavor $i$ as follows
\begin{eqnarray*}
p_{i}(m_i,T,\mu_i,\alpha_s) &=& p_{i}(m_i,T,\mu_i,0)
\\
\\
&+& \left[ p_{i}(0,T,\mu_i,\alpha_s) - p_{i}(0,T,\mu_i,0) \right]
\\
\\
&=& p_{i}(m_i,T,\mu_i,0)
\end{eqnarray*}
\begin{equation}
- \left[ \frac{7}{60} T^4 \pi^2 \frac{50 \alpha_s}{21 \pi}
+ \frac{2 \alpha_s}{\pi} \left(\frac{1}{2} T^2 \mu_i^2
+ \frac{\mu_i^4}{4 \pi^2} \right) \right] .
\label{p_f}
\end{equation}
The last two terms in Eq.(\ref{p_f}) are taken from analytical expressions in
\citet{FarhiJaffe:1984}, where $p_{i}(m_i,T,\mu_i,0)$ can be calculated 
by numerically solving the Fermi integrals.
The energy density $\epsilon_i$, number density $n_i$, and entropy density
$s_i$ can be calculated in a similar way.
We will apply this procedure only for the strange quarks, for which we choose
a mass of $m_s = 100$~MeV in accordance with the range of
$m_s \sim 70$--$130$~MeV and the weighted average of
$105_{-1.3}^{+1.5}$~MeV from \citet{Amsler:2008}.
The up and down quarks have masses of several~MeV and can be treated
as massless, that is $m_u = m_d =0$.
The remaining quark flavors are too heavy to appear in supernova and
neutron star environments.

The value of the bag constant is an active subject of research and expected
to be in the range of $B^{1/4}\sim145$--$235$~MeV, from hadron fitting
\citep[][]{Detar83}.
An upper limit is difficult to define.
However, concerning compact stars a value of $B^{1/4} \lesssim 200$~MeV
allows the presence of a small pure quark matter core in the star interior
\citep[see e.g.][]{Schertler00}.
The lower limit for the bag constant depends on whether strange quark
matter is considered as the ground state of nuclear matter or not.
This absolute stability was introduced by \citet{Witten84} and is
based on the idea that  hadronic matter is a metastable state while strange
quark matter has a lower energy per baryon than $^{56}$Fe and is therefore
the true ground state of nuclear matter.
This so-called {\em Witten hypothesis} leads to the existence of strange stars,
which are composed of absolutely stable strange quark matter.
These can either represent all compact stars or be in co-existence with
hadronic stars, as was recently discussed by \citet{Bauswein09}.

In the simple bag model with $m_u = m_d = 0$ and $m_s=100$~MeV,
the absolute stability of strange quark matter sets in for
$B^{1/4} \lesssim 161$~MeV.
For finite $\alpha_s$ and different quark masses, \citet{FarhiJaffe:1984}
have mapped out the limiting values of $B$
\citep[see also][for different $\alpha_S$]{Weissenborn:etal:2011}.
In the present article, we will not assume that strange quark matter
is absolutely stable.
However, since our aim is to probe low critical densities for the quark matter
phase transition in the early post-bounce phase of supernovae, we chose the
parameter sets of $B^{1/4}=162$~MeV (EOS1) and
$B^{1/4}=165$~MeV (EOS2) both with $\alpha_s = 0$
and $B^{1/4}=155$~MeV with $\alpha_s = 0.3$ (EOS3).
%(see Table~\ref{tab-eos})
%
%
%
%\begin{table}[h]
%\centering
%\caption{The hybrid EoSs.}
%\begin{tabular}{ c c c }
%\hline \hline
%Label
%&
%bag constant [MeV]
%&
%$\alpha_S$
%\\
%\hline 
%EOS1 & 162 & 0 \\
%EOS2 & 165 & 0 \\
%EOS3 & 155 & 0.3 \\
%\hline
%\end{tabular}
%\label{tab-eos}
%\end{table}
%

\subsection{The hybrid EoS}

\begin{figure*}
\centering
\subfigure[]{
\includegraphics[width=0.48\textwidth]{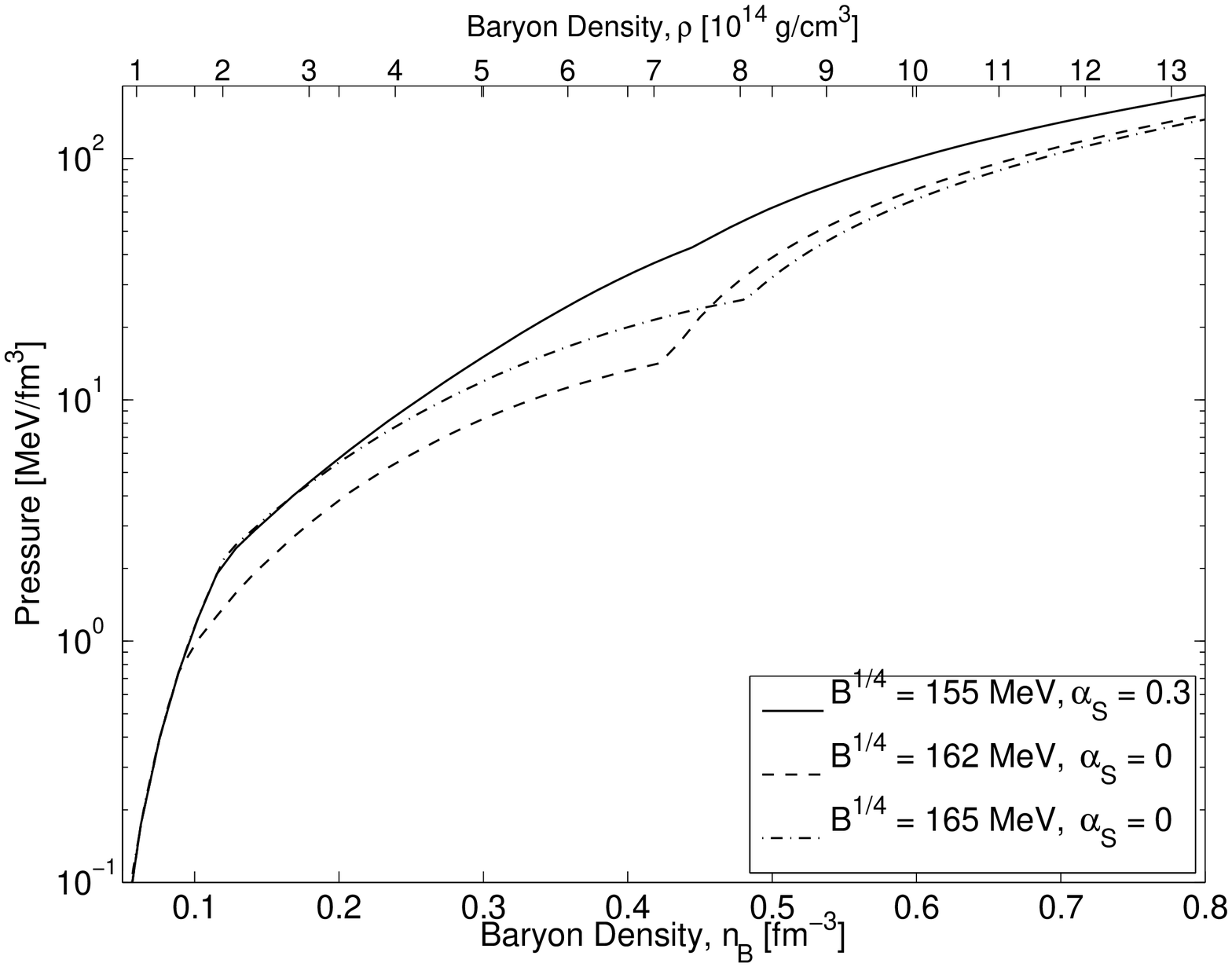}
\label{entropy_p_as}}
\hfill
\subfigure[]{
\includegraphics[width=0.48\textwidth]{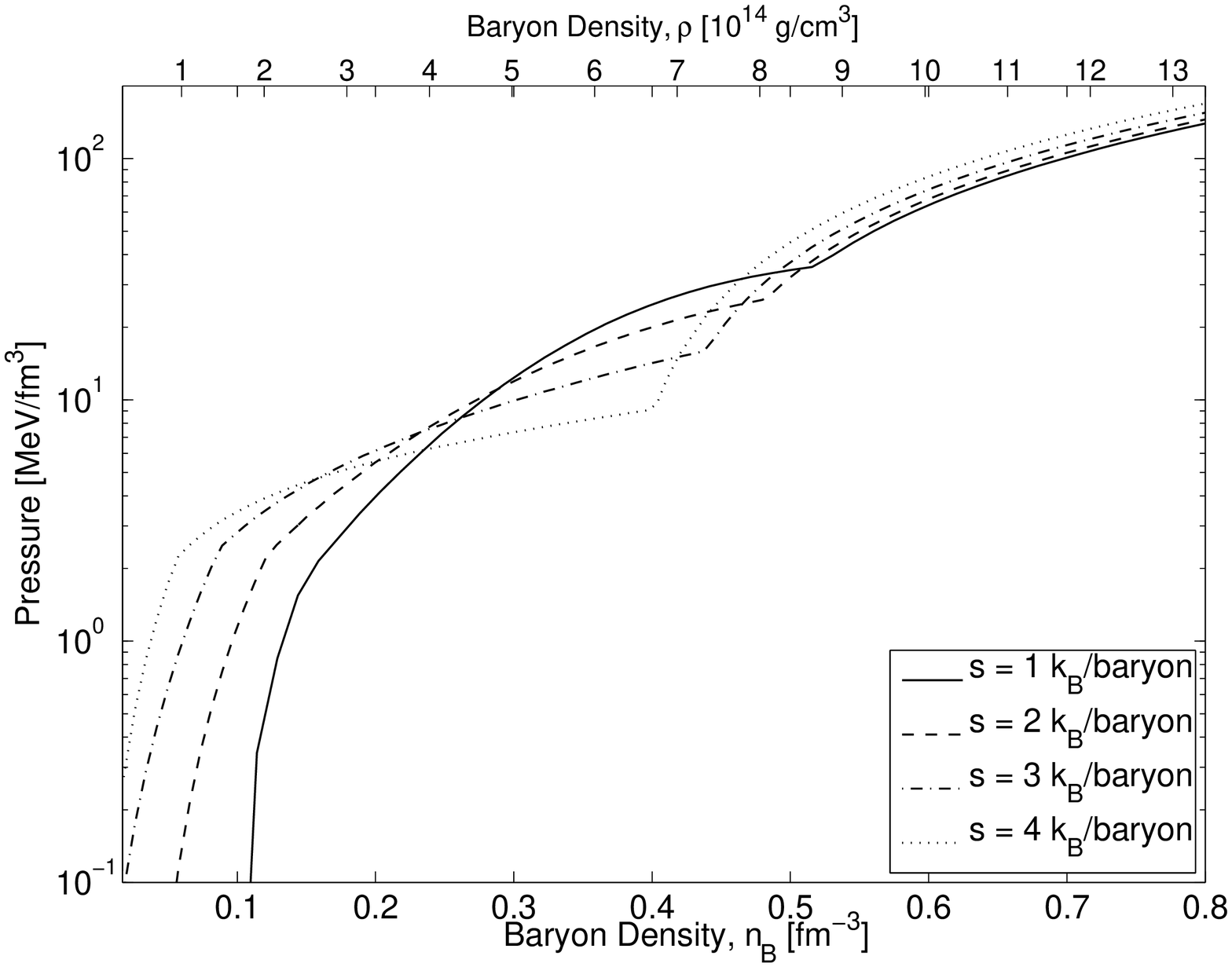}
\label{entropy_p}}
\caption{Hybrid EoS with a global proton fraction of $Y_p=0.3$.
(a) shows the pressure versus density for fixed entropy per baryon
of $s=2$~k$_B$, for different bag constants $B$ and $\alpha_s$.
(b) shows pressure verses density for different entropies per
baryon at fixed bag constant of $B^{1/4}=165$~MeV and
$\alpha_S=0$.
Due to the smaller critical density, the EoS in the co-existence
region is softer for higher values of $s$.}
\end{figure*}

To probe the appearance of quark matter in supernova environments
we implement the quark-hadron phase transition into the pure hadronic
EoS from \cite{Shen:etal:1998}. 
The critical density for the onset of quark matter and the properties
of the mixed phase depend strongly on the choice of global or local
conservation laws.
In this work we will apply the Gibbs approach as discussed in
\citet{Glendenning_book}, where conservation laws are always fulfilled
globally and the pressure in the mixed phase is a smooth function of the
density (see Figs.~\ref{entropy_p_as} and \ref{entropy_p}).

Another frequently used method to calculate phase transition is the
Maxwell construction.
Here only the baryon number is a global quantity and other conservation
laws must be fulfilled locally in the quark and hadronic phases.
As a consequence of this restriction, the onset of the mixed phase for the
Maxwell condition is generally later (i.e. at higher density).
Furthermore, for an isothermal phase transition the pressure is
constant throughout the entire mixed phase, with respect to the
baryon density.

Despite their different behaviors, the Maxwell and the Gibbs approaches
both represent first order phase transitions.
Gibbs falls in the class of non-congruent  phase transitions whereas the
restriction to only one globally conserved quantity in the case of a
Maxwell construction is congruent
\citep[for details on (non)congruent phase transitions, see e.g.][]{Iosilevskiy10}.
Similar to the pasta phases in nuclear matter at the liquid-gas phase transition,
finite size effects from e.g Coulomb interactions and surface tension of quark
or hadronic matter can lead to the formation of structures in the mixed phase,
like spheres, rods, and planes.
It was shown by e.g. \citet{Yasutake09}, that such a scenario can make the
quark-hadron phase transition very similar to the one given by a Maxwell
construction.
However, in the present article we neglect any finite-size effects and
Coulomb contributions and treat the phases in the thermodynamic limit.

Both the Maxwell and the Gibbs approaches are used in astrophysics
and nuclear physics frequently.
However, they represent only two extreme scenarios for equilibrium
conditions at the phase transition.
It was recently shown by \citet{Hempel:etal:2009} that the actual variety
of possible combinations of global and local conservation laws is much
richer and can open up interesting new scenarios for e.g. the early evolution
of protoneutron stars \citep[][]{Pagliara:etal:2009}.

Chemical potentials which correspond to a globally conserved quantity
$k$, are equal in the quark and hadron phases.
Together with thermal and mechanical equilibrium of quark and 
hadronic matter, the conditions for the co-existence region are:
\begin{eqnarray}
T^{H} & \equiv & T^{Q}
\label{THTQ}\\
\mu_{k}^Q & \equiv &\mu_{k}^H
\label{muHmuQ}\\
p^H & \equiv & p^Q.
\label{pHpQ}
\end{eqnarray}
The pressures $p^H$ and $p^Q$, in the hadronic and quark phases,
respectively, are in principle sums over contributions from all present
particles, including electrons and neutrinos.
As discussed in \S 2.1, the high temperatures and densities when quark
matter sets in, correspond to conditions where neutrinos are completely
trapped.
Hence, neutrinos are in weak equilibrium with nuclear matter at given lepton
fraction $Y_L$ \citep[see e.g.][]{Steiner00, Pagliara:etal:2009},
together with the electrons and positrons.
However, within the Gibbs approach the electron and neutrino chemical
potentials are equal in the quark and hadronic phases and so are their
contributions to the thermodynamic quantities such as the pressure, which
consequently cancel in the equilibrium condition of Eq.(\ref{pHpQ}).
In this context, it was recently discussed by \citet{Hempel:etal:2009} that
neutrino contributions do not in fact have to be taken into account,
neither for the Gibbs nor the Maxwell construction if the EoS is
provided for given temperature $T$, baryon density $n_B$ and
proton fraction  $Y_p$.

The chemical potentials for the up and down quarks can be obtained from the equilibrium conditions at deconfinement as follows
\begin{eqnarray}
p  \rightleftarrows 2 u + d,  \: \: \: n  \rightleftarrows  2 d + u,
\end{eqnarray}
where
\begin{eqnarray}
\mu_u  =  \frac{2}{3} \mu_p - \frac{1}{3} \mu_n, \: \: \: 
\mu_d  =  \frac{2}{3} \mu_n - \frac{1}{3} \mu_p.
\label{weak_eq1}
\end{eqnarray}
The dynamical timescales considered in supernova simulations
can be as small as milliseconds.
On the other hand, weak reactions producing strangeness operate
much faster on the order of $10^{-6}$~s
\citep[from kaon decays;][]{Amsler:2008}.
The latter should be equilibrated in nuclear matter and hence be
present either already in the hadronic phase in form of hyperons
\citep[see e.g.][]{Ishizuka08} or be produced by a series of weak
reactions in the quark phase, that can be schematically expressed
by the following effective reaction
\begin{eqnarray}
u + d \rightleftarrows u + s \,\,,
\end{eqnarray}
which leads to $\mu_s = \mu_d$ for the strange quark
chemical potential.
Consequently, we consider quark matter composed of the three flavors
up, down, and strange.
We did not implement hyperons or kaons in the hadronic phase, because
our main goal is to study the effects from quark matter in core-collapse
supernovae.
However, such inclusion of strange hadrons, also in connection with
strangeness conserving quark matter phase transitions, is of high interest
and should be addressed in future studies.

The mixed phase is characterized by the fraction of quark matter $\chi$,
as follows
\begin{eqnarray}
\chi = \frac{V^Q}{V^Q + V^H},
\end{eqnarray}
where $V^Q$ and $V^H$ are the volume fractions of the quark and
hadronic phases, respectively.
Consequently,
\begin{eqnarray*}
&\chi& = 0 \,\,\,\,\,\,\,\,\,\,\text{(hadronic phase),} \\
0<&\chi&<1 \,\,\,\,\,\,\,\,\,\,\text{(mixed phase),} \\
&\chi&=0 \,\,\,\,\,\,\,\,\,\,\text{(quark phase).}
\end{eqnarray*}
For $0 < \chi < 1$,  the baryon number density $n_B$, the energy density
$\epsilon$, and the entropy density $s$ are calculated as a sum of the quark
and the hadron contributions via the following expressions
\begin{eqnarray}
n_B & =&  (1- \chi) n_B^H + \chi n_B^Q \,\,, \\ 
\epsilon &=&  (1- \chi) \epsilon^H + \chi \epsilon^Q \,\,, \\
s &=&  (1- \chi) s^H + \chi s^Q \,\,.
\end{eqnarray}
Furthermore, within the Gibbs approach, the proton fraction $Y_p$ is a
global quantity and composed of the charge fractions in 
the hadronic and the quark phases as follows
\begin{eqnarray}
Y_p n_B =  (1-\chi) Y_c^H n_B^H + \chi Y_c^Q n_B^Q,
\end{eqnarray}
where $Y_c^H = n_p/n_B^H$ (hadronic matter) and
$Y_c^Q =n_c^Q /n_B^Q$ (quark matter) are calculated from the
positively charged up and negatively charged down and strange
quarks via $n_c^Q = (2 n_u -  n_d -  n_s)/3$.

\subsection{Discussion of the Hybrid EoS}

Fig.~\ref{entropy_p_as} shows the pressure with respect to the baryon
density for different values of $B$ and $\alpha_S$ as well as for
different entropies per baryon.
The EoSs are calculated for a global proton fraction of $Y_p = 0.3$
at fixed entropy $s = 2$ k$_B$/baryon.
The onset of the mixed phase happens for all parameter sets around
nuclear saturation density.
It can be identified by a softening of the EoS
\footnote{The softness/stiffness of the EoS is measured via the
adiabatic index $\gamma = \partial\ln p/\partial\ln\rho$.},
while the appearance of the pure quark matter phase is accompanied
by a stiffening.
In the following, we will discuss the origin of these changes in the pressure
slopes while their effects on the supernova dynamics will be addressed in
\S 3.

\begin{figure}
\centering
\includegraphics[width=0.48\textwidth]{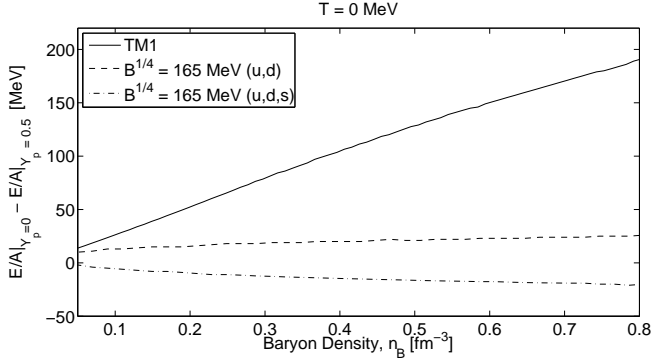}
\caption{Symmetry energy of hadronic matter from \citet{Shen:etal:1998}
(solid line), 2 flavor quark matter (dashed line) and
three flavor quark matter (dash-dotted line) within the quark bag
model with $B^{1/4}=165$~MeV.} 
\label{symmetry_en}
\end{figure}

For the simple bag model, without $\alpha_s$ corrections, the smaller
bag constant $B^{1/4}=162$~MeV leads to an earlier onset of the mixed
phase and therefore also to a lower pressure in the co-existence region
than for $B^{1/4}=165$~MeV.
However, as given in Eq.(\ref{e_p}) and can be seen from
Fig.~\ref{entropy_p_as}, for pure quark matter the larger bag constant
gives a lower total pressure, while the inclusion of first order $\alpha_s$
corrections stiffens the quark EoS \citep[see also][]{Alford:etal:2005}.
Hereby, the accompanying shift in the critical density to higher values
can be decreased again by lowering the bag constant
(in our case $B^{1/4}=155$~MeV) \citep[see][]{Sagert10}.
However, the higher pressures in the mixed and the quark phases
due to $\alpha_s$ corrections remain.
With increasing entropy, the critical density shifts to lower values,
and furthermore the pressure in the mixed phase is reduced
(see Fig.~\ref{entropy_p}).

The global proton fraction $Y_p$ also impacts the critical
density due to the symmetry energy $E_{sym}$ of hadronic and quark
matter.
For nucleons, $E_{sym}$ is composed of a potential part which is
caused by isospin dependent nucleon-nucleon interactions and a kinetic
part stemming from the proton and neutron Fermi energies.
In contrast, quark matter in the bag model is sensitive to its isospin state
only by the Fermi energies of the quarks and has therefore a smaller
symmetry energy than hadronic matter.
This can be seen in Fig.~\ref{symmetry_en}, which shows $E_{sym}$ for
hadronic matter, for two-flavor and for three-flavor quark matter with
$B^{1/4}=165$~MeV (see also \cite{PagliaraSchaffnerBielich:2010}).
The symmetry energy can be defined by the difference of the energy per baryon
of pure neutron matter, that is $Y_p =0$, and isospin symmetric matter with
$Y_p = 0.5$ \citep[see][]{Li08}
\begin{eqnarray}
E_{sym} (n_B) \propto \left.\frac{E}{A}\right\vert_{n_B,Y_p=0} -
\left.\frac{E}{A}\right\vert_{n_B,Y_p=0.5}.
\label{symmetry}
\end{eqnarray}
Because the down and  strange quarks have negative charge, they
can replace the electrons as agents for charge neutrality.
In Fig.~\ref{symmetry_en}, $\beta$-equilibrium\footnote{
$\beta$-equilibrium is given by the following relation between
the chemical potentials $\mu_e+\mu_p=\mu_n$
\citep[see e.g.][]{Weber:1999}} was assumed,
which leads to compositions with small values
of $Y_e$ or $Y_c^Q$, respectively.
As a consequence, we see in Fig.~\ref{symmetry_en} that $E_{sym}$
is even negative for strange quark matter in contrast to two flavor quark
matter and hadronic matter.

\begin{figure}
\centering
\subfigure[$B^{1/4} = 165$~MeV, $\beta$-equilibrium]{
\includegraphics[width=0.48\textwidth]{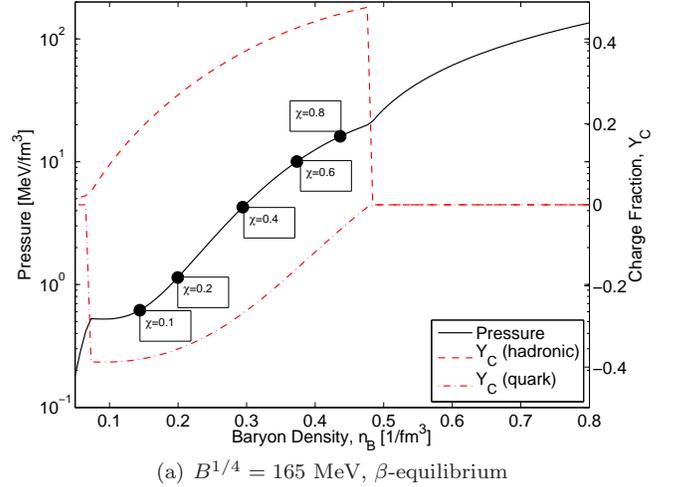}
\label{p_nb_beta}}
\\
\centering
\subfigure[$B^{1/4} = 165$~MeV, $Y_p = 0.3$]{
\includegraphics[width=0.48\textwidth]{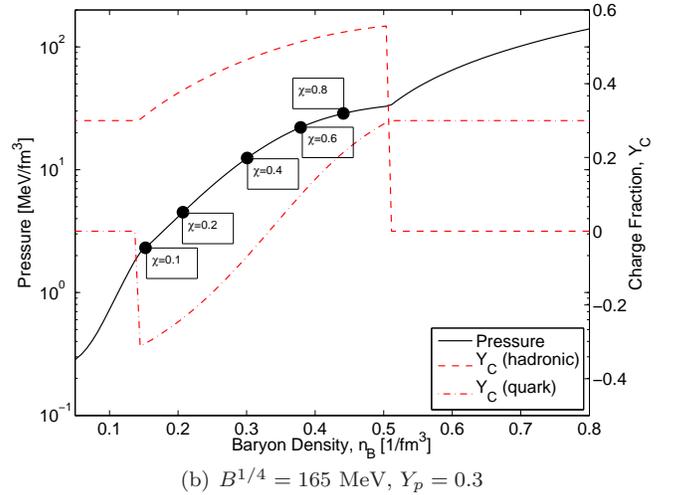}
\label{p_nb_yp03}}
\caption{Pressure (black solid lines) and charge fractions in the hadronic
and quark phase, $Y_c^Q$ (red dash-dotted lines) and $Y_c^H$
(red dashed lines), respectively, for matter in $\beta$-equilibrium
in graph (a) and at a fixed global proton fraction of $Y_p=0.3$
in graph (b).
Both calculations are done for zero temperature and show the
corresponding pressure-density curves, together with different
quark volume fractions $\chi$ in the mixed phase.}
\end{figure}
\begin{figure*}
\centering
\subfigure[EOS1, (u,d,s)-quark matter]{
\includegraphics[width=0.45\textwidth]{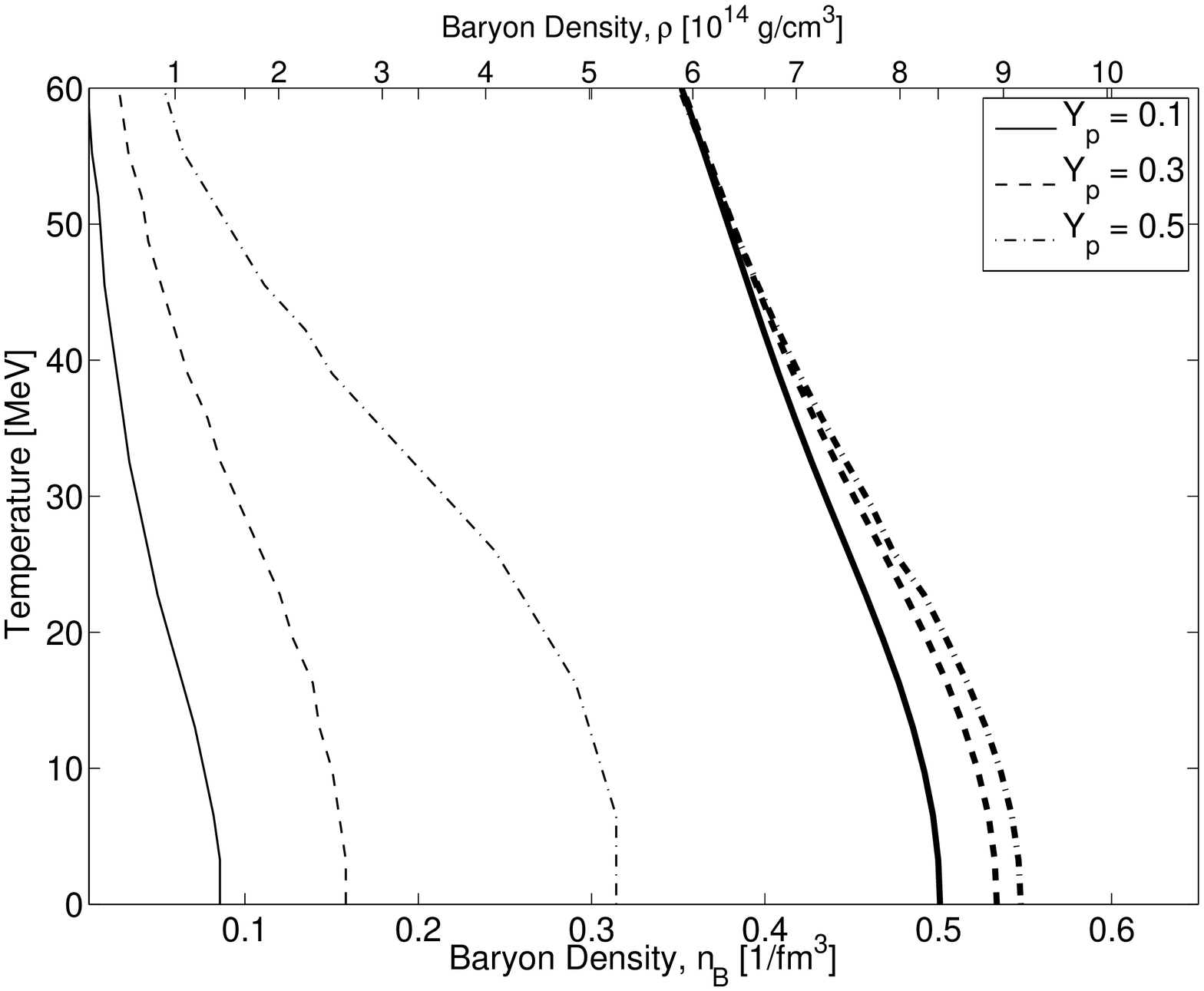}
\label{phase_diag165}}\hfill
\subfigure[EOS2, (u,d,s)-quark matter]{
\includegraphics[width=0.45\textwidth]{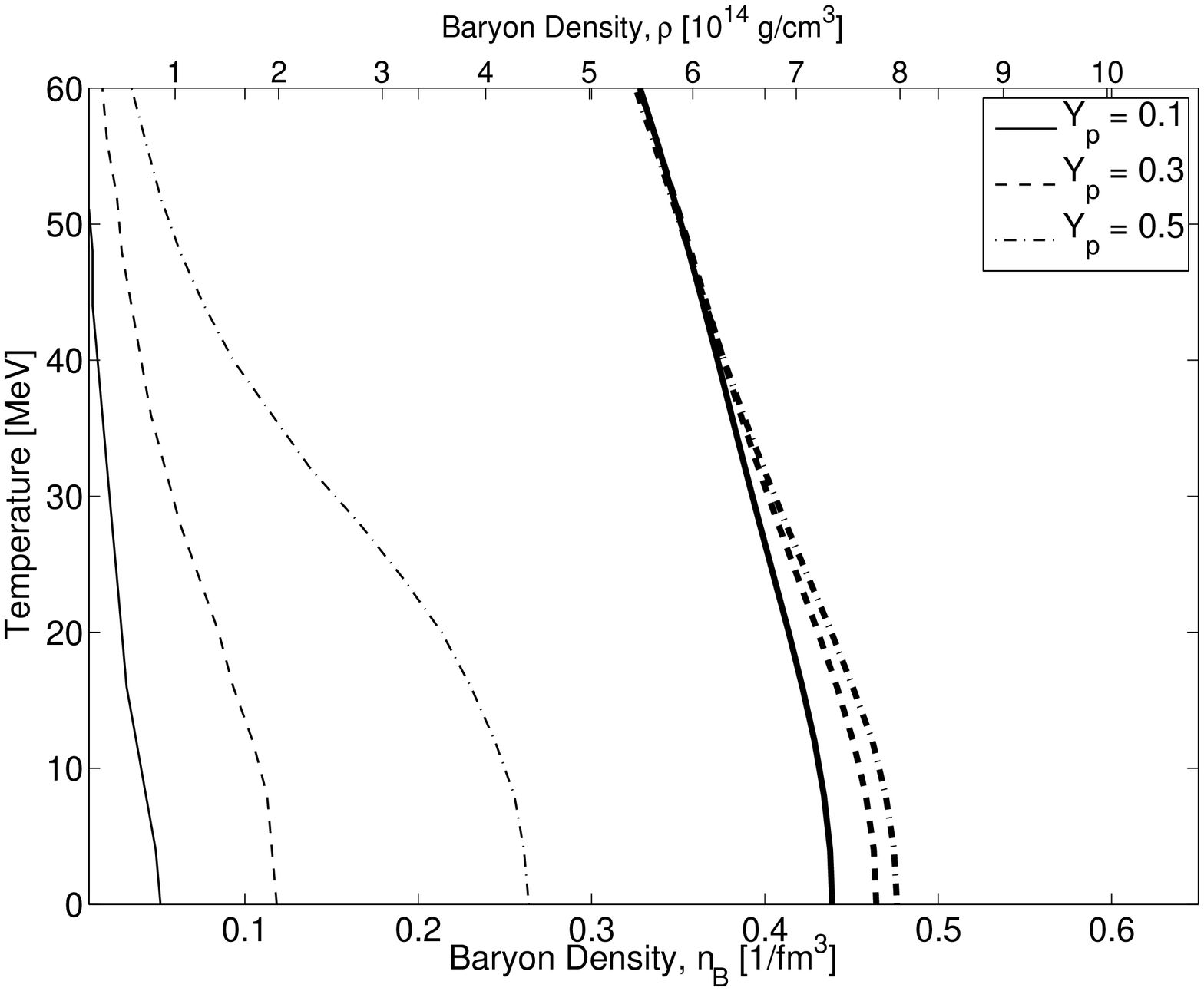}
\label{phase_diag162}}
\\
\subfigure[EOS3, (u,d,s)-quark matter]{
\includegraphics[width=0.45\textwidth]{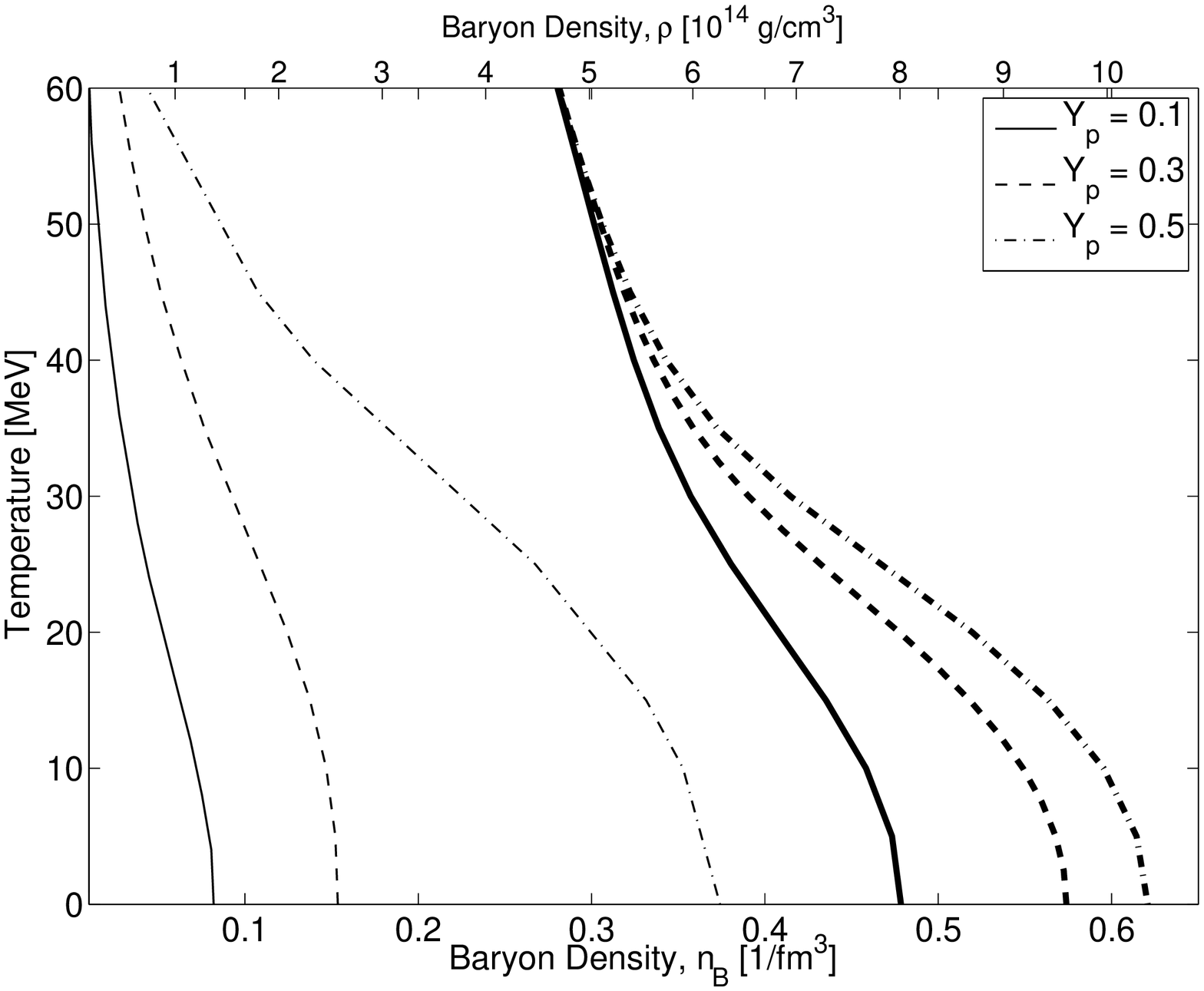}
\label{phase_diag155}}\hfill
\subfigure[EOS2,
comparing (u,d,s)-quark matter with $Y_p=0.3$
and (u,d)-quark matter with $Y_p=0.4,0.5$,
which represents the proton-to-baryon ratio
of matter in heavy-ion collisions.]{
\includegraphics[width=0.45\textwidth]{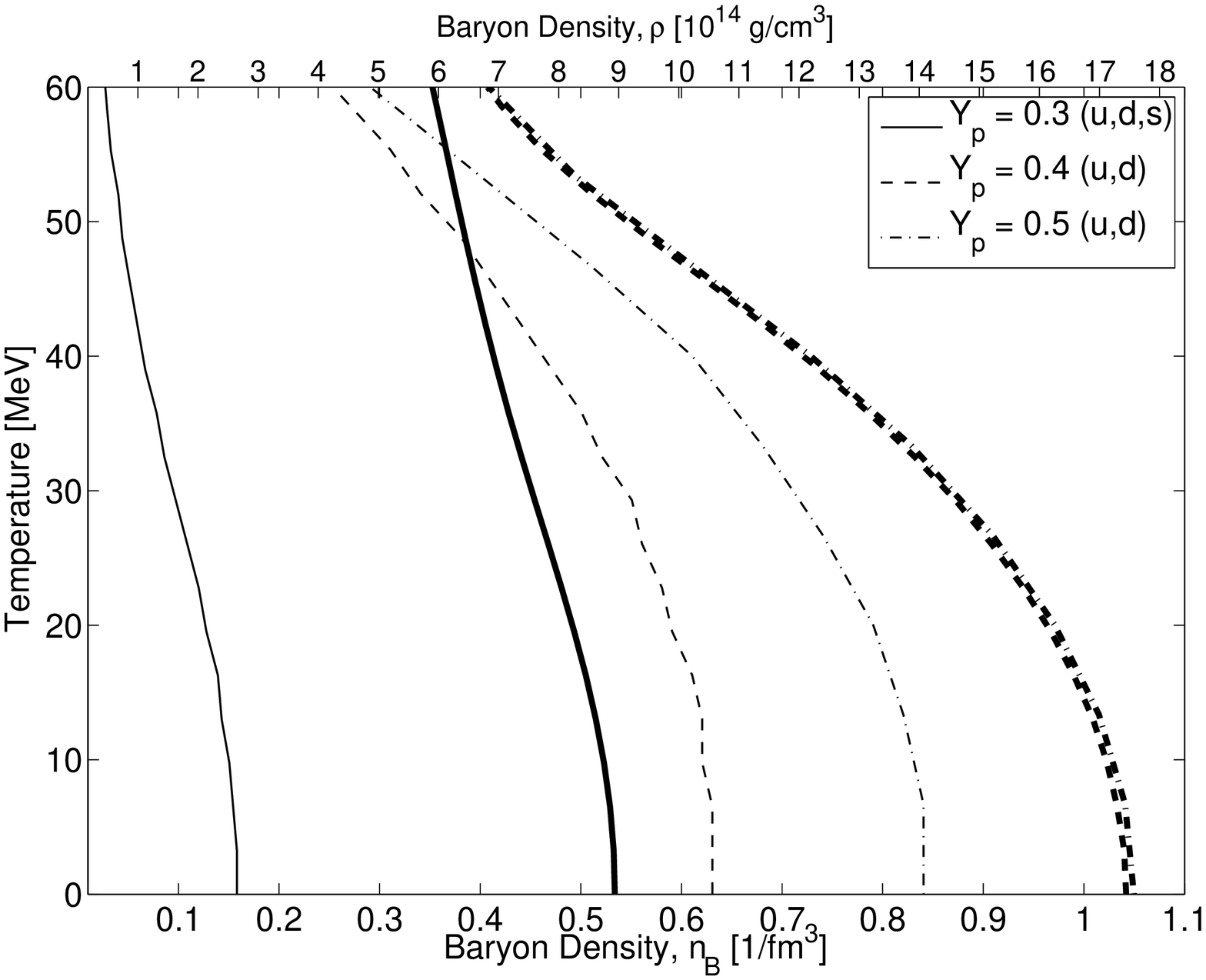}
\label{phase_diag165_ud}}
\caption{Phase diagrams for transitions from hadronic matter based on
\citet{Shen:etal:1998} (RMF, TM1) to strange quark matter with fixed local
proton fractions of $Y_p=0.1$, $0.3$, and $0.5$ and different quark EoS
parameter choices in the graphs (a)-(c).
In addition, Fig. (d) compares a phase diagram for the conditions in
supernova environments where $Y_p=0.3$ (strange quark matter) with
heavy-ion collisions where the proton fraction is close to isospin symmetry
and the phase transition takes place from hadronic to up and down quark
matter.
The thin lines show the onset of quark matter and the thick lines
the pure quark phase.}
\end{figure*}

In Fig.\ref{p_nb_beta}, the charge fractions $Y_c^H$ and $Y_c^Q$ are
plotted as functions of the baryon density for the mixed phase as well as 
the regions with pure hadronic and quark matter.
The quark EoS is calculated using $B^{1/4}=165$~MeV in
$\beta$-equilibrium at $T=0$.
As described by \citet{Glendenning_book}, the large asymmetry energy
in hadronic matter can be relieved by arranging the up, down, and strange
quark fractions in the mixed phase to negative values of $Y_c^Q$.
This is also mirrored in the behavior of the EoS which experiences
a prompt softening at the beginning of the mixed phase, seen in
Fig.~\ref{p_nb_beta}. 
With growing amount of quarks and decreasing contribution from hadronic
matter, the latter can approach its energetically more favorable
state with $Y_c^H \sim 0.5$, while the charge fraction in the quark phase
reaches the discussed small value of $Y_c^Q \sim 0$.

\begin{figure*}
\centering
\subfigure[$Y_p = 0.1$ ($B^{1/4}=165$~MeV, $\alpha_S=0$)]{
\includegraphics[width=0.48\textwidth]{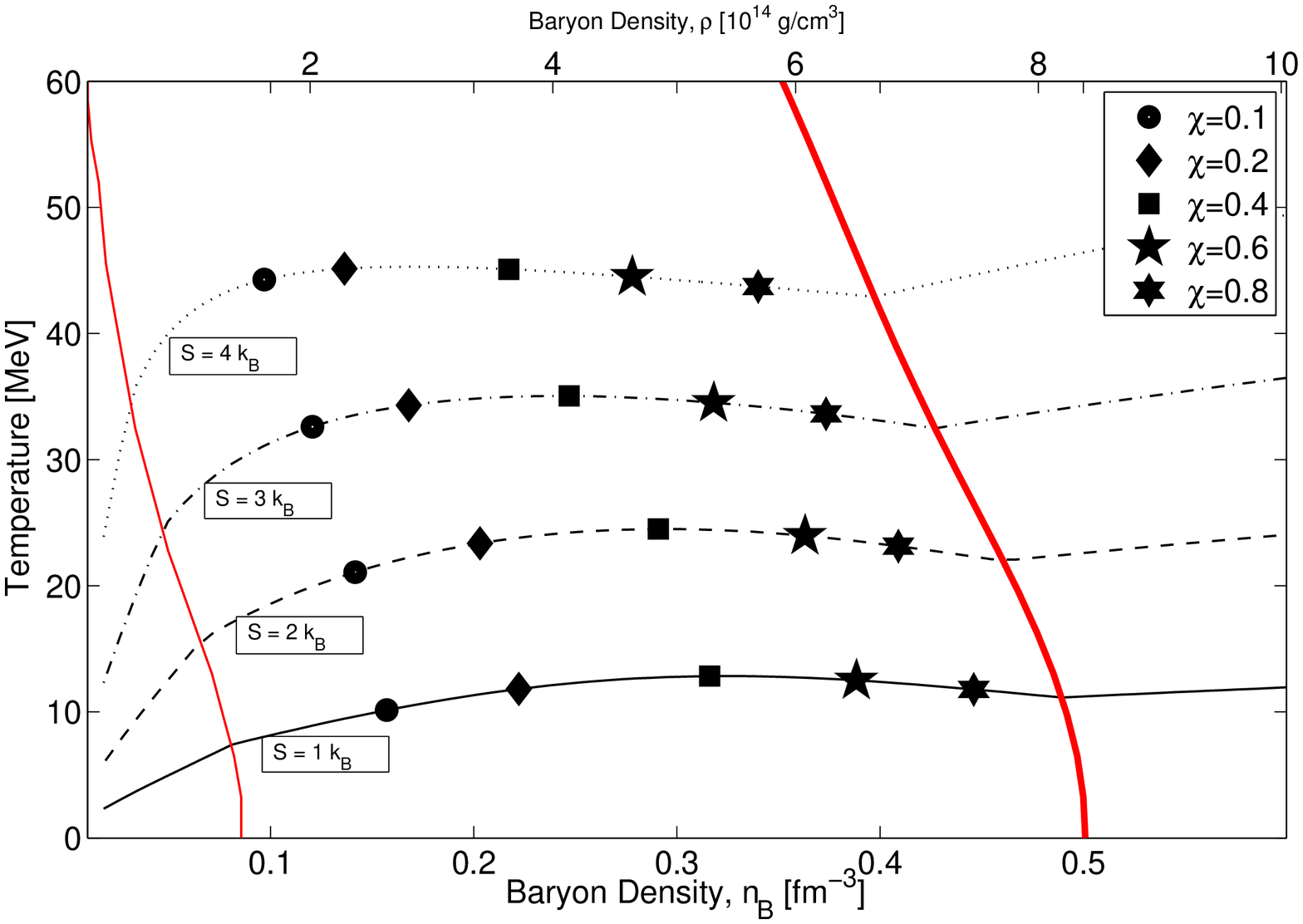}
\label{entropy_01}}
\hfill
\subfigure[$Y_p = 0.3$ ($B^{1/4}=165$~MeV, $\alpha_S=0$)]{
\includegraphics[width=0.48\textwidth]{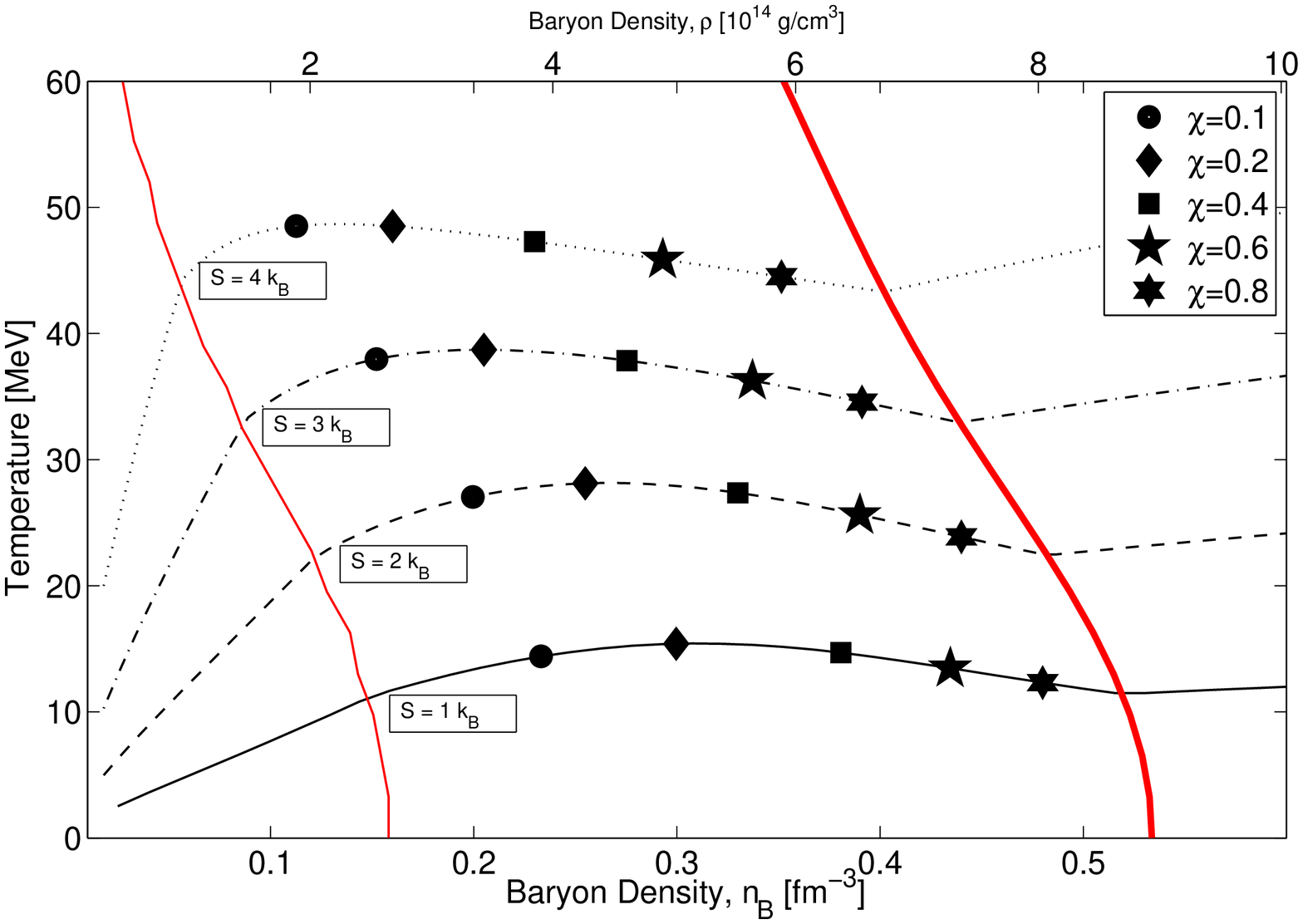}
\label{entropy_03}}
\\
\subfigure[$Y_p = 0.5$ ($B^{1/4}=165$~MeV, $\alpha_S=0$)]{
\includegraphics[width=0.48\textwidth]{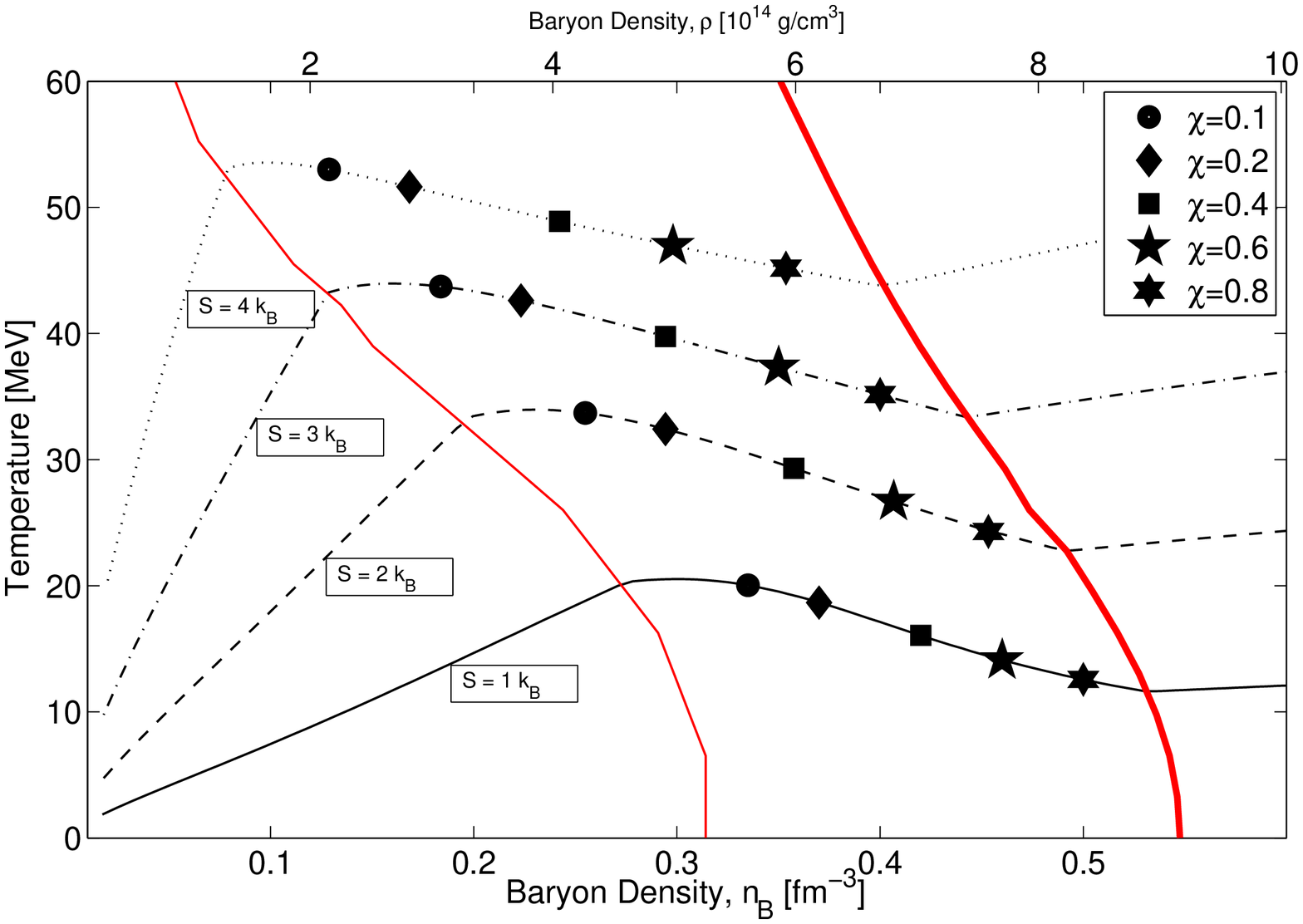}
\label{entropy_05}}\hfill
\subfigure[$Y_p = 0.3$ ($B^{1/4} = 200$~MeV, $\alpha_S = 0$)]{
\includegraphics[width=0.48\textwidth]{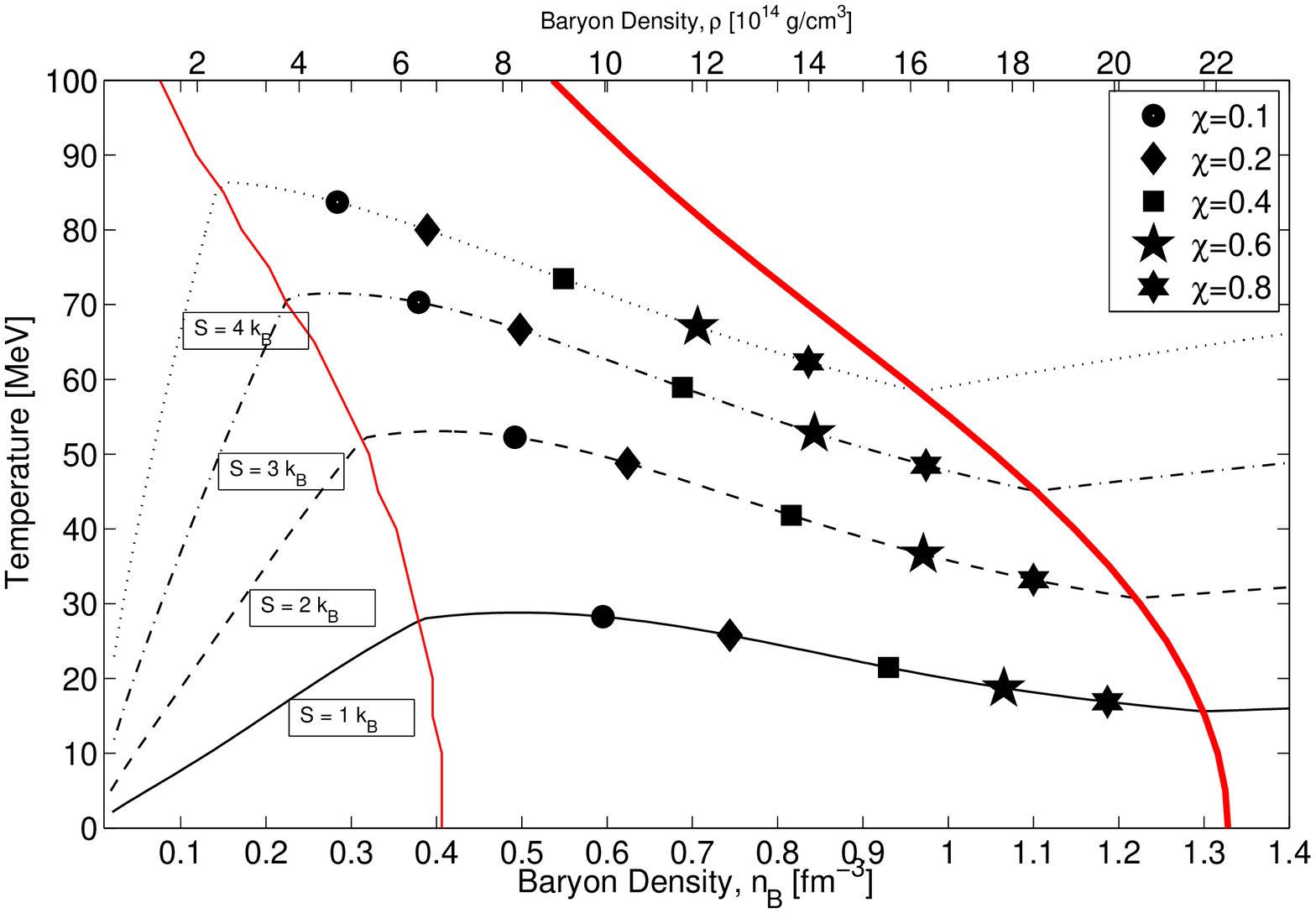}
\label{entropy_200}}
\caption{Temperature evolution in the mixed phase for the different
entropies per baryon $s=1,2,3,4$ k$_B$.
The graphs (a) - (c) show calculations for global proton fractions of 
$Y_p=0.1$, $Y_p=0.3$ and $Y_p=0.5$ with $B^{1/4}=165$~MeV.
The red solid lines show the onset of the mixed phase (thin lines)
and the beginning of the pure quark phase (thick lines) and are the
same as in Fig.~\ref{phase_diag165}.
For comparison, the graph (d) shows the temperature evolution in the mixed
phase for $Y_p=0.3$ and $B^{1/4}=200$~MeV.
}
\end{figure*}

As shown in Fig.~\ref{p_nb_yp03}, for $Y_p=0.3$ the EoS does not
experience a significant softening at the onset of the phase transition,
but stays similar to the one of pure hadronic matter.
The reason is that at $Y_p = 0.3$ nucleonic matter is already close to its
energetically favorable isospin symmetric state.
Only for larger $\chi$ the lower pressure of quark matter starts to
dominate and the EoS in the mixed phase softens.

Due to the dependence of the hadronic EoS on $Y_p$, the critical
density for the phase transition is also sensitive to the proton fraction.
The earlier onset of the mixed phase for smaller values of $Y_p$ was
already presented by \citet{Drago99} and is seen in the
Figs.~\ref{phase_diag165}, \ref{phase_diag162} and \ref{phase_diag155}
for $B^{1/4}=165$~MeV, $B^{1/4}=162$~MeV and $B^{1/4}=155$~MeV 
with $\alpha_s=0.3$, respectively \citep[see also][]{Fischer:etal:2010a}.

\begin{figure*}
\centering
\subfigure[Mass-radius relations for the hadronic star using the EoS
from \citet{Shen:etal:1998} (TM1,RMF) and hybrid stars with different
bag constants using the Gibbs construction for the phase transition.
The horizontal lines show three different observational constraints.]{
\includegraphics[width=0.48\textwidth]{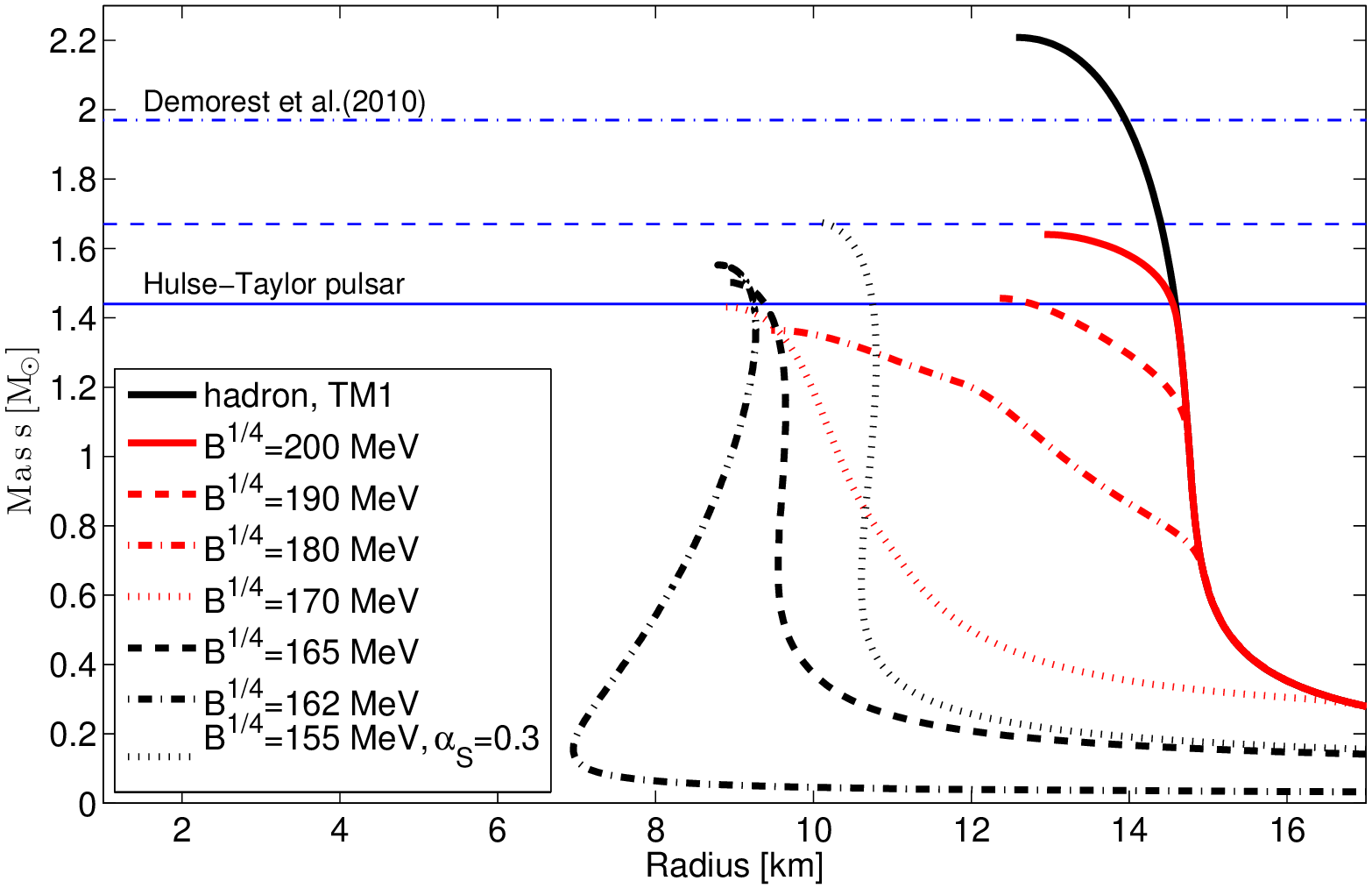}
\label{plot_mr_as03_wf}}
\hfill
\subfigure[Neutron star density profiles (thick blue lines) including the
quark volume fractions (thin red lines).]{
\includegraphics[width=0.48\textwidth]{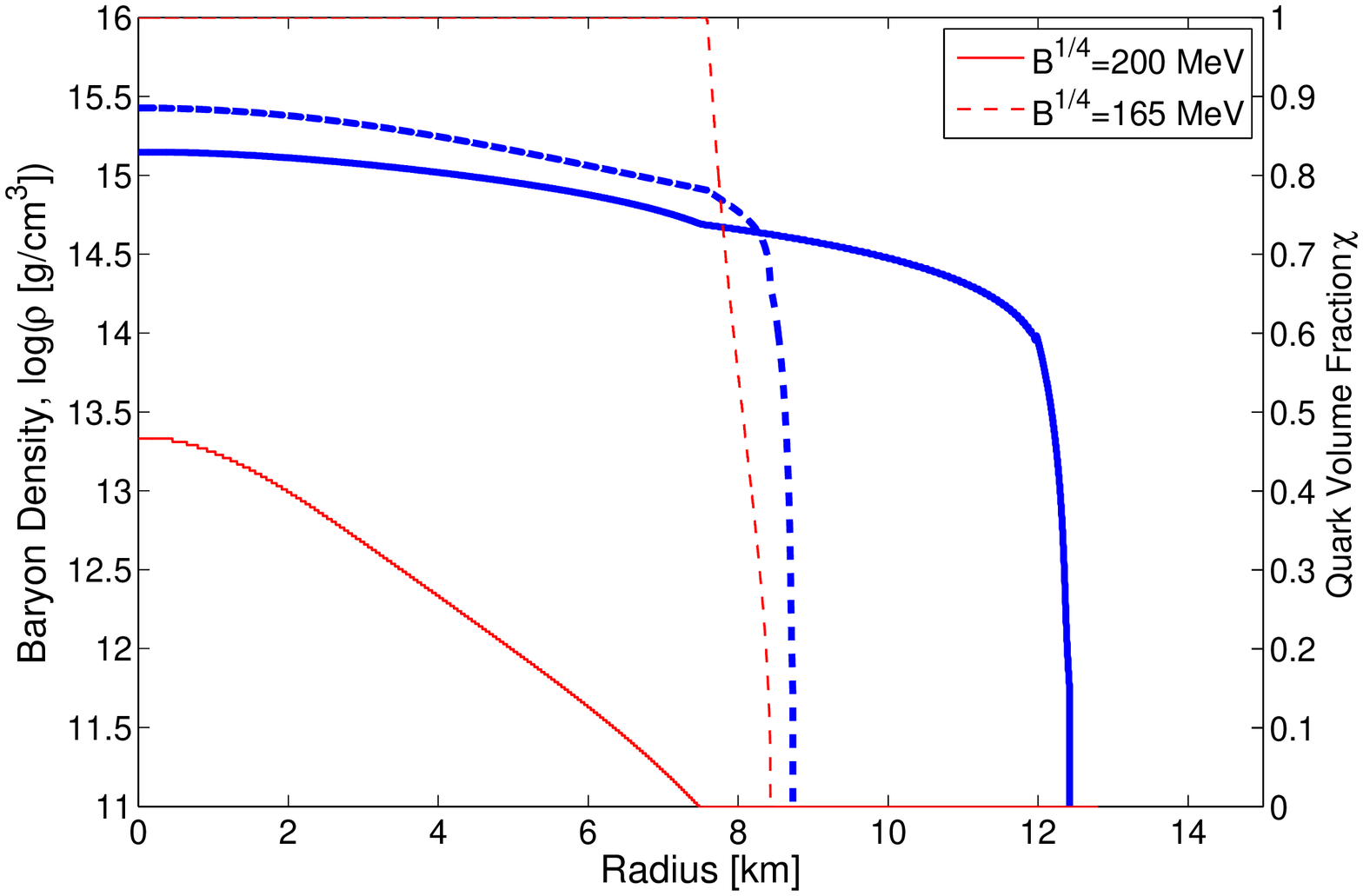}
\label{plot_profile_max}}
\caption{Different mass-radius relations, comparing different bag
constants, in graph (a) and neutron star density profiles,
for two selected bag constants, in graph (b).
The stiffening of the hybrid EoS when corrections from the strong coupling
constant $\alpha_s$ are included, leads to higher maximum masses 
of the hybrid star branches.
Small values of $B^{1/4} \lesssim 170$~MeV as well as large bag constants
of $B^{1/4} > 190$~MeV can reach the mass limit of $1.44$~M$_\odot$
(horizontal solid line in graph (a)).
Additionally,  mass limits of
1.65~M$_\odot$
\citep[PSR J1903+0327,][horizontal dashed line]{Freire:2010}
and
$1.97\pm0.04$~M$_\odot$
\citep[PSR J1614-2230,][horizontal dash-dotted line]{Demorest:etal:2010}
are also shown in graph (a).
The pure quark phase in the maximum mass configuration is large for small
bag constants, while for $B^{1/4} = 200$~MeV
quark matter is present only in form of a  mixed phase.}
\end{figure*}

Another important aspect for the phase transition is the different number
of degrees of freedom of hadronic and quark matter.
As discussed above, it is appropriate to assume the quark-hadron
phase transition to strange quark matter in supernovae.
The strange quarks are produced and equilibrated by weak
interactions in supernova  environments due to the timescales
which are on the order of milliseconds.
In heavy-ion collisions, the hot fireball of nuclear matter expands within
$10^{-23}$ seconds \citep[][]{Kolb00} and thereby prevents the
appearance of net strangeness via weak reactions.
Hence, the quark-hadron phase transition occurs only to up- and
down-quark matter in heavy-ion collision experiments.
Furthermore, for supernovae and neutron stars we can regard
phase transitions from hadronic matter to strange quark matter at
low global proton fractions $Y_p\ll0.5$.
In heavy-ion collisions, the colliding nuclei are close to isospin
symmetry with a proton fraction of $Y_p\simeq0.5$.
In the case of only two flavor quark matter, the stiffening of the quark EoS
due to the fewer degrees of freedom and the softness of hadronic matter
makes a phase transition to quark matter less favorable and shifts the
critical density to higher values.
As shown in Fig.~\ref{phase_diag165_ud}, a low critical density close to
normal nuclear matter density in supernova and neutron star environments
is compatible with a high critical density (five times nuclear matter density)
in heavy ion experiments, based on the simple bag model for the
description of quark matter.

The Figs.~\ref{entropy_01} - \ref{entropy_200} show the evolution of the
temperature in the pure and the mixed phases along different isentropes
(i.e. curves of constant entropy per baryon) for fixed proton fraction
and three-flavor quark matter.
Hereby, the Figs.~\ref{entropy_01} -- \ref{entropy_05} give the phase
diagrams for $B^{1/4} = 165$~MeV for different values  of $Y_p$ while the
Fig.~\ref{entropy_200} shows the temperature evolution for
$B^{1/4} = 200$MeV and $Y_p = 0.3$.
\cite{Steiner00} and \cite{Nakazato:etal:2010b} discuss such a
temperature evolution along isentropes for $B^{1/4} \sim 200$~MeV
at a fixed lepton fraction $Y_L$.
The authors find a temperature decrease in the mixed phase due to the
larger number of degrees of freedom and the different heat capacity for
relativistic quarks.

However, \citet{Drago99} showed that the behavior of $T$ in the
mixed phase depends on the critical density and the temperature
at the onset and at the end point of the mixed phase,
which is set by $Y_p$.
The temperature behavior based on the quark bag model can be seen
from the Figs.~\ref{entropy_01} and \ref{entropy_03} in comparison
to \ref{entropy_05} and \ref{entropy_200}.
For early phase transitions, the temperature curves rise in the beginning of
the mixed phase, while for high critical densities the temperature in the
hadronic phase has already reached large values and the onset of quark
matter causes a prompt temperature decrease along the isentropes.
These results are in qualitative agreement with the previous
studies of \citet{Steiner00} and \citet{Nakazato:etal:2010b}.

Dynamical simulations (e.g. in astrophysics and heavy-ion collisions)
that use microscopic EoSs raise the question about convexity,
in particular if phase transitions are included.
In the context of phase transitions, one very often discusses the convexity
of the thermodynamic potential
\citep[in our case the Helmholtz free-energy, see e.g.,][]{MuellerSerot:1995}:
If the thermodynamic potential is a convex function of its
arguments, the system is thermodynamically stable.
Due to the construction of the quark-hadron mixed phase and the
stability of each phase separately, our hybrid EoS is stable and thus convex.
However, another definition of convexity can be found in the literature
in the context of the hydrodynamic Riemann problem, which refers to the
curvature of the pressure as a function of density along isentropes.
This property of the EoS is particularly relevant for shock waves traversing
the coexistence region of a first order phase transition
\citep[see e.g.][]{Rischke:etal:1990,Bugaev:etal:1989}
as unusual wave patterns and effects like wave splitting can occur.
Within this definition of convexity,  \citet{Heuze:etal:2008} explored whether
finite volume methods give a reasonable description of a Riemann problem
with a non-convex EoS.
After performing such a test we can conclude that
AGILE belongs to the class of hydrodynamics codes, which show a
satisfactory agreement with the analytical solutions.
\begin{figure*}[htp!]
\centering
\subfigure[10.8~M$_\odot$ ]{
\includegraphics[width=0.485\textwidth]{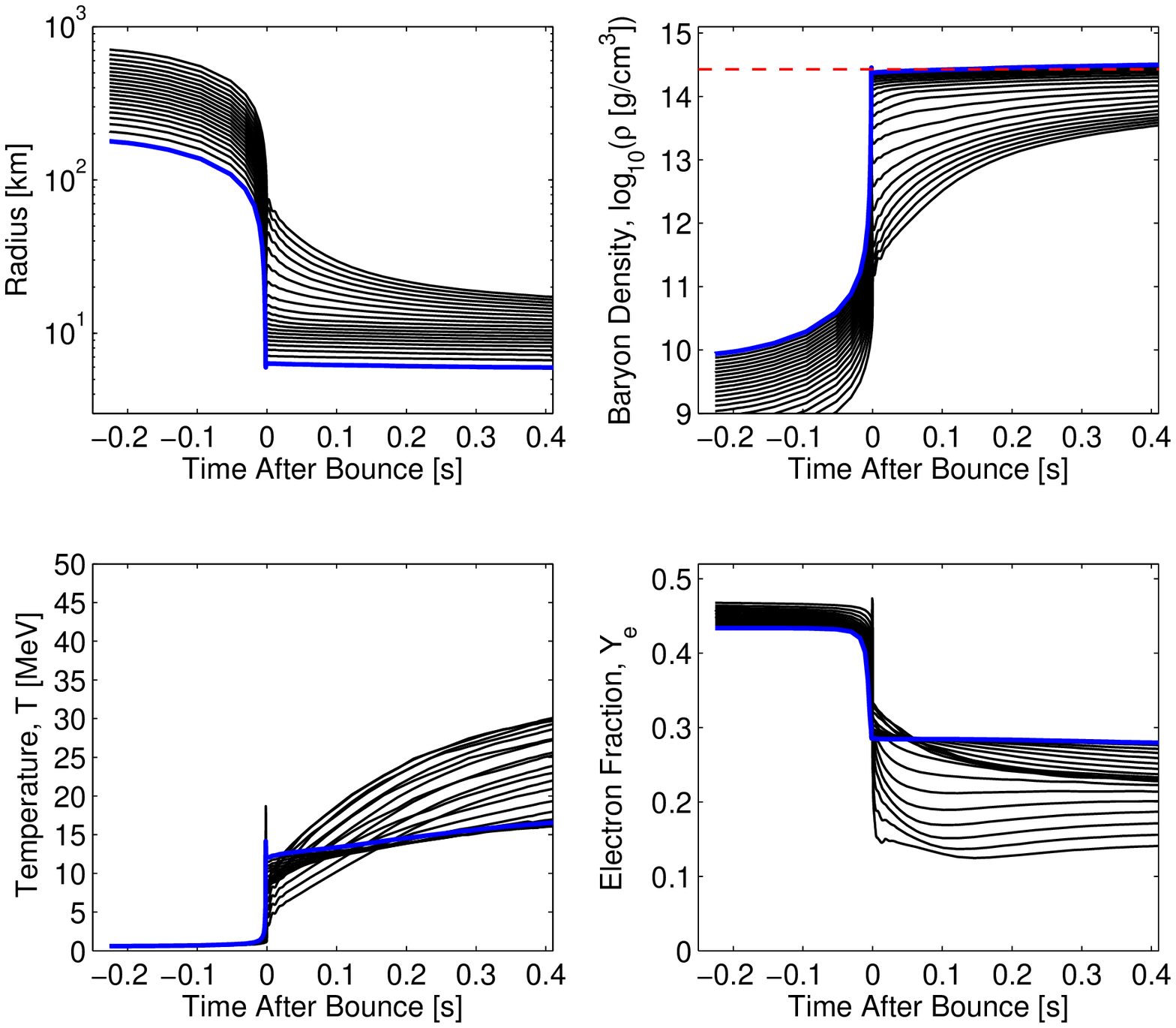}}
\hfill
\subfigure[40~M$_\odot$]{
\includegraphics[width=0.485\textwidth]{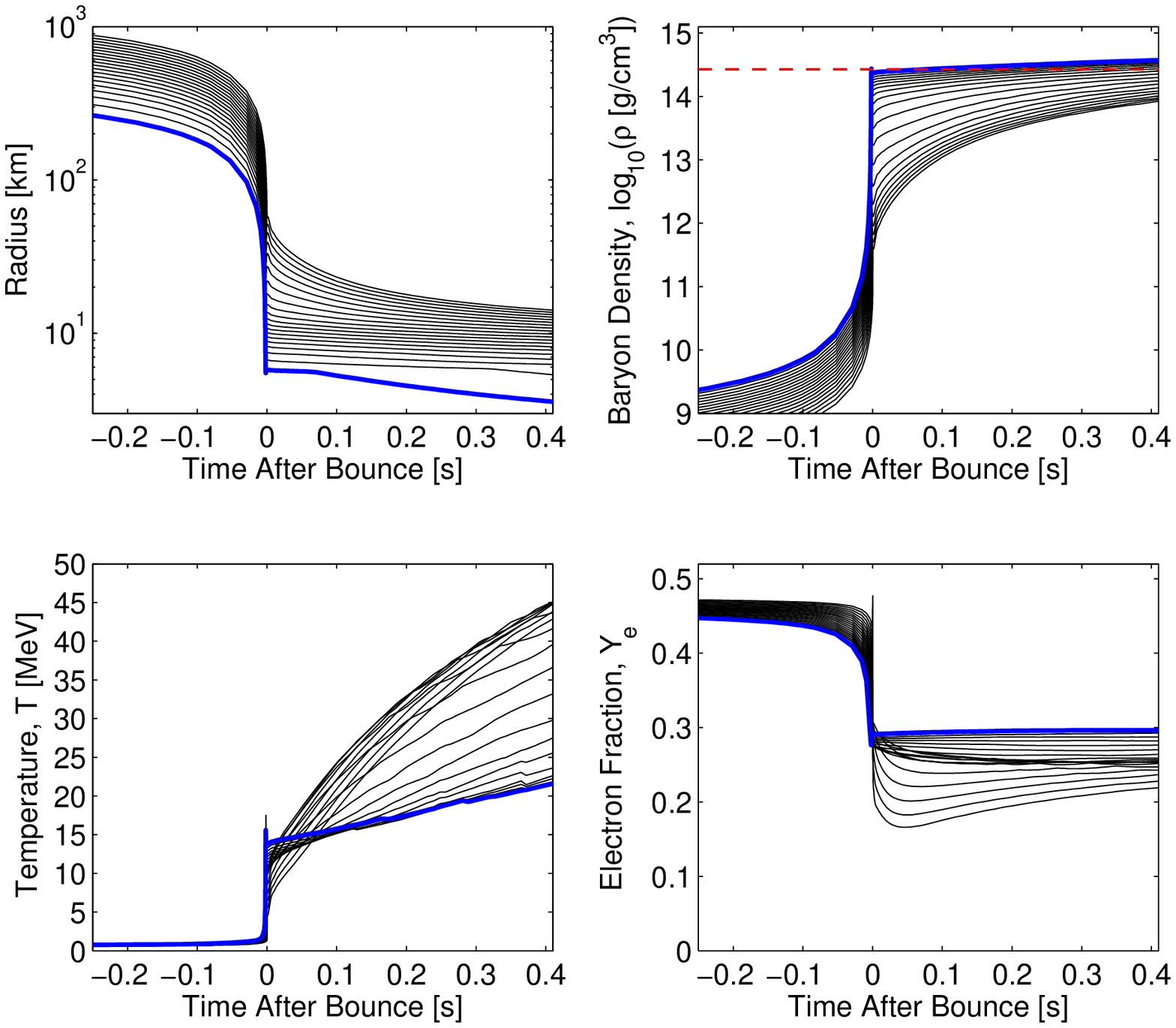}}
\caption{Evolution of selected properties of the innermost mass trajectories
in core-collapse supernova simulations of massive progenitor stars at two
examples.
The red dashed lines in the density graphs illustrate nuclear
saturation density and the solid blue lines show the central data.}
\label{fig-tracer-eos}
\end{figure*}

\subsection{Masses and Radii of Hybrid Stars}

Fig.~\ref{plot_mr_as03_wf} shows hybrid star mass-radius relations for the
hadron EoS from \citet{Shen:etal:1998} and the bag model for different
values of $B$ and $\alpha_S$.
For comparison, we indicate the mass of the Hulse-Taylor pulsar of
$M \sim1.44$~M$_\odot$
\footnote{
The Hulse-Taylor pulsar is the highest precisely known
mass of compact stars, at present.
It represents therefore a minimal limit for masses which should be
reachable with realistic nuclear EoSs.
}.
For simple quark bag models where the phase transition is constructed
applying the Gibbs conditions, only the small values of
$B^{1/4} \lesssim 170$~MeV as well as large bag constants of
$B^{1/4} > 190$~MeV can reach this limit.
However, as shown in Fig.~\ref{plot_profile_max}, hybrid stars constructed
using $B^{1/4} = 200\:$MeV, do not contain pure quark matter but only the
mixed phase with quark matter fractions of $\chi \lesssim 0.5$.
In contrast, for the small bag constant $B^{1/4} = 165$~MeV the hybrid star
at maximum mass is almost completely composed of pure quark matter,
with only a thin hadronic crust with a thickness of up to one kilometer.
As will be discussed in the following, the small value of the 
critical density and the resulting early phase transition to quark matter,
together with the stiffening of the EoS in pure quark matter, lead
to very specific dynamical consequences for the core-collapse supernova
which may not be reached for large values of $B$ due to the high critical
density and absence of the stiffening of the hybrid EoS.

Recent mass limits for physical EoSs come from the millisecond pulsars
J1903+0327 and J1614-2230 which were announced to have masses of
$M=1.67 \pm 0.01$~M$_\odot$ \citep[][]{Freire:2010} and
$M=1.97 \pm 0.04$~M$_\odot$ \citep[][]{Demorest:etal:2010},
respectively.
As shown in Fig.~\ref{plot_mr_as03_wf}, within the simple bag model such
masses can only be reached for bag constants of $B^{1/4}>200$~MeV.
Further corrections from the strong interaction coupling constant can stiffen 
the normal bag EoS and therefore lead to higher maximum masses of
compact stars \citep[see e.g.][]
{Schertler00,Alford:etal:2005,Sagert10a, Weissenborn:etal:2011}.
It can be seen from Fig.~\ref{plot_mr_as03_wf} that the chosen parameter
set $B^{1/4}=155$~MeV and $\alpha_S=0.3$ gives a maximum mass of
hybrid stars of $M_\text{max} \sim 1.67$~M$_\odot$, which would be compatible
with the observation of J1903+0327 \citep[][]{Freire:2010}.
The recent mass measurement by \citet{Demorest:etal:2010} points to an
even higher neutron star mass of $1.97 \pm 0.04$~M$_\odot$.
However, as discussed in e.g.
\citet{Kurkela10,Alford:etal:2005,Alford:etal:2007,Ozel:etal:2010},
such high values exclude neither the presence of quark
matter in compact star interiors nor low critical densities.
The stiffening effect of strong interactions on the quark EoS
can be modeled in our approach by e.g. according
choices of larger values of $\alpha_S$ and will
be tested in the future.

\subsection{Neutrino interaction treatment in quark matter}

Based on the parameters chosen for the bag model as discussed
in the previous subsections, quark matter occurs at nuclear
matter densities.
At such high densities, the neutrino mean free paths are extremely
small and neutrinos are trapped.
Hence, we always assume chemical, weak and thermal equilibrium
when determining the neutrino chemical potentials in quark matter.
%

%......................................................................
\section{Simulation results}

In this section, we report on results from core-collapse supernova
simulations in spherical symmetry where we apply the quark-hadron
hybrid EoSs introduced in \S 2.3.
The quark-hadron phase transition was found to occur during the
early post bounce phase within the first $500$~ms for all cases
considered in this investigation.
This is the central property of the quark matter description applied
here due to the selective choice of parameters, which lead to
the onset of the quark-hadron phase transition
close to nuclear saturation density.
Consequences with respect to the dynamical evolution will be
further discussed below.
The simulations are launched from low  and intermediate mass
Fe-core progenitors of 10.8, 13 and 15~M$_\odot$ from
\citet{Woosley:etal:2002}.
First results from the 10.8 and the 15~M$_\odot$ simulations
have already been discussed in \citet{Sagert:etal:2009}
and \citet{Fischer:etal:2010a}.
Here, we extend their analysis and present a deeper investigation
with respect to the appearance of quark matter in PNS interiors
and discuss consequences with respect to the neutrino signal emitted.
\begin{figure*}
\centering
\subfigure[At Bounce]{
\includegraphics[width=0.32\textwidth]{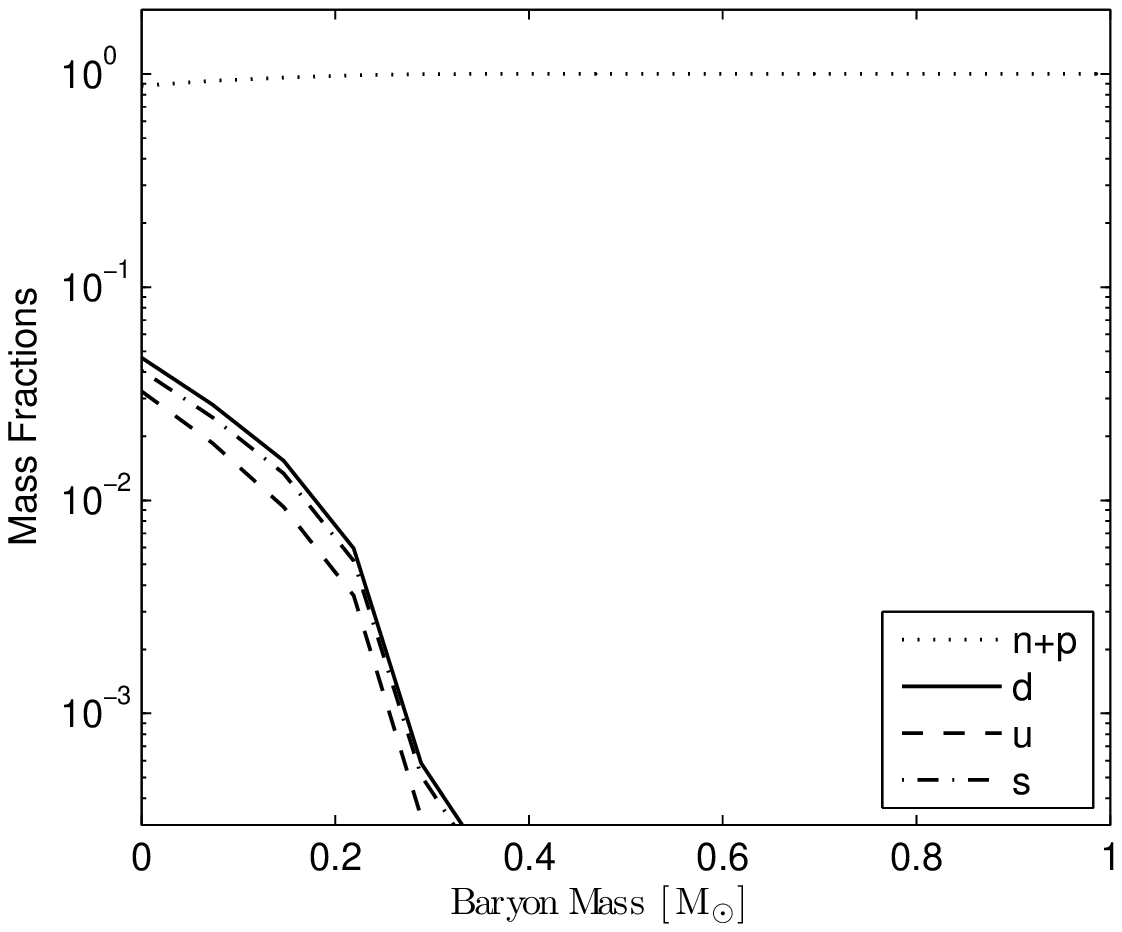}}
\hfill
\subfigure[100~ms post bounce]{
\includegraphics[width=0.32\textwidth]{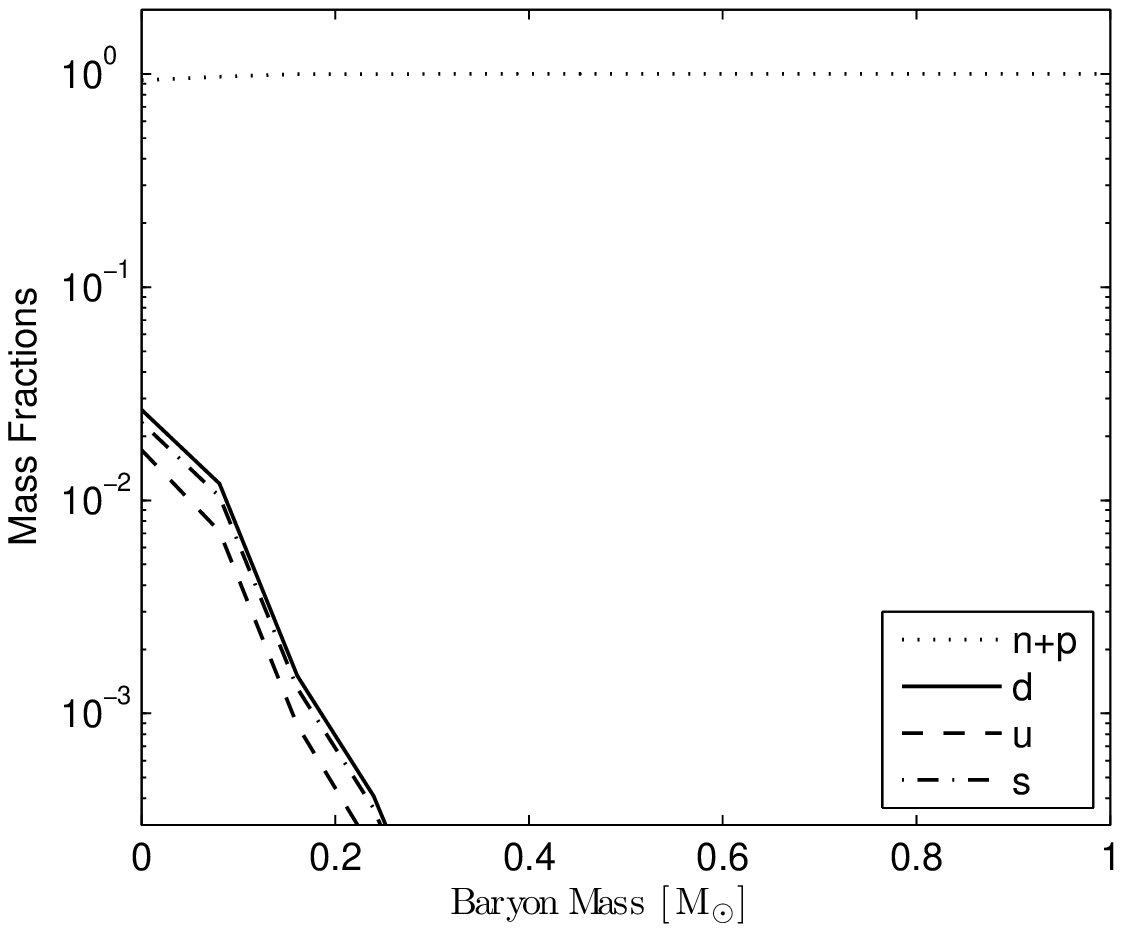}}
\hfill
\subfigure[400~ms post bounce]{
\includegraphics[width=0.32\textwidth]{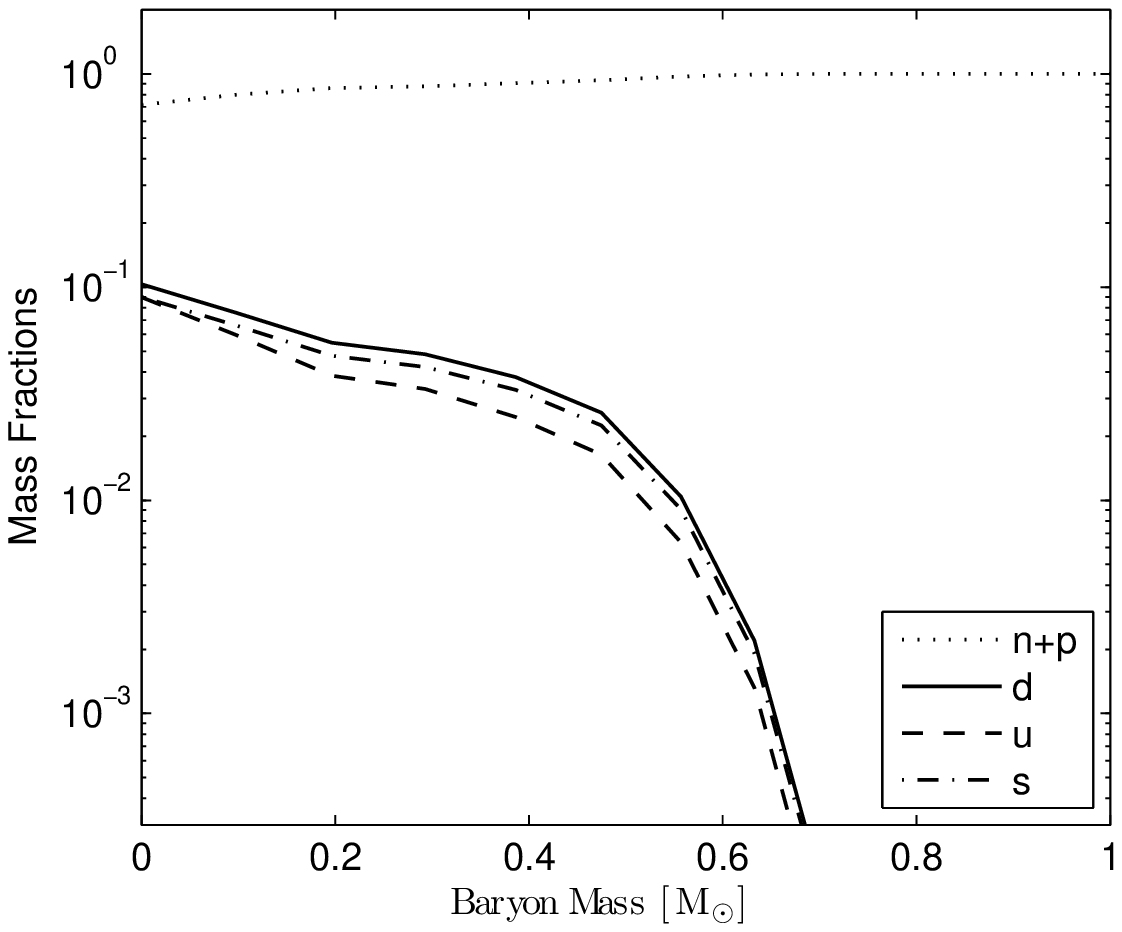}}
\caption{Evolution of the mass fractions of nucleons and
the different quark flavors at bounce (left panel) and at the two selected
post bounce times 100~ms (middle panel) and 400~ms (right panel)
for the 10.8~M$_\odot$ reference model.}
\label{fig-composition-h10n-postbounce}
\end{figure*}
\begin{figure}
\centering
\includegraphics[width=.48\textwidth]{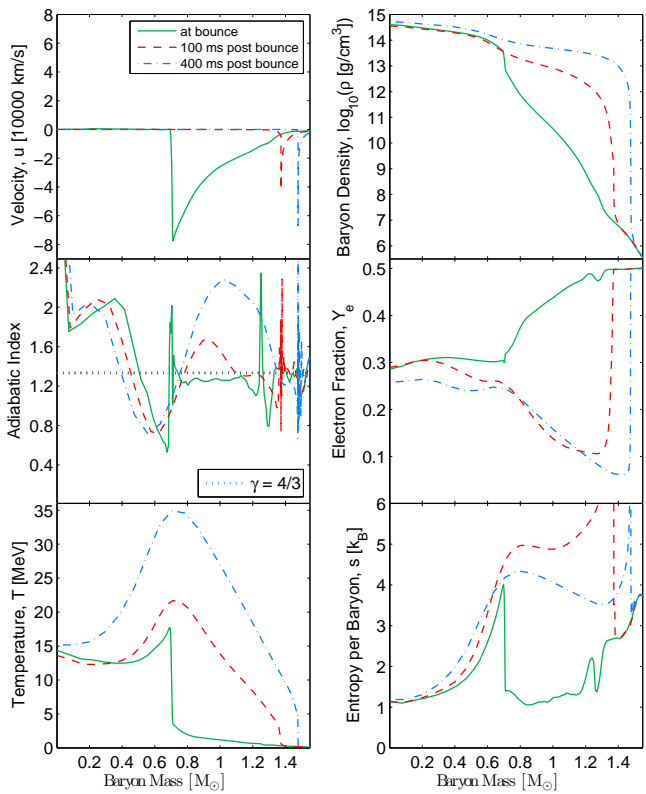}
\caption{Radial profiles of selected hydrodynamic quantities
during the appearance of quark matter for the 10.8~M$_\odot$
reference model at bounce (solid lines) and at the two selected post
bounce times 100~ms (dashed lines) and 400~ms (dash-dotted lines).
The horizontal dotted line in the entropy graph illustrates the critical
adiabatic index of $\gamma=4/3$.}
\label{fig-fullstatehydro-h10n-postbounce}
\end{figure}

\subsection{The early post bounce phase}

%.the standard thermodynamic conditions obtained post bounce........
The early post bounce phase of core-collapse supernovae of massive
Fe-core progenitors is determined by an extended mass accretion period,
during which the central density and temperature increase above several
times nuclear density and above several tens of~MeV.
This behavior is shown in Fig.~\ref{fig-tracer-eos} for the example of
the low-mass 10.8~M$_\odot$ and the high-mass 40~M$_\odot$
progenitor models, both using the (pure hadronic) EoS from
\citet{Shen:etal:1998}.
Illustrated is the time evolution of selected mass trajectories,
which belong to the PNS interiors, during the Fe-core collapse
(see the radial evolution), bounce and the first $400$~ms post bounce.
Both models reach several times nuclear saturation density
(red dashed lines in the density graphs) at bounce (i.e. at $t=0$).
They proceed in a similar fashion during the post bounce
compression after the expanding bounce shock has come to halt.
The central densities increase slowly above nuclear saturation
density and reach $3.36\times10^{14}$ g/cm$^3$ (0.201~fm$^{-3}$)
for the 10.8~M$_\odot$ progenitor model and
$3.99\times10^{14}$ g/cm$^3$ ($0.238$ fm$^{-3}$)
for the $40$~M$_\odot$ progenitor model within the first
$500$~ms after bounce.
Furthermore, the central electron fraction reaches low values,
between $Y_e\simeq0.28$ (at the center) and $Y_e\simeq0.1$
near the neutrinospheres at intermediate densities around
$10^{11}-10^{13}$ g/cm$^3$.
This highly deleptonized region originates from the launch of the
deleptonization burst at about $5$~ms post bounce (dominated by $\nu_e$).
Differences between these two progenitor models occur in the
temperatures obtained during the first $400$~ms post bounce evolution.
The 10.8~M$_\odot$ progenitor model in Fig.~\ref{fig-tracer-eos} (a)
reaches temperatures of $T=15-35$~MeV and the $40$~M$_\odot$
model in Fig.~\ref{fig-tracer-eos} (b) between $20-60$~MeV.
This effect is due to the more compact PNS obtained at bounce
for the $40$~M$_\odot$ progenitor model and the larger mass accretion
rate at the PNS surface.

\subsection{The quark-hadron phase transition}

%.appearance of quark matter post bounce.......................................................
We apply the hybrid EoS introduced in \S~2.3 to core-collapse supernova
simulations of massive Fe-core progenitor stars in spherical symmetry,
where no explosions are obtained using standard hadronic EoSs
from e.g. \citet{LattimerSwesty:1991} and \citet{Shen:etal:1998}.
In the following paragraphs, we illustrate the dynamical evolution
obtained during the quark-hadron phase transition.
We select as standard reference model, the 10.8~M$_\odot$
progenitor model applying EOS2.

For this choice of parameters, quark matter appears at densities
close to nuclear saturation density, where the critical density depends
strongly on the temperature and the electron fraction obtained
(see the discussion in \S~2.3).
The conditions permit quark matter already at core bounce, as
illustrated in Fig.~\ref{fig-composition-h10n-postbounce}
(left panel).
The corresponding hydrodynamic properties obtained are
illustrated in Fig.~\ref{fig-fullstatehydro-h10n-postbounce}.
Quark matter appears at the center ($\leq0.3$~M$_\odot$),
at a mass fraction of about $\chi$=0.01--0.1
(see the radial bounce profiles in
Fig.~\ref{fig-composition-h10n-postbounce} (left panel) and
Fig.~\ref{fig-fullstatehydro-h10n-postbounce}
for baryon density, electron fraction and temperature).
The figures also show that down-quarks are the most abundant species,
followed by strange-quarks and up-quarks, which can be understood
as a consequence of charge neutrality and the low electron fraction
at the center.
This hierarchy remains during the later post bounce evolution.
The small central fraction of quark matter reduces
during the early post bounce expansion
(see Fig.~\ref{fig-composition-h10n-postbounce} (middle panel) and
the radial profiles at bounce and at 100~ms
post bounce in Fig.~\ref{fig-fullstatehydro-h10n-postbounce}).
Only after the dynamic bounce shock has stalled and the subsequent
evolution is determined by mass accretion onto the PNS and heating
behind the standing accretion shock (SAS), does the central density
increase and the central electron fraction decrease.
This in turn causes the quark matter volume fraction to rise
(see Fig.~\ref{fig-composition-h10n-postbounce} (right panel) and
Fig.~\ref{fig-fullstatehydro-h10n-postbounce} at 400~ms post bounce).
The timescale for the quark matter fraction increase is given by the
central density and temperature increase as well as by the
electron fraction decrease.
Hence it is determined by the
mass accretion rate, which in turn depends on the progenitor model
and the hadronic EoS (before the EoS is dominated by the quark
contributions).
For the massive Fe-core progenitors under investigation using the
stiff EoS from \citet{Shen:etal:1998}, the timescale is on the order of
100~ms, where a central density of $5.2\times10^{14}$~g/cm$^3$
and a temperature of 15~MeV are obtained
(see Fig.~\ref{fig-fullstatehydro-h10n-postbounce}).
During the post bounce evolution, the adiabatic index $\gamma$
decreases slowly below the critical value of 4/3 on the same
timescale, illustrated in Fig.~\ref{fig-fullstatehydro-h10n-postbounce}.
Furthermore, the appearance of quark matter at the PNS interior
proceeds adiabatically as shown via the entropy per baryon
profiles in Fig.~\ref{fig-fullstatehydro-h10n-postbounce}.
The PNS configuration is gravitationally stable as long as the
maximum mass (given by the hybrid EoS) is not reached.

\begin{figure}
\centering
\includegraphics[width=0.49\textwidth]{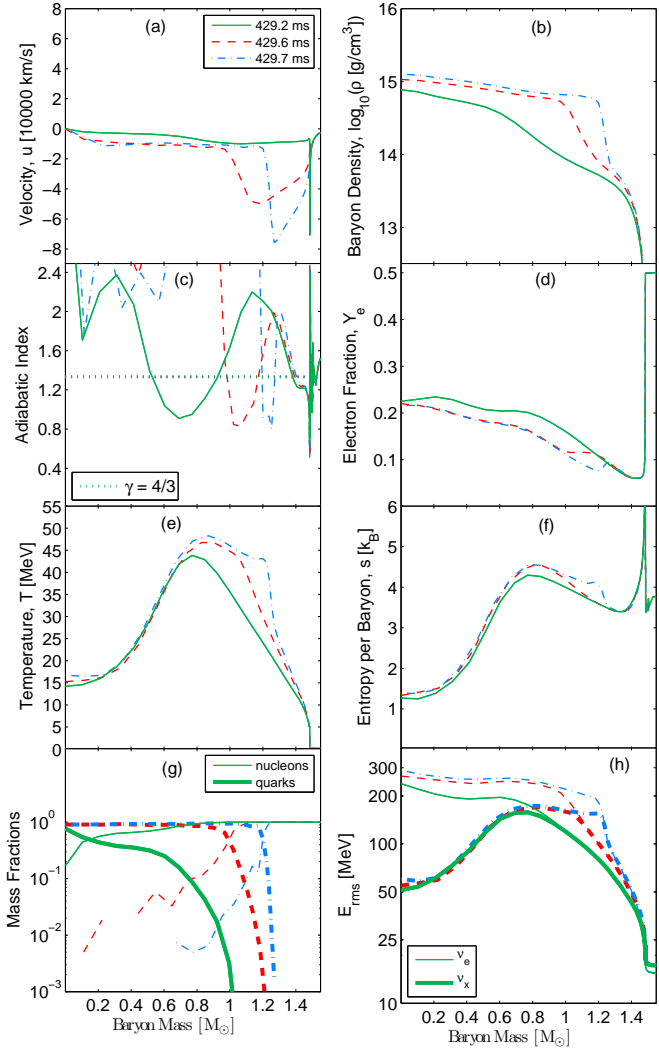}
\caption{PNS collapse of the 10.8~M$_\odot$ progenitor model
due to the presence of quark matter, at three different post bounce times
(solid lines: $429.2$~ms, dashed lines: $429.6$~ms, dash-dotted lines:
$429.7$~ms)
after the PNS has become gravitationally unstable.
The same configuration as Fig.~\ref{fig-composition-h10n-postbounce}
but in addition graphs (g) and (h) show the mass fractions
of nucleons (thick red lines) and quarks (thin blue lines)
and the mean energies for $\nu_e$ (thin lines) and for
$(\mu/\tau)$-(anti)neutrinos (thick lines), respectively.}
\label{fig-composition-h10n-collapse}
\end{figure}
\begin{figure*}
\centering
\subfigure[The initial explosion phase.]{
\includegraphics[width=.9\textwidth]{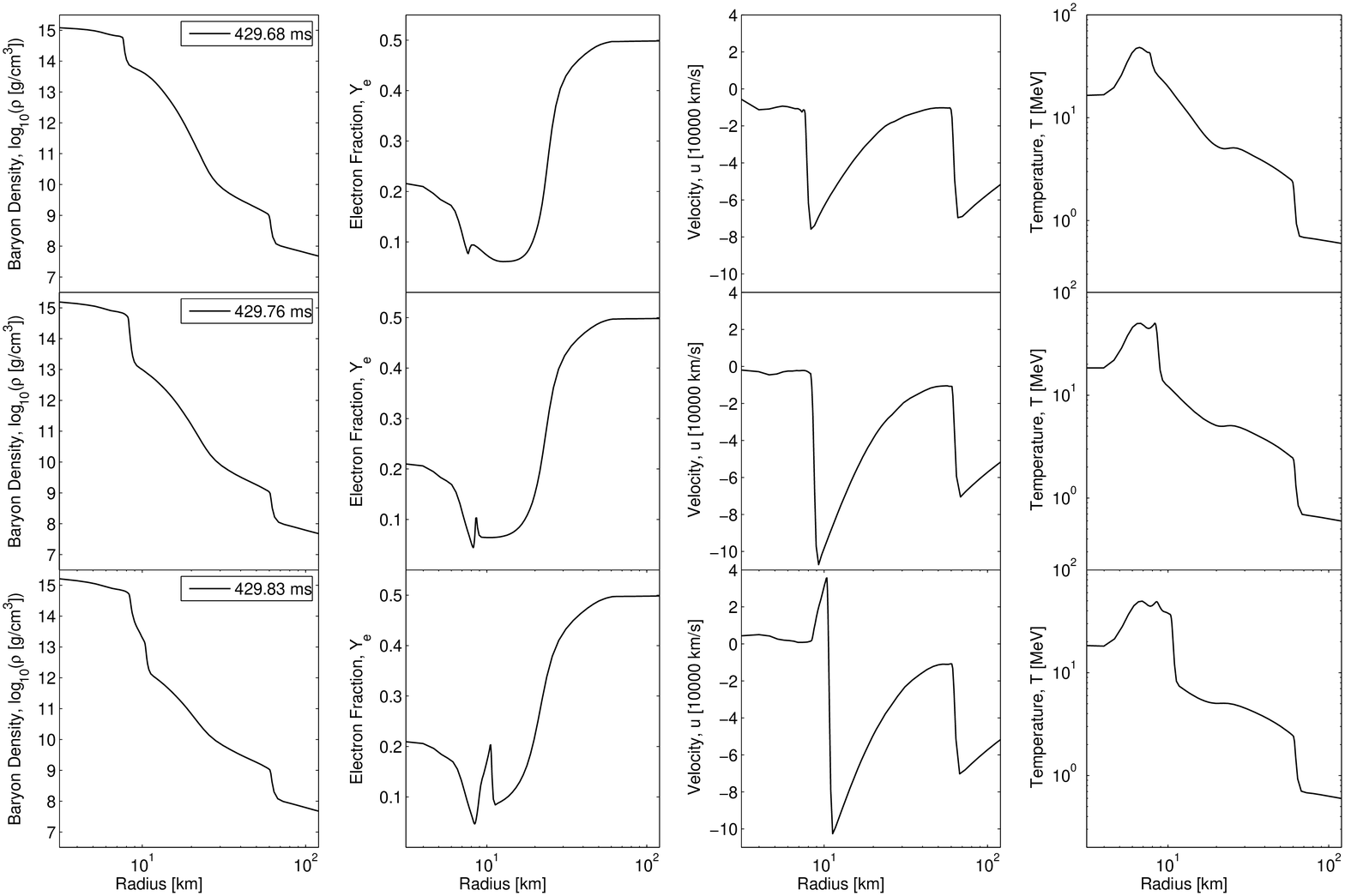}
\label{fig-composition-h10n-expl-a}}
\\
\subfigure[The shock expansion along the decreasing density at the PNS
surface.]{
\includegraphics[width=.9\textwidth]{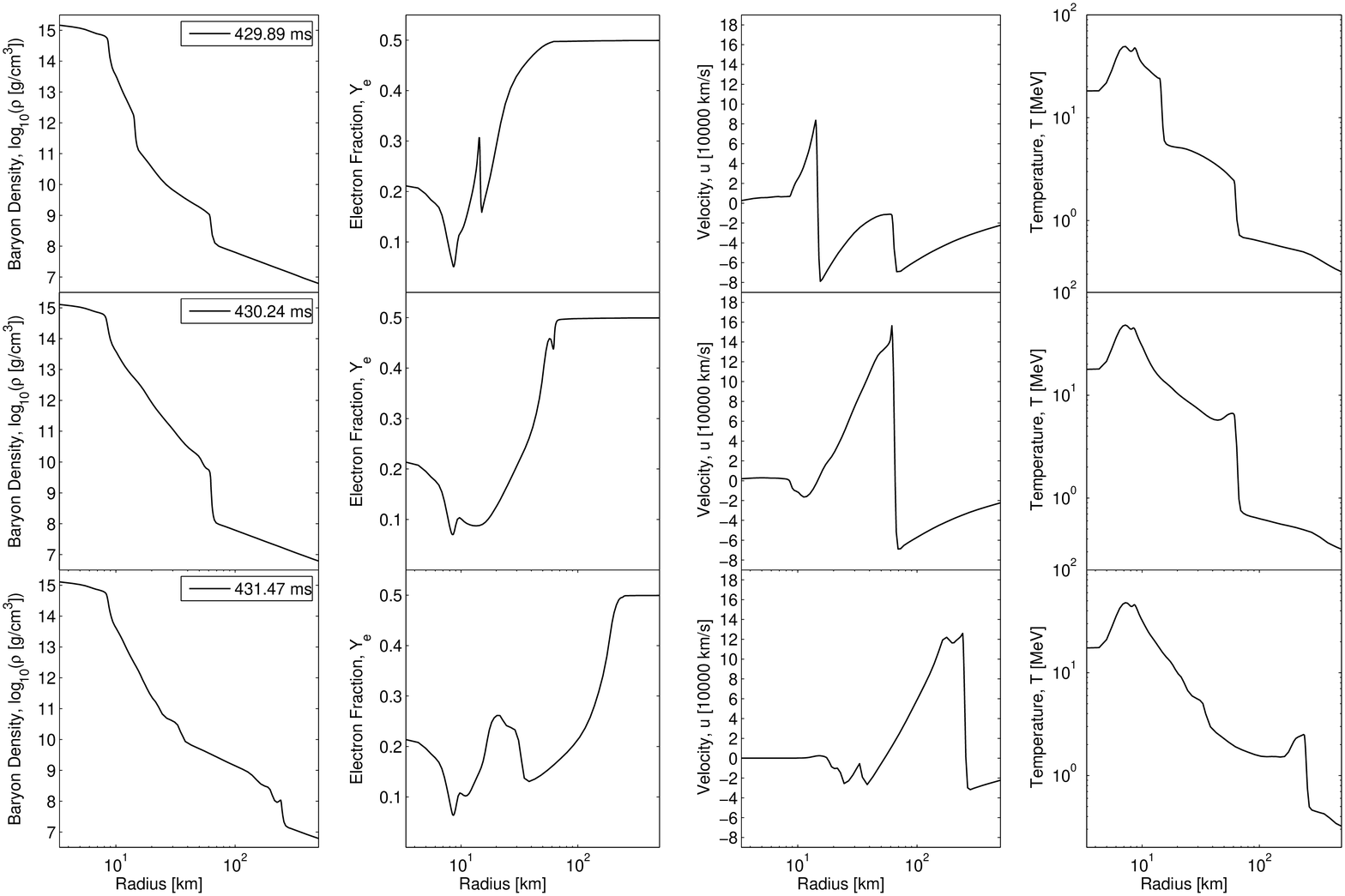}
\label{fig-composition-h10n-expl-b}}
\caption{Radial profiles of selected hydrodynamic variables during the
explosion of the 10.8~M$_\odot$ reference model due to the quark
hadron phase transition, at selected post bounce times.}
\end{figure*}
%

%.PNS collapse..................................................................................................
A quark matter volume fraction between $0$ and $1$ translates to matter
in the quark-hadron mixed phase, for which the adiabatic index
is reduced (see \S2.3).
This behavior is illustrated in Fig.~\ref{fig-composition-h10n-collapse}
(c) at selected post bounce times, after the PNS exceeds its maximum
stable mass and starts to collapse.
The last stable configuration (it is not illustrated for simplicity)
corresponds to the post bounce time of $428.5357$~ms.
The consequent softening of the EoS for matter in the mixed phase
leads to a dramatic dynamical evolution.
The reduced adiabatic index causes the mixed phase of the
PNS to contract, as shown via the radial velocity profiles in
Fig.~\ref{fig-composition-h10n-collapse} (a)
during the collapse at three selected post bounce times.
During the contraction, the central density and temperature increase
(see Fig.~\ref{fig-composition-h10n-collapse} (b) and (e)).
Furthermore, weak-equilibrium is established at a lower value of the
electron fraction due to the changed thermodynamic conditions
(see Fig.~\ref{fig-composition-h10n-collapse} (d)).
The contraction accelerates and proceeds into an adiabatic
collapse, during which the central entropy per baryon stays
constant as shown in Fig.~\ref{fig-composition-h10n-collapse} (f).
The central part of the PNS collapses sub-sonically and the outer
part collapses super-sonically (similar to the Fe-core collapse).
The timescale reduces from $100$~ms to milliseconds.
The compression results in significantly higher densities and
temperatures, which in turn favors quark matter over hadronic matter.
In this sense, the quark-hadron phase transition which started
slowly on timescale of $100$~ms proceeds rapidly during the
PNS collapse until the pure quark matter phase at the PNS center
grows and a quark core of about $1.3$~M$_\odot$ is formed
(see the composition in Fig.~\ref{fig-composition-h10n-collapse} (g)).
The quark fractions of up-, down- and strange-quarks rise
equally whereas in the pure quark-matter phase down-quarks and
strange-quarks are favored over up-quarks
(see Fig.~\ref{fig-composition-h10n-postbounce}).
This is due to the fact that the chemical potentials are equal
for s-quarks and d-quarks (see \S~2.3) and matter is neutron-rich.
The different symmetry energy for quark matter, together with the
higher densities and temperatures obtained during the PNS
collapse, result in a different weak-equilibrium where
$Y_e$ is generally lower.
The lower $Y_e$ obtained (see Fig.~\ref{fig-composition-h10n-collapse}
(d)) between $0.8-1.3$~M$_\odot$ reduces the pressure of the
degenerate electron gas and hence softens the EoS additionally,
which in turn supports the collapse.
In addition, the mean neutrino energies are also shifted to higher
values, where the electron flavor neutrinos are most sensitive
to density variations.
As shown in Fig.~\ref{fig-composition-h10n-collapse} (h),
the central mean energies of the electron neutrinos increase from
about $250$~MeV before the PNS collapse to about $300$~MeV.
The largest increase of the mean electron neutrino energy was found
for the infalling material where density and temperature increase most,
from $10^{14}$ g/cm$^3$ to $10^{15}$ g/cm$^3$ and from
15~MeV to $45$~MeV
(see Fig~\ref{fig-composition-h10n-collapse} (b) and (e)).
There, the mean neutrino energy increases from about $75$~MeV
to about $200$~MeV for $\nu_e$ and to about $150$~MeV
for the $(\nu_{\mu/\tau},\bar{\nu}_{\mu/\tau})$.
A less pronounced increase of the mean neutrino energies
can be found for the $(\mu/\tau)$-(anti)neutrinos, which rise from
about $50$~MeV to about $60$~MeV at the center.
The mean free paths of all neutrinos are extremely small,
on the order of few $10-1000$ cm.
Hence the neutrinos are highly trapped and cannot escape
on the timescales on the order of milliseconds up to hundreds of
milliseconds that are found during the post bounce PNS evolution.

%.PNS interior and shock formation.......................................................................
As discussed in \S 2.3, the adiabatic index rises again in the
pure quark phase.
This behavior is illustrated in Fig.~\ref{fig-composition-h10n-collapse}
(c) during the PNS collapse.
The obtained quark core and the consequent stiffening of the EoS
for matter in the pure quark phase at the PNS interior,
causes the PNS collapse to halt.
A strong hydrodynamic shock forms, which can be identified in
the radial velocity profile at $t=429.6$~ms post bounce in
Fig.~\ref{fig-composition-h10n-collapse} (a) and the
entropy increase in Fig.~\ref{fig-composition-h10n-collapse} (f).
The shock wave forms at about $1.25$~M$_\odot$.
%.not a second bounce
The system does not overshoot its hydrostatic equilibrium configuration,
because quark stars (or hybrid stars with extended strange quark cores)
are energetically self bound objects, the total internal energy becomes
larger than the gravitational binding energy, at the densities considered here
\citep[for a detailed discussion, see e.g. \S~18 in][]{Weber:1999}.
Hence, the scenario cannot be considered a second bounce.
Furthermore, when the shock reaches the sonic point
it still remains a pure accretion front with no matter outflow,
as shown in the Figs.~\ref{fig-composition-h10n-collapse}
and \ref{fig-composition-h10n-expl-a}.

%.Why does the shock front propagate outwards?
The shock forms due to the stiffening of the EoS in the pure quark
phase and due to the supersonically infalling outer part, which does not
know about the halted collapse at the center.
Information about the central stiffening cannot propagate
outward across the sonic point.
Initially, the second shock wave appears as a standing accretion
front with no matter outflow
(see Fig.~\ref{fig-composition-h10n-collapse} (a)).
The shock evolution is given by the balance of ram-pressure
ahead of the shock from the infalling nucleons and the thermal
pressure from the dissociated quarks behind the shock.
The shock position determines the phase boundary between the
quark-hadron mixed and the pure hadronic phases, due to the large
density and temperature jumps at the shock front.
The dissociated nuclear matter, which accumulates onto the
PNS surface, has been shock heated at the SAS previously.
As this dissociated material crosses through the second accretion
shock, it is converted from hadronic matter to quark matter.
%Differently to the scenario after the Fe-core bounce, this process
%liberates energy and causes a significant rise of the internal energy
%and thermal pressure behind the second accretion shock.
For all our models, the second accretion shock was found to
propagate outward towards the PNS surface
(see the radial velocity profiles in Fig.~\ref{fig-composition-h10n-expl-a}
(a) at the example of the 10.8~M$_\odot$ progenitor reference model).
Thereby, the outwards propagating shock wave remains
a pure accretion front with no matter outflow, as shown in
Fig.~\ref{fig-composition-h10n-collapse} (a)
between $429.6$~ms and $429.7$~ms post bounce and in
Fig.~\ref{fig-composition-h10n-expl-a}
at $429.68$~ms and $429.83$~ms.

\begin{figure*}
\centering
\subfigure[At 429.68~ms, 429.76~ms and 429.83~ms post bounce
(from the top to the bottom), which correspond to the post bounce times
shown in Fig.~\ref{fig-composition-h10n-expl-a}.]{
\includegraphics[width=0.42\textwidth]{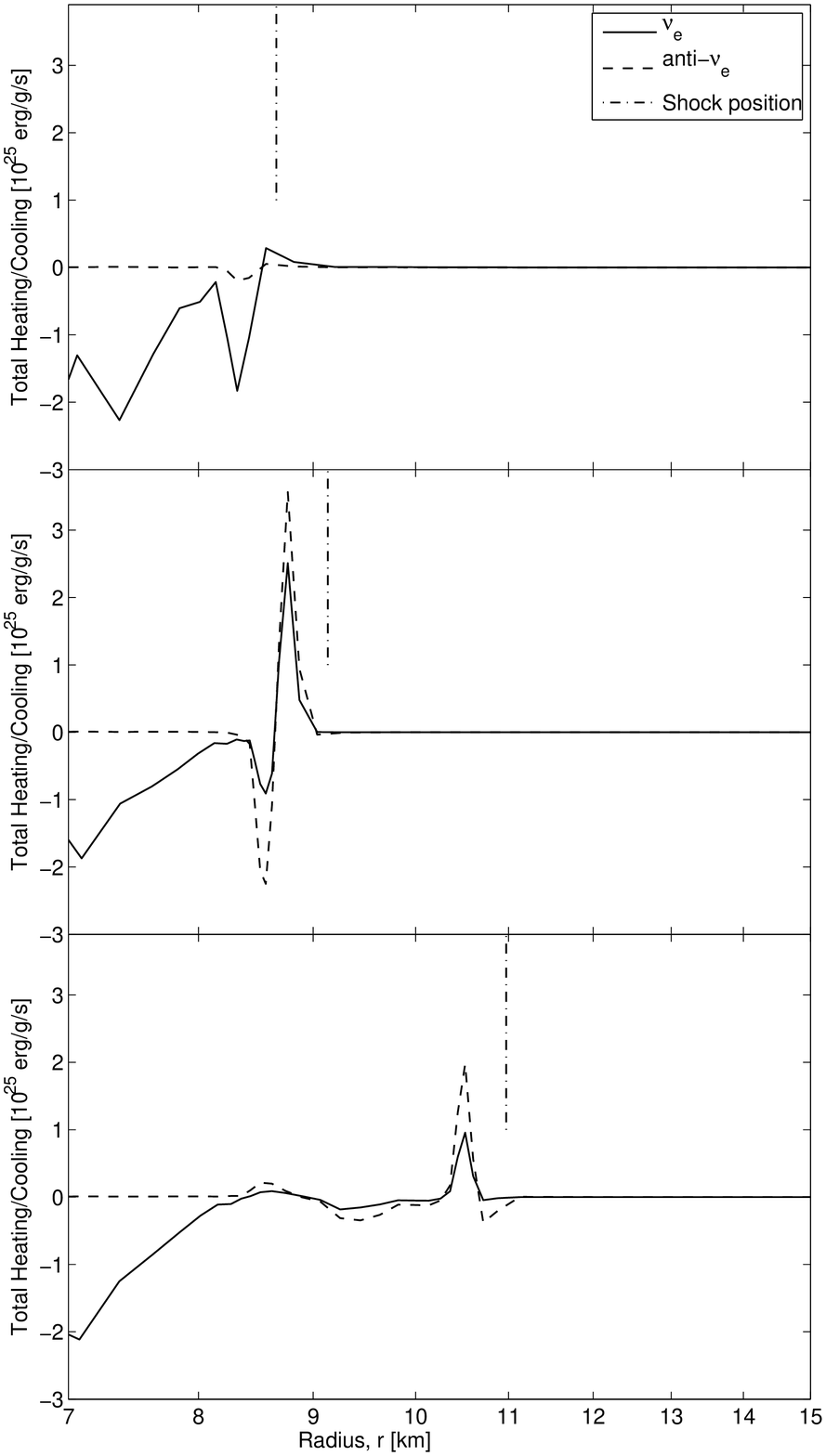}
\label{fig-heatplot-h10n-a}}
\hspace{10mm}
\subfigure[At 429.89~ms, 430.24~ms and 431.47~ms post bounce
(from the top to the bottom), which correspond to the post bounce times
shown in Fig.~\ref{fig-composition-h10n-expl-b}.]{
\includegraphics[width=0.42\textwidth]{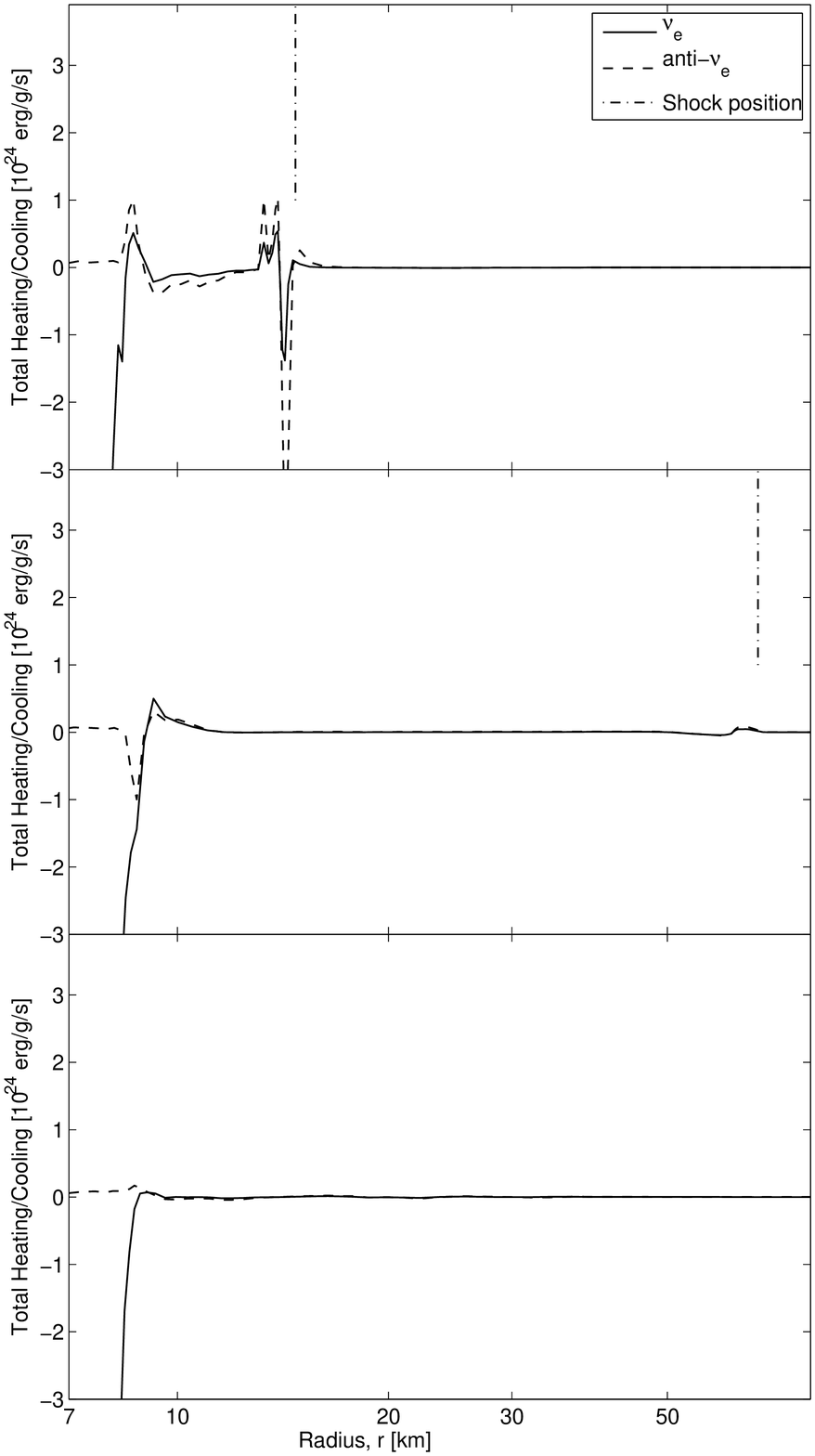}
\label{fig-heatplot-h10n-b}}
\caption{Total net heating/cooling rates for $\nu_e$ (solid lines)
and $\bar{\nu}_e$ (dashes lines) at selected post bounce times.
Negative (positive) values correspond to a net cooling (heating).
The vertical dash-dotted lines show the position of the
expanding explosion shock.}
\end{figure*}

%.explosion scenario.........................................................
The density and temperature jumps at the shock increase,
as long as it remains a pure accretion front.
Mass from the outer gravitationally unstable part of the PNS
continues to fall onto the shock front, increasing the
infall velocities at the shock
(see the velocities in the middle panel of
Fig.~\ref{fig-composition-h10n-expl-a}).
The material accumulates onto the central quark core.
The increasing thermal pressure behind the standing shock front
drives the shock slowly outwards, supported by the increasing
heating right behind the standing accretion shock illustrated in
Fig.~\ref{fig-heatplot-h10n-a}.
In the presence of the largely enhanced $\bar{\nu}_e$-luminosity,
$\bar{\nu}_e$-absorption at protons dominates over $\nu_e$-absorption
at neutrons, even though the conditions are generally neutron-rich.
The sharp rise of the heating rates just behind the shock corresponds to
the increasing $Y_e$ illustrated in the Fig.~\ref{fig-composition-h10n-expl-a}.
Behind the heating region, small (in comparison to the heating at
the shock) cooling rates dominate
(see Fig.~\ref{fig-heatplot-h10n-a}).

As the density of the infalling material decreases it becomes too
low for hadronic matter to be converted into quark matter.
At the phase boundary between the quark-hadron mixed phase and
the pure hadronic phase, the density decreases from about
$10^{15}$ g/cm$^3$ to about $10^{14}$ g/cm$^3$
(see the lower panel of Fig.~\ref{fig-composition-h10n-expl-a}).
Eventually, the shock wave detaches from the surface of the PNS
and accelerates.
Positive matter velocities are obtained, where where the accretion
front turns into a dynamic shock wave.
Because the density at the PNS surface decreases also at larger radii,
from $10^{14}$ g/cm$^3$ at the phase boundary between the mixed
and the hadronic phases to $10^{10}$ g/cm$^3$ at the low density
envelope, the shock continues to accelerate
(see Fig.~\ref{fig-composition-h10n-expl-b}).
Velocities on the order of $1-1.5\times10^{5}$ km/s are obtained
during the shock passage across the low density envelope.
The scenario is again different from the early shock propagation
of the bounce shock, where the shock suffers immediately after its
formation from the dissociation of infalling heavy nuclei which causes
an energy deficit of about $8$~MeV per baryon.
Here, the infalling matter is already dissociated and composed
of only free nucleons and light nuclei.
In this sense, the shock does not loose energy during the
initial propagation.
Even more, during the initial shock expansion neutrino heating still
deposits energy behind the shock wave
(see
Fig.~\ref{fig-heatplot-h10n-a} and compare with
Fig.~\ref{fig-composition-h10n-expl-a}
as well as
Fig.~\ref{fig-heatplot-h10n-b} and compare with
Fig.~\ref{fig-composition-h10n-expl-b}).

The expanding dynamic shock wave finally merges with the SAS
from the Fe-core bounce (see
Fig.~\ref{fig-composition-h10n-expl-b} at $t=430.24$~ms post bounce),
which remained unaffected from the events inside the PNS
at a radius of about $80$ km.
Maximum matter outflow velocities of $1.6\times10^{5}$ km/s 
are obtained.
The initial shock expansion slows down when the dynamic
shock wave reaches infalling heavy nuclei from the outer layers after
merging with the standing accretion shock from core bounce.
The matter velocities decrease continuously during the later evolution
to typically $4-6\times 10^4$ km/s
(depending on the progenitor model).
However, for all models under investigation the expanding shock
wave was never found to stall again at later times.
Hence, the quark-hadron phase transition triggers not only the
formation of a strong additional shock wave but also launches
an explosion in core-collapse supernova simulations where otherwise
no explosions could be obtained.

\subsection{The neutrino observables}

First, we describe the standard neutrino signal obtained from
a failed core-collapse supernova explosion of 10.8~M$_\odot$
using the pure hadronic EoS from \citet{Shen:etal:1998}.
Below, we will compare these neutrino spectra with
the spectra obtained taking QCD degrees of freedom into account.

\begin{figure*}
\centering
\subfigure[The standard case using the hadronic EoS from
\citet{Shen:etal:1998}.]{
\includegraphics[width=1.\textwidth]{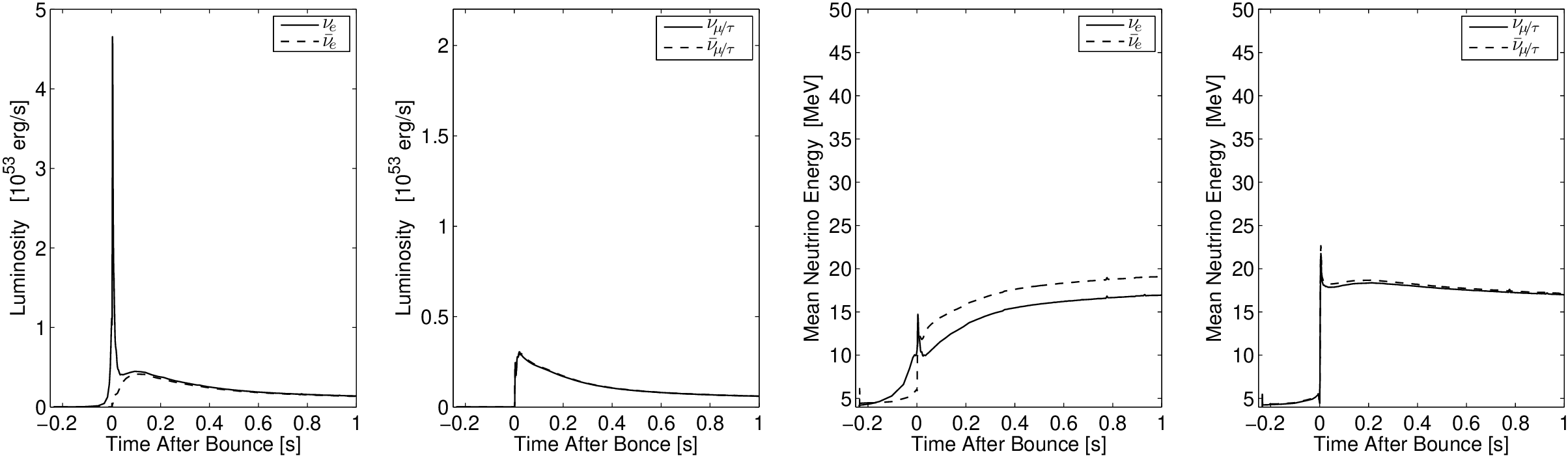}}\\
\subfigure[Simulation using the hybrid EOS2.]{
\includegraphics[width=1.\textwidth]{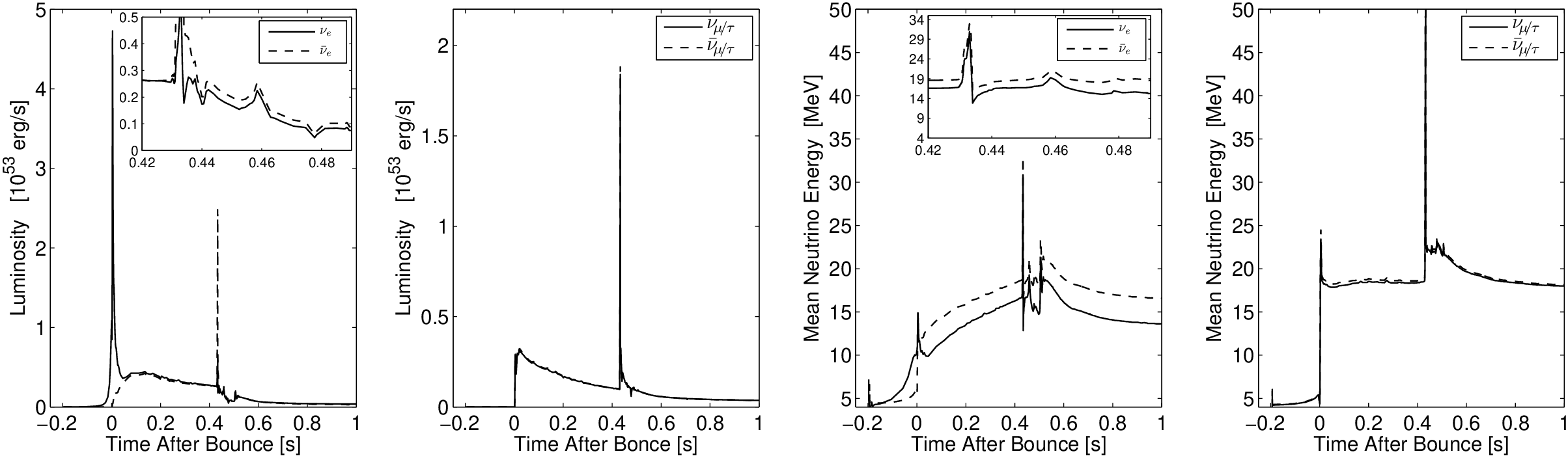}}
\caption{Evolution of the neutrino luminosities and mean neutrino
energies for the 10.8~M$_\odot$ progenitor model with respect
to time after bounce, comparing the standard hadronic EoS from
\citet{Shen:etal:1998} and the hybrid EOS2.
The data are measured at a distance of $500$ km in the
co-moving reference frame.}
\label{fig-lumin-h10n}
\end{figure*}

The evolution of the neutrino luminosities and root-mean-square energies
are shown in Fig.~\ref{fig-lumin-h10n} (a) for the 10.8~M$_\odot$
progenitor model using the standard hadronic description of nuclear
matter, measured at a distance of $500$ km in the co-moving reference
frame
\citep[for a definition of the neutrino observables, see][]{Fischer:etal:2010b}.
During the deleptonization burst at bounce, the electron neutrino
luminosity rises up to several $10^{53}$~erg/s on a timescale of
$5-20$~ms post bounce.
$\bar{\nu}_e$ are produced only after bounce due to the
decreased degeneracy which allows, next to electrons, also for
positrons and the corresponding charged current reactions
as well as pair-processes.
The same holds for $(\mu/\tau)$-neutrinos, which are produced
only via pair-processes after bounce.
The early (on timescale of $100$~ms post bounce) evolution of the
neutrino spectra is given by the balance of mass accretion and diffusion
at the neutrinospheres.
After the deleptonization burst has been launched, the electron-flavor
neutrino luminosities increase slightly until about $150$~ms post bounce,
due to the moderately large mass accretion rate.
The $(\mu/\tau)$-neutrino luminosities decrease constantly with time
because their spectra are given by diffusion rather than by accretion.
The electron flavor luminosities follow the same decreasing behavior
after the mass accretion rate at the neutrinospheres drops below a certain
threshold (depending on the progenitor model).
At $400$~ms post bounce, the electron-flavor ($(\mu/\tau)$-neutrino)
luminosities reach values of about $5\times10^{52}$ ($3\times10^{52}$)
erg/s.
The mean energies of the electron-flavor neutrinos increase continuously
with respect to time after bounce, from $10$ (15)~MeV after the
deleptonization burst has been launched to $14$ ($17$)~MeV
at $400$~ms post bounce.
The mean energies of the $(\mu,\tau)$-(anti)neutrinos decrease slightly
with time after bounce and reach about  $19$~MeV at $400$~ms
(see Fig.~\ref{fig-lumin-h10n} (a)).

The evolution of the neutrino observables from the simulations applying
the hybrid EoS introduced in \S2.3,
is shown in Fig.~\ref{fig-lumin-h10n} (b) for the 10.8~M$_\odot$
progenitor reference model.
Compared to the standard hadronic scenario, we find very similar
spectra for post bounce times before the PNS collapse
(see Fig.~\ref{fig-lumin-h10n} (b)).
Differences occur in slightly higher neutrino luminosities, which are
due to the more compact PNS where quark matter is present.

\begin{figure}
\subfigure[The early shock propagation.]{
\includegraphics[width=0.49\textwidth]{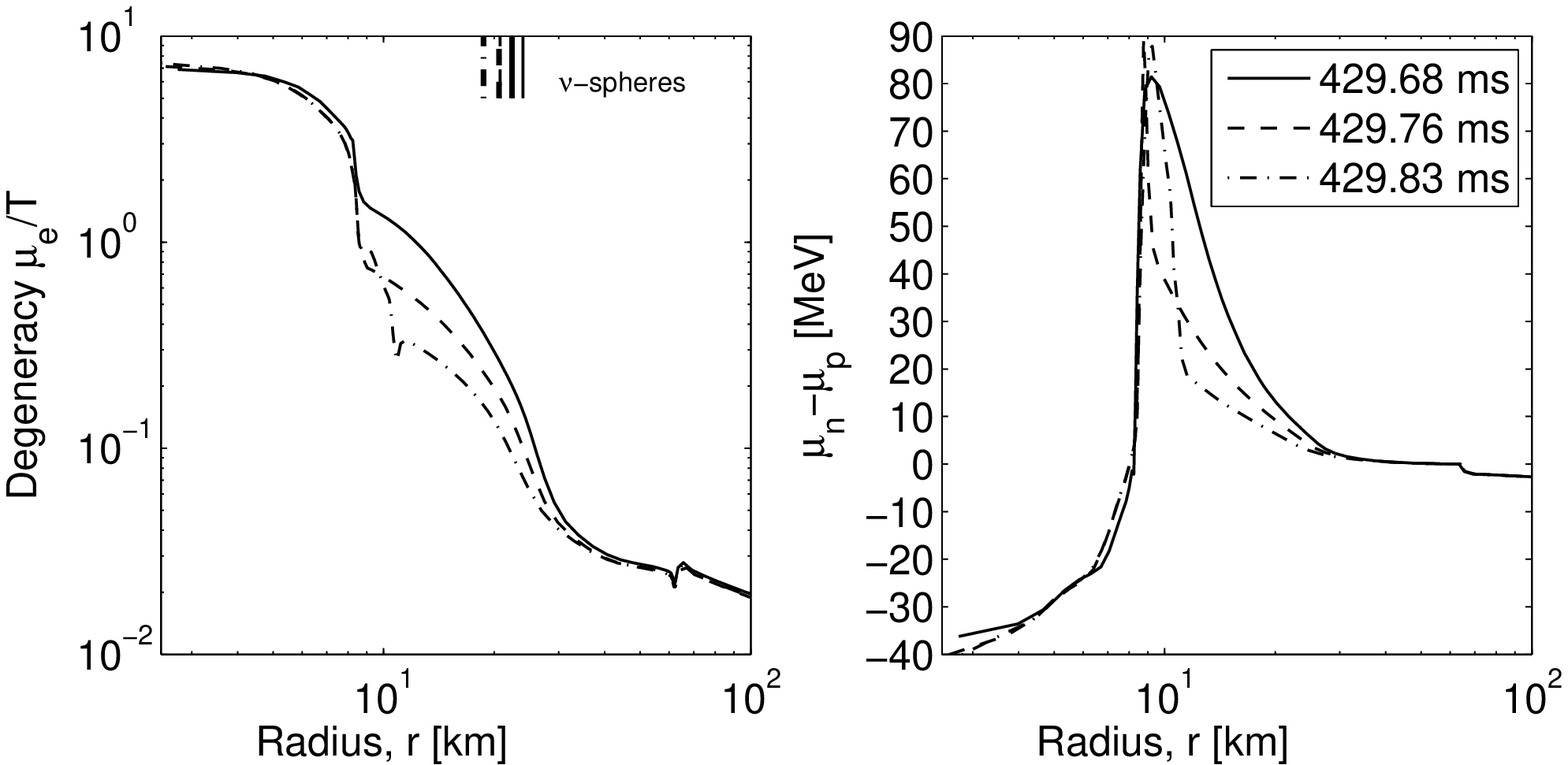}}
\\
\subfigure[The ongoing explosion.]{
\includegraphics[width=0.49\textwidth]{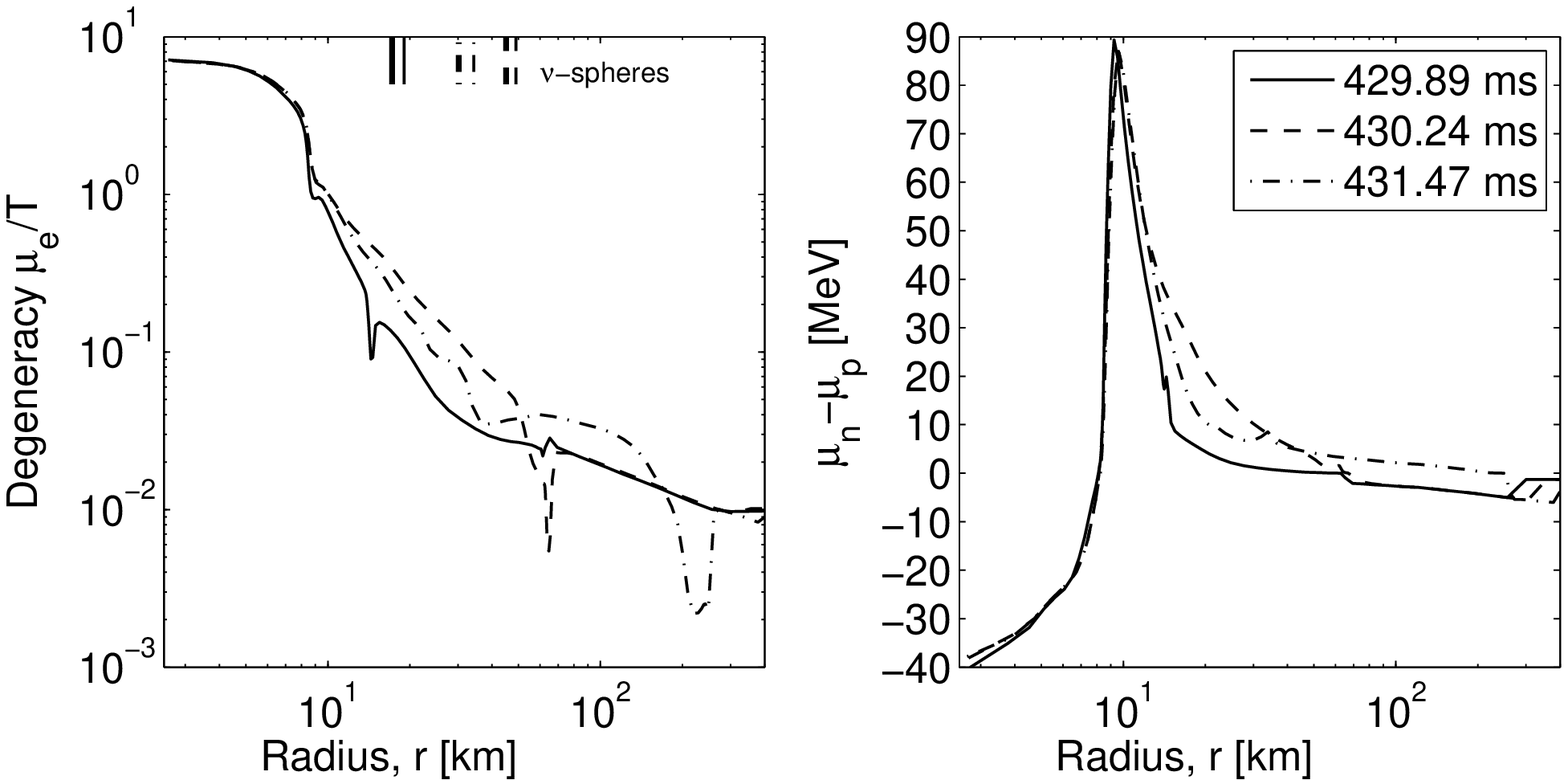}}
\caption{Radial profiles for the degeneracy parameter $\eta=\mu_e/T$ and
charged chemical potential $\mu_n-\mu_p$ for the selected post bounce
times of Figs.~\ref{fig-composition-h10n-expl-a}
and \ref{fig-composition-h10n-expl-b}.}
\label{fig-chemstate-h10n}
\end{figure}

The propagation of the second shock wave across the
neutrinospheres releases an additional burst of neutrinos,
where neutrinos of all flavors carry away energy on the order of
several $10^{53}$~erg/s on a short timescale on the order of
a few milliseconds.
The scenario is similar to the bounce shock propagation across
the neutrinospheres where the deleptonization burst is released
which, however, appears only in $\nu_e$.
In order to understand the release and the nature of the second
burst as shown in Fig.~\ref{fig-lumin-h10n} (b) at about $430$~ms
after bounce, we must consider what happens when the expanding
second shock wave passes through the formerly dissociated and
now again shock heated nucleons.
Material at the PNS surface has an electron fraction of $Y_e\simeq 0.1$.
This low $Y_e$ stems from the deleptonization burst shortly after core
bounce (see Fig.~\ref{fig-composition-h10n-expl-a}).
The heating by the shock lifts the electron degeneracy
(compare Figs.~\ref{fig-composition-h10n-expl-a}
and \ref{fig-composition-h10n-expl-b}).
It allows for the creation of electron-positron pairs followed
by positron captures on neutrons that increase the electron fraction.
This can be clearly identified via the reduced degeneracy of the
infalling material ahead of the shock during the initial shock
expansion, as long as the shock front remains a standing accretion
shock (see Fig.~\ref{fig-chemstate-h10n} (a)).
The situation changes when the shock accelerates at the PNS surface
to positive velocities (see Figs.~\ref{fig-composition-h10n-expl-a}
and \ref{fig-composition-h10n-expl-b}).
The degeneracy of the infalling material reduces significantly
at the shock front (see Fig.~\ref{fig-chemstate-h10n} (b)).
On the other hand, the degeneracy of the matter that accumulates
behind the shock increases.
Hence, the electron fraction of the infalling material decreases
while the electron fraction behind the shock increases
(see Fig.~\ref{fig-composition-h10n-expl-a}).
Note that the same behavior for the electron degeneracy discussed here
and shown in the Figs.~\ref{fig-chemstate-h10n} (a) and (b) (left panel),
holds for the charged chemical potential $\mu_n-\mu_p$,
which is shown in the Figs.~\ref{fig-chemstate-h10n} (a) and (b)
(right panel).
The charged chemical potential decreases ahead of the expanding
shock front and increases behind it.

The shock passage across the neutrinospheres (see the vertical lines
in Fig.~\ref{fig-chemstate-h10n}) releases the second neutrino burst.
The increasing temperature enhances the production of electron-positron
pairs, where most of the positrons are captured by neutrons producing
a burst of electron antineutrinos.
Some positrons interact via pair process (7) from
Table~\ref{table-nu-reactions}, contributing to all neutrino species.
This explains why the second neutrino burst is dominated by
$\bar{\nu}_e$ while $\nu_e$ and $(\nu_{\mu/\tau},\bar{\nu}_{\mu,\tau})$
have similar luminosities (see Fig.~\ref{fig-lumin-h10n} (b)).
The electron (anti)neutrino luminosities increases from
$0.2611\times10^{53}$ ($0.2708\times10^{53}$)~erg/s to
$0.55\times10^{53}$ ($2.2\times10^{53}$)~erg/s on a short
timescale on the order of milliseconds.
The same increase in luminosity holds for the $(\mu/\tau)$-neutrinos,
which reach about $1.5\times10^{53}$~erg/s.
The sharp peak of the neutrino luminosities is accompanied by
a sharp rise of the mean neutrino energies.
The electron (anti)neutrino mean energies increase from
$16.5$ ($18.4$)~MeV
to $29.38$ ($32.2$)~MeV, and the mean $(\mu/\tau)$-neutrino energies
rise from about $18.5$~MeV to $53.7$~MeV on the same short timescale on
the order of milliseconds.

\begin{figure*}
\centering
\includegraphics[width=1.\textwidth]{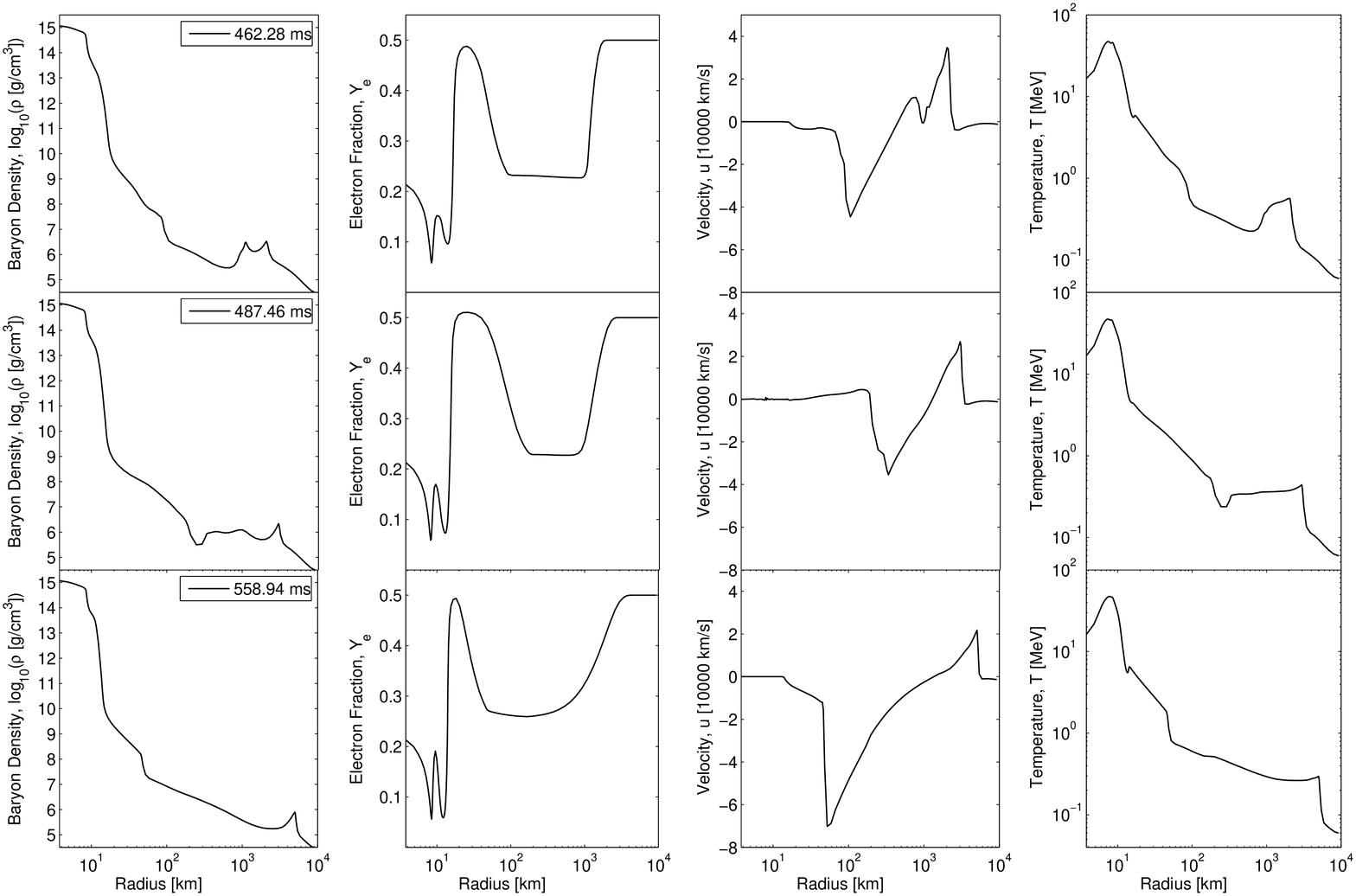}
\caption{Selected radial profiles for the 10.8~M$_\odot$ reference model
during the ongoing explosion phase at three late times post bounce.
Illustrating is the back and forth propagating accretion shock
at the PNS surface established via to neutrino cooling/heating.}
\label{fig-hydrostate-h10n-expl3}
\end{figure*}

The later evolution of the PNS surface is shown in
Fig.~\ref{fig-hydrostate-h10n-expl3} via the radial profiles of $\rho$,
$Y_e$, velocity and $T$ at selected post bounce
times during the ongoing explosion phase.
Neutrino cooling after the shock propagation across the neutrinospheres
(see Fig.~\ref{fig-heatplot-h10n-b})
leads to the establishment of matter infall between the expanding
explosion shock and the PNS surface already at about 430.24~ms
after bounce in Fig.~\ref{fig-composition-h10n-expl-b}
between 10--20 km.
As matter continues to fall onto the PNS surface with even supersonic
velocities, an additional accretion shock forms at the PNS surface.
This additional accretion shock on top of the PNS surface
expands due to neutrino heating and falls back due to cooling
(see therefore the large cooling rates between 7--10~km in
Fig.~\ref{fig-heatplot-h10n-b})
on timescales on the order of 10~ms, during which the
neutrinospheres expand and contract correspondingly.
The same phenomena apply as discussed in the paragraph above.
The shock propagation towards the neutrinospheres increases
the temperature at the neutrinospheres, where the enhanced opacities
cause the propagation of the neutrinospheres to lower densities.
Neutrino cooling, as well as the obtained matter expansion at the
neutrinospheres, shifts the neutrinospheres to higher
densities.
The consequently increasing and decreasing mass accretion rate
is reflected in the electron flavor neutrino luminosities and mean
energies, shown in the insets in Fig.~\ref{fig-lumin-h10n} (b) after the
second neutrino burst between 432--500~ms.

The additional standing accretion shock, propagating back and forward
at the PNS surface, settles down to a quasi-stationary state at about
$550$~ms post bounce. 
After that, the evolution of the neutrinospheres is determined by mass
accretion on the order of few $0.1$~M$_\odot$/s at a timescale on
the order of $100$~ms up to seconds.
The neutrino luminosities decrease to values below $10^{52}$~erg/s
within the first $500$~ms after the onset of the explosion.
The mean neutrino energies follow the same behavior.
The mean electron (anti)neutrino energies decrease from about
$18$ ($20$)~MeV at $550$~ms post bounce to $12$ ($18$) at
$1$ second post bounce (see Fig.~\ref{fig-lumin-h10n} (b)).
The mean $(\mu/\tau)$ neutrino energies decrease only slightly
from $22$~MeV at $550$~ms post bounce to about $19$~MeV
at $1$ second post bounce.
Unaffected by the behavior of the standing accretion shock at the
PNS surface, the explosion shock continues to expand to larger
radii with matter velocities on the order of several $10^4$ km/s
(see Fig.~\ref{fig-hydrostate-h10n-expl3}).
%

%....................................................................................
\section{Discussion}

In this section, we will compare the results obtained using the
10.8~M$_\odot$ reference model with the remaining
simulations that have been performed where different progenitor
models as well as different quark EoS parameters are used.
Further below, we will discuss possible applications with respect
to the PNS structure and the formation of magnetars as well as
with respect to possible nucleosynthesis investigations.
\begin{table}[htp]
\centering
\caption{Summary of the models under investigation.}
\begin{tabular}{ c c c c c c }
\hline \hline
Prog. 
& EoS
& $t_\text{pb}$ 
& $\rho_\text{c}$\footnote{
Selected central properties of the PNSs at the onset of collapse ($t_{pb}$).} 
& $T_\text{c}^a$
& $Y_e^a$
\\
$[$M$_\odot]$
&
& $[$ms$]$ 
& $[10^{14}$ g/cm$^3 ]$
& $[$MeV$]$
&  
\\
\hline 
10.8 & EOS1 & 240 & 6.607 & 13.14 & 0.234 \\
10.8 & EOS2 & 428 & 6.457 & 14.82 & 0.237 \\
13    & EOS1 & 235 & 6.493 & 13.32 & 0.235 \\
13    & EOS2 & 362 & 7.228 & 16.38 & 0.191  \\
15    & EOS1 & 172 & 7.523 & 17.15 & 0.170  \\
15    & EOS2 & 275 & 7.586 & 16.25 & 0.187 \\
15    & EOS3 & 308 & 5.511& 17.67 & 0.197 \\
\hline
\end{tabular}
\label{tab-summary1}
\end{table}

\subsection{Explosions in spherical symmetry}

In addition to the low-mass reference model of 10.8~M$_\odot$,
we apply the hybrid EoSs introduced in \S~2.3
to core-collapse supernova simulations of the intermediate mass
Fe-core progenitor stars of 13 and 15~M$_\odot$ from
\citet{Woosley:etal:2002}.
The results are summarized in Table~\ref{tab-summary1}.
We list the characteristic properties of the simulations for
the different progenitor models and the hybrid EoSs with
the two different values of the bag constant.
The values $t_\text{pb}$ are the post bounce times  when
the PNSs become gravitationally unstable and start to collapse.
The corresponding central conditions, i.e. density $\rho_c$,
temperature $T_c$ and electron fraction $Y_e$, are also
listed\footnote{Note that the values of $\rho_c$, $T_c$ and $Y_e$ are
not the critical conditions for the appearance of quark matter.}.
The models using the larger bag constant, which corresponds
to a higher critical density, reach typically lower quark volume
fractions at equal evolutionary states post bounce.
In order to reach similar central densities, the PNSs have to
accrete more mass which leads to a longer accretion
time post bounce.
This effect is compensated by the lower maximum stable mass
of the configurations using the larger bag constant.
Hence, the central densities required for the PNSs to become
gravitationally unstable and collapse are rather similar for
the same progenitor model.
On the other hand, due to the longer mass accretion period for the
models using the larger bag constant on the order of $100-200$~ms,
the PNSs are more compact before collapsing.
In addition, the central temperatures obtained are higher by
about $2-3$~MeV.
This in turn favors quark matter over hadronic matter
(see the phase diagrams in Fig.~\ref{p_nb_beta}--\ref{phase_diag155}
in \S~2.3),
where weak-equilibrium is established at a lower
value of the electron fraction (see Table~\ref{tab-summary1}).

\begin{figure}
\centering
\includegraphics[width=.49\textwidth]{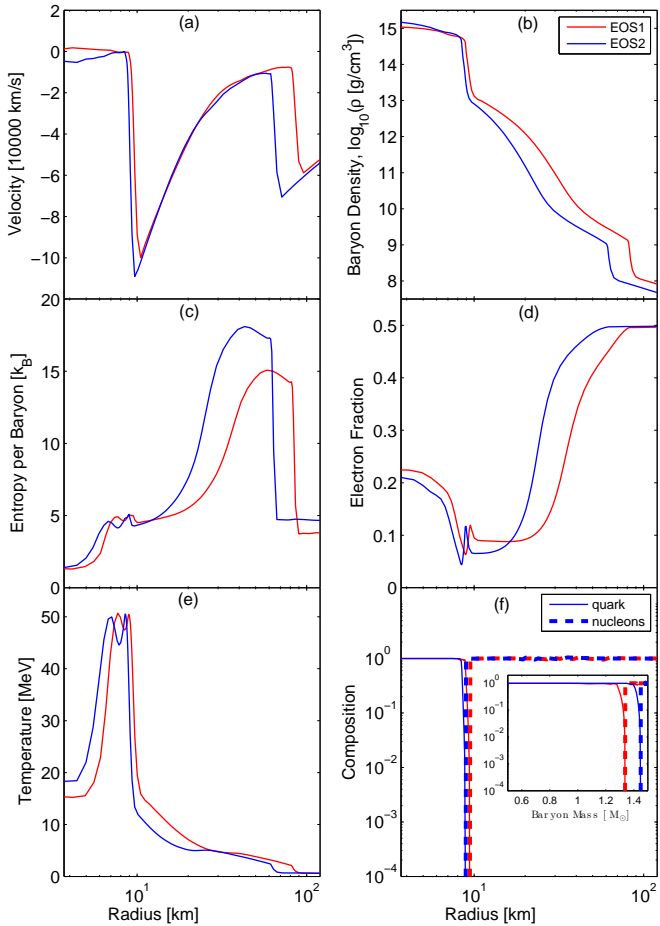}
\caption{Selected radial profiles for the 10.8~M$_\odot$ progenitor model,
comparing the hybrid EOS1 (red lines, at 248.57~ms post bounce)
and EOS2 (blue lines, at 429.76~ms post bounce),
both at the moments just before shock breakout
along the decreasing density at the PNS surfaces.
Graph (f) shows the composition, i.e. quark volume fraction (solid lines)
and the hadronic matter (dashed lines).
In addition to the radial profile in graph (f), the inset shows the
composition with respect to the baryon mass.}
\label{fig-quarkstate}
\end{figure}

All models listed in Table~\ref{tab-summary1} evolve in a similar
fashion during the quark-hadron phase transition.
The explosions are obtained due to the formation of the strong
second hydrodynamic shock front inside the PNSs.
The shocks accelerate at the PNS surfaces, i.e. the shock breakout,
which triggers the explosions where otherwise no explosions could
have been obtained in spherical symmetry.
The explosion energy estimates $E_\text{expl}$ and approximate
neutron star masses $M_\text{NS}$
\citep[for a definition of the mass cut and the explosion energy,
see][following a suggestion by S.~Bruenn]{Fischer:etal:2010b},
are listed in Table~\ref{tab-summary2}.
Moderate explosion energies of about $1\times10^{51}$~erg
could be obtained for the reference model, i.e. 10.8~M$_\odot$
using the hybrid EOS2.
The models with an early PNS collapse and hence the early onset
of the explosion reach smaller explosion energies.
The shorter mass accretion period post bounce lead to less compact
PNS configurations and therefore less steep density gradients at the
PNS surfaces until they become gravitationally unstable and collapse
due to the presence of quark matter in the interiors.
The comparison of the two hybrid EOS1 and EOS2
($\alpha_S=0$) is shown in Fig.~\ref{fig-quarkstate} for the
10.8~M$_\odot$ progenitor model.
The PNS obtained for $B^{1/4}=165$~MeV (EOS2) is more compact,
indicated by the more massive quark core in Fig.~\ref{fig-quarkstate} (f)
of about $0.15$~M$_\odot$ at the moment of shock breakout.
It results in the higher density and
temperature shown in the Figs.~\ref{fig-quarkstate} (e) and (b).
Furthermore, the longer mass accretion time for the model using
$B^{1/4}=165$~MeV (EOS2) relates to lower densities and a steeper
density gradient, surrounding the central quark core of the PNS
(see Fig.~\ref{fig-quarkstate} (b)) at the moment of shock breakout.
This allows for a stronger shock acceleration of the second shock wave
expanding along the decreasing density gradient at the PNS surface.
These differences are the origin of the higher explosion energy estimates
using EOS2 in comparison to EOS1.
Hence, the acceleration of the formed hydrodynamic shock
at the PNS surface is less intense for the models using the low
bag constant.
Lower explosion energies are also obtained for the more
massive progenitors.
The more massive envelopes of the PNSs with significantly higher
densities prevents a similar strong shock acceleration.
In general, the explosion energy estimates may be
lifted to larger values if multi-dimensional phenomena
(e.g. convection and rotation) are taken into account.

\begin{figure*}[htp]
\centering
\includegraphics[width=1.\textwidth]{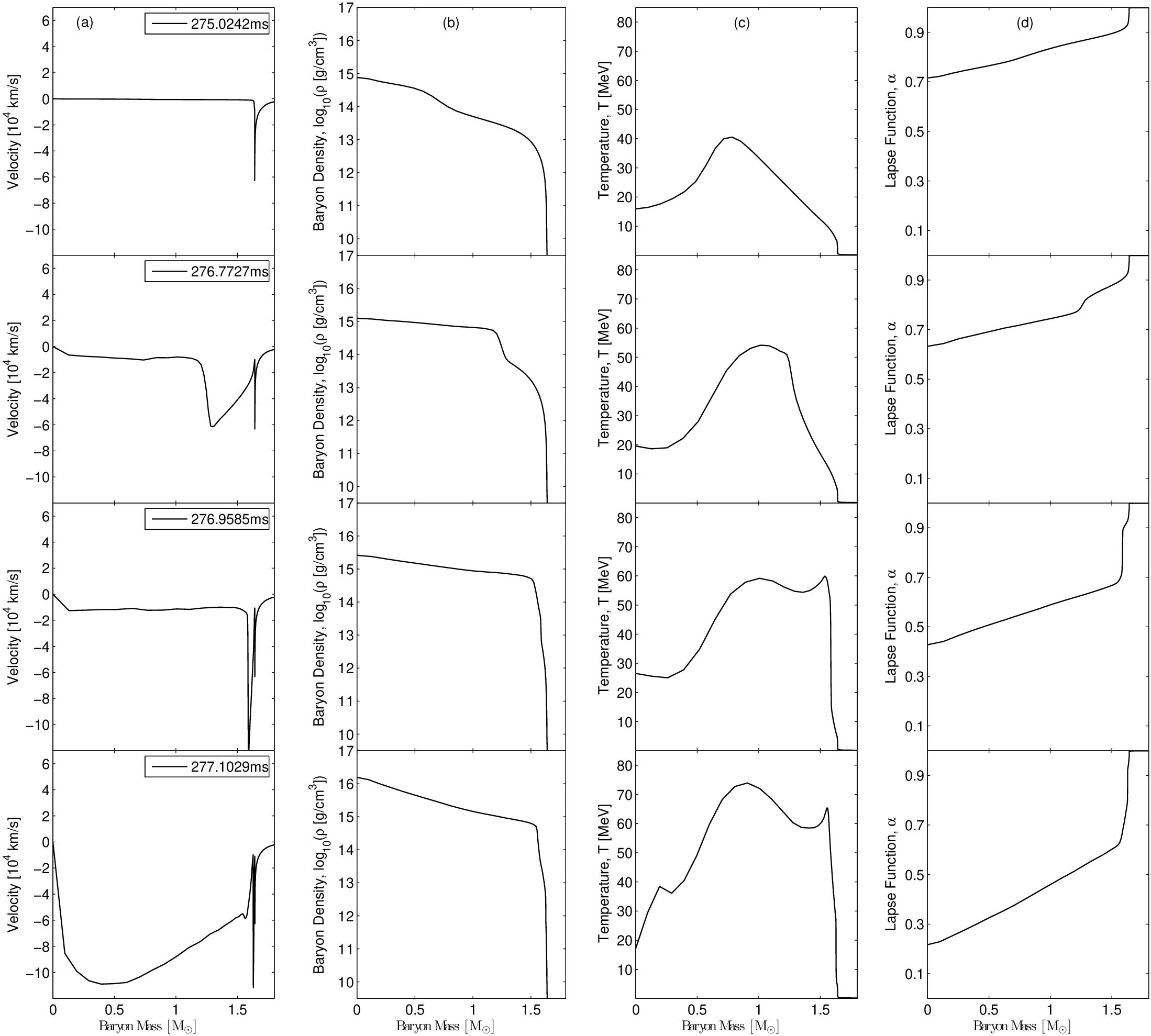}
\caption{PNS collapse of the 15~M$_\odot$ progenitor model using
EOS2, at selected post bounce times during the PNS collapse,
shock formation and shock propagation as well as
the black hole formation.}
\label{fig-PNS-collapse}
\end{figure*}

\subsection{PNS collapse to a black hole}

A special model is the 15~M$_\odot$ progenitor, using the hybrid
EOS2 (see Table~\ref{tab-summary1}).
The evolution of this model is shown in Fig.~\ref{fig-PNS-collapse}
via radial profiles of velocity, density and temperature in the graphs
(a), (b) and (c) respectively at four selected post bounce times.
The graphs illustrate the last stable configuration as well as the
PNS collapse.
The evolutionary scenario, i.e. the PNS collapse due to the softening
of the EoS in the mixed phase and the subsequent shock formation
due to the stiffening of the EoS in the pure quark phase
(clearly identified via the velocity, density and temperature profiles 
in Fig.~\ref{fig-PNS-collapse} (a), (b) and (c) respectively
at the first three post bounce times), is in qualitative
agreement with the reference case discussed in \S~3.
During the shock propagation towards the PNS surface,
mass accretion onto the second shock wave continuously
increases the central density and hence the quark core of
the PNS grows in mass.
When the maximum stable mass (given by the hybrid EoS) of the
configuration is reached, the PNS collapses a second time as shown
in Fig.~\ref{fig-PNS-collapse} (dotted-lines).
Relativistic effects become more and more important.
This is illustrated in Fig.~\ref{fig-PNS-collapse} (d)
via the metric function $\alpha(t,a)=g_{tt}$, i.e. the
lapse function, which approaches zero.
It indicates the appearance of the event horizon and hence the
formation of a black hole.
By our choice of a co-moving coordinate system, stable solutions
for the evolution equations of energy and momentum cannot be obtained.
Hence, the simulations cannot be continued beyond that point.

The evolution scenario of black hole formation induced via
the PNS collapse found via the hybrid EoS here is in qualitative
agreement with the results obtained by \citet{Nakazato:etal:2008b}
and \citet{Nakazato:etal:2010b} for very massive stars of 100~M$_\odot$
and for the 40~M$_\odot$ progenitor from \citet{WoosleyWeaver:1995}.
Furthermore, the results agree qualitatively with the pure
hadronic scenario of black hole formation discussed by
\citet{Sumiyoshi:etal:2007} and \citet{Fischer:etal:2009}
for several massive progenitor stars in the mass range of
40--50~M$_\odot$.
Differences occur in the larger maximum masses for the hadronic EoS.
\begin{figure*}[htp]
\centering
\subfigure[EOS1]{
\includegraphics[width=0.42\textwidth]{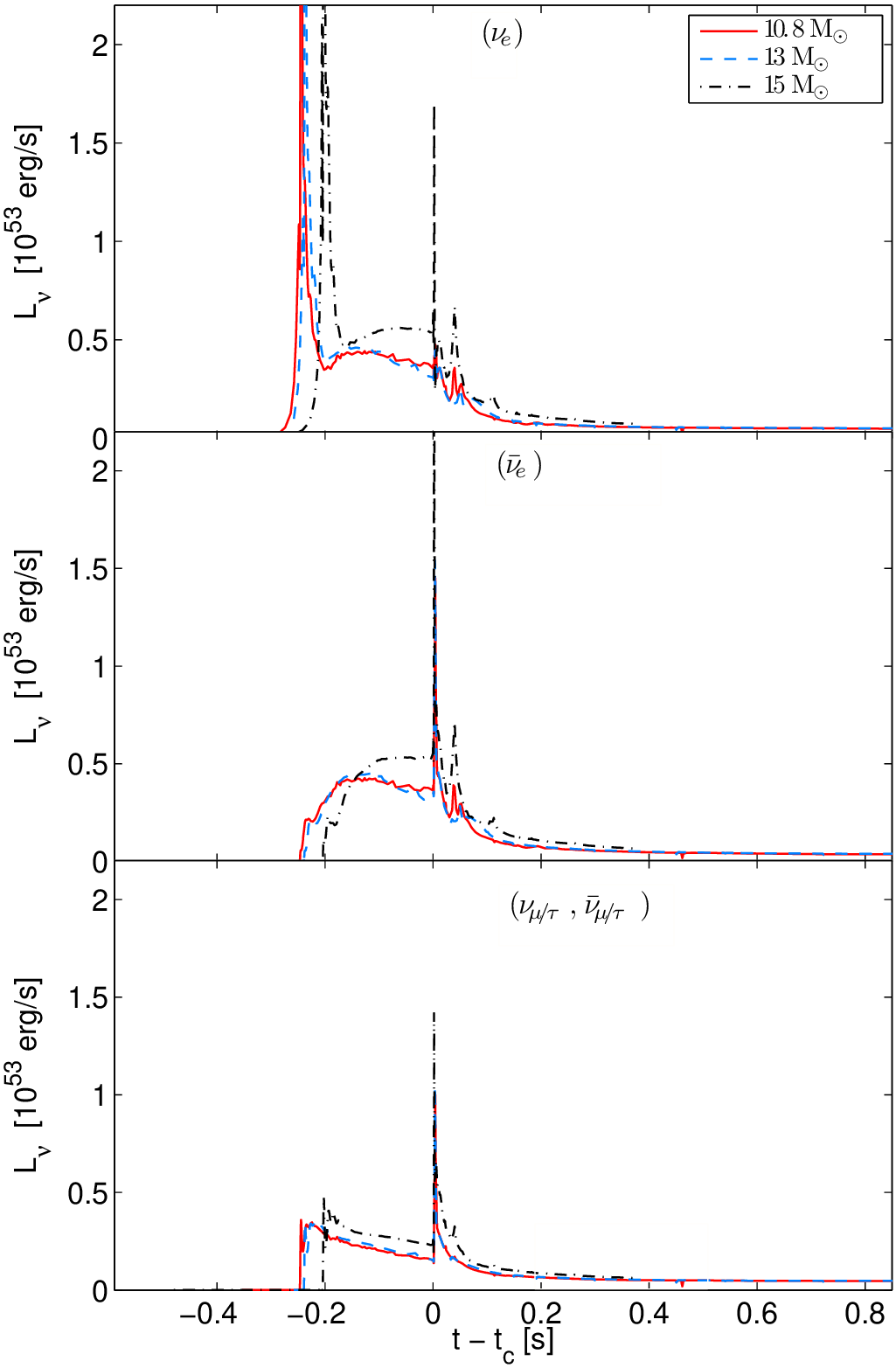}}
\hspace{10mm}
\subfigure[EOS2]{
\includegraphics[width=0.42\textwidth]{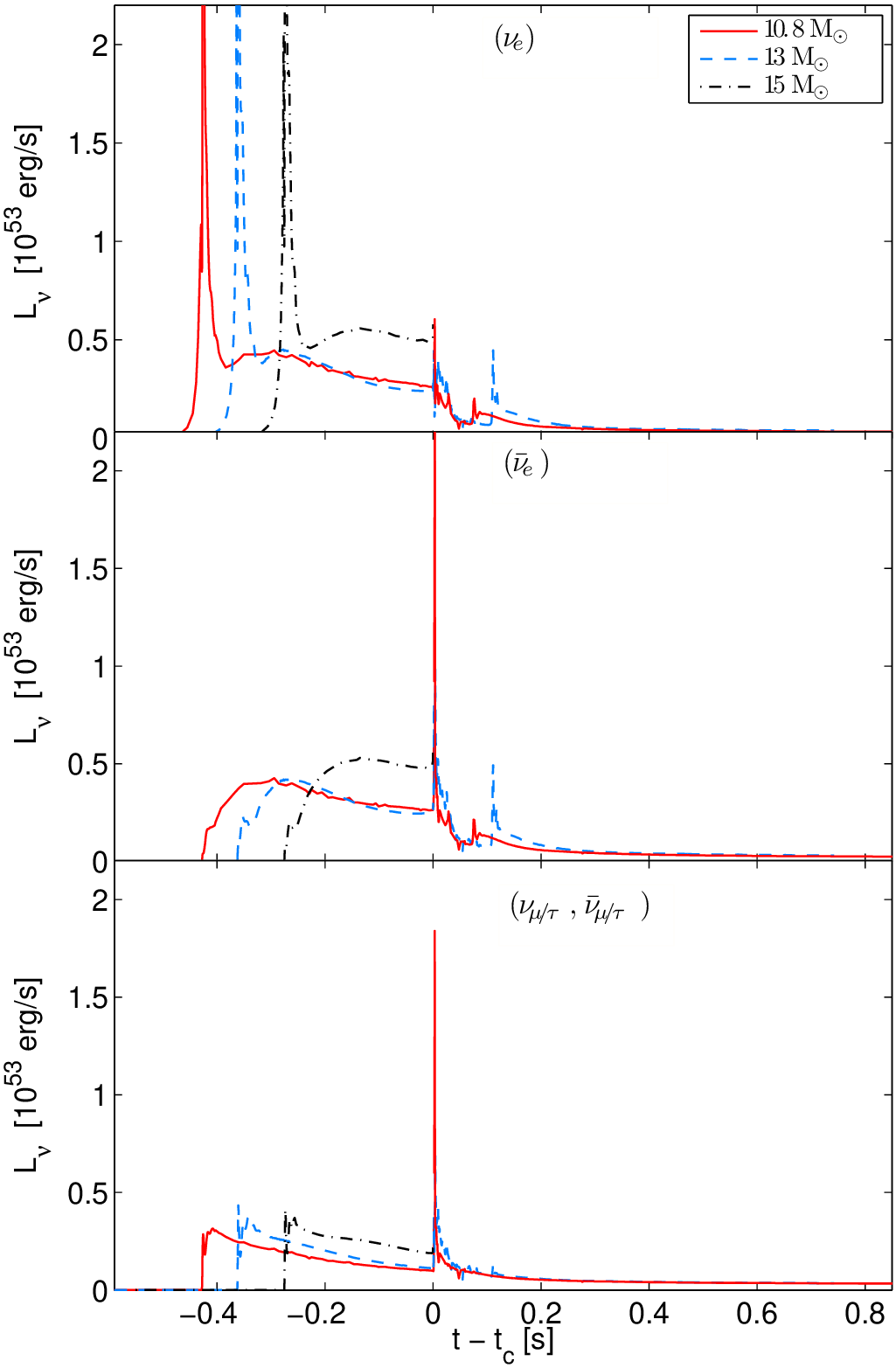}}
\caption{Neutrino luminosities for the different progenitor models
under investigation.
The time gauges $t_c$ correspond to the moment when the
maximum central densities are obtained for the different models
during the PNS collapses.}
\label{fig-lumin-comp}
\end{figure*}
\begin{figure*}[htp]
\centering
\subfigure[EOS1]{
\includegraphics[width=0.42\textwidth]{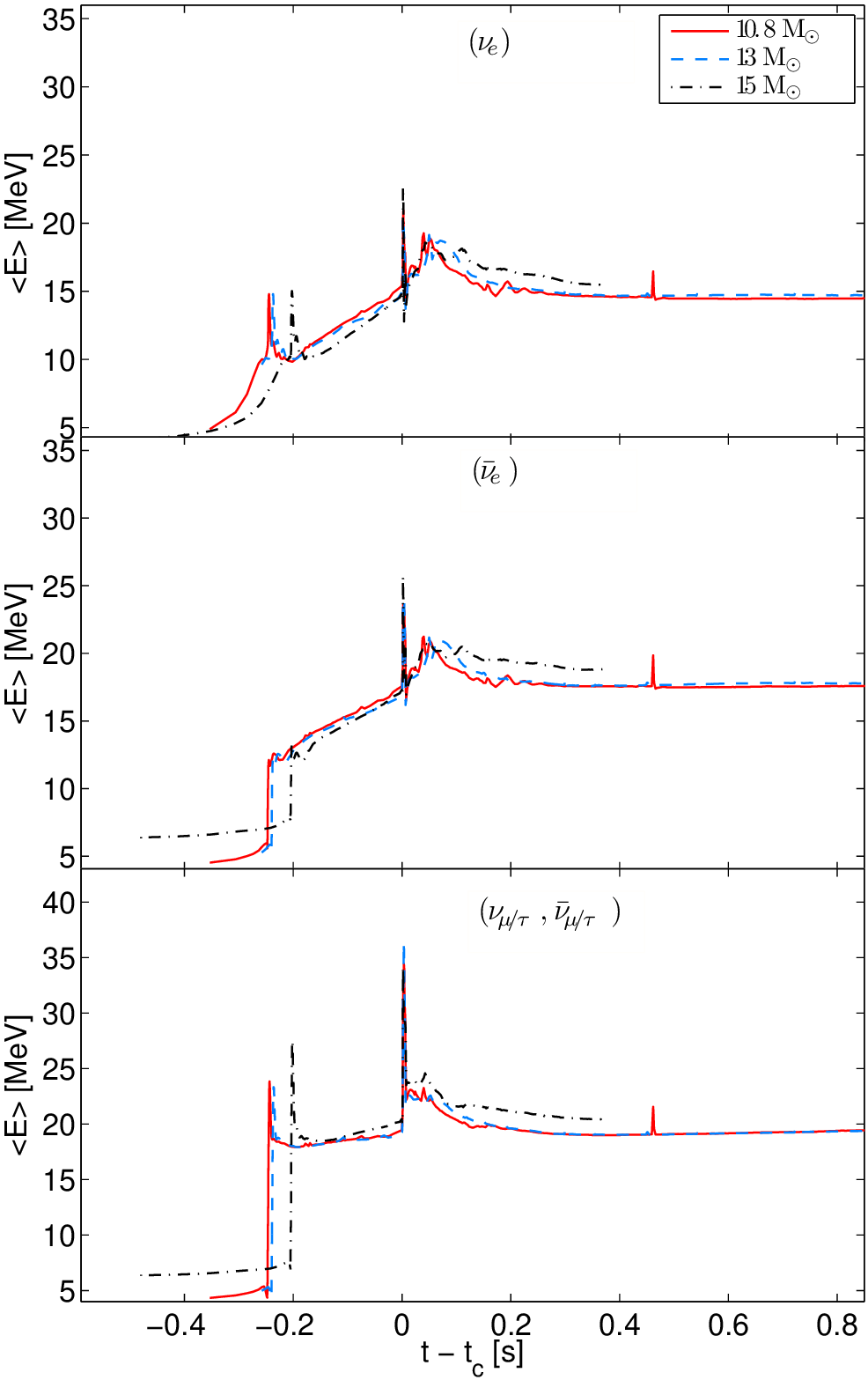}}
\hspace{10mm}
\subfigure[EOS2]{
\includegraphics[width=0.42\textwidth]{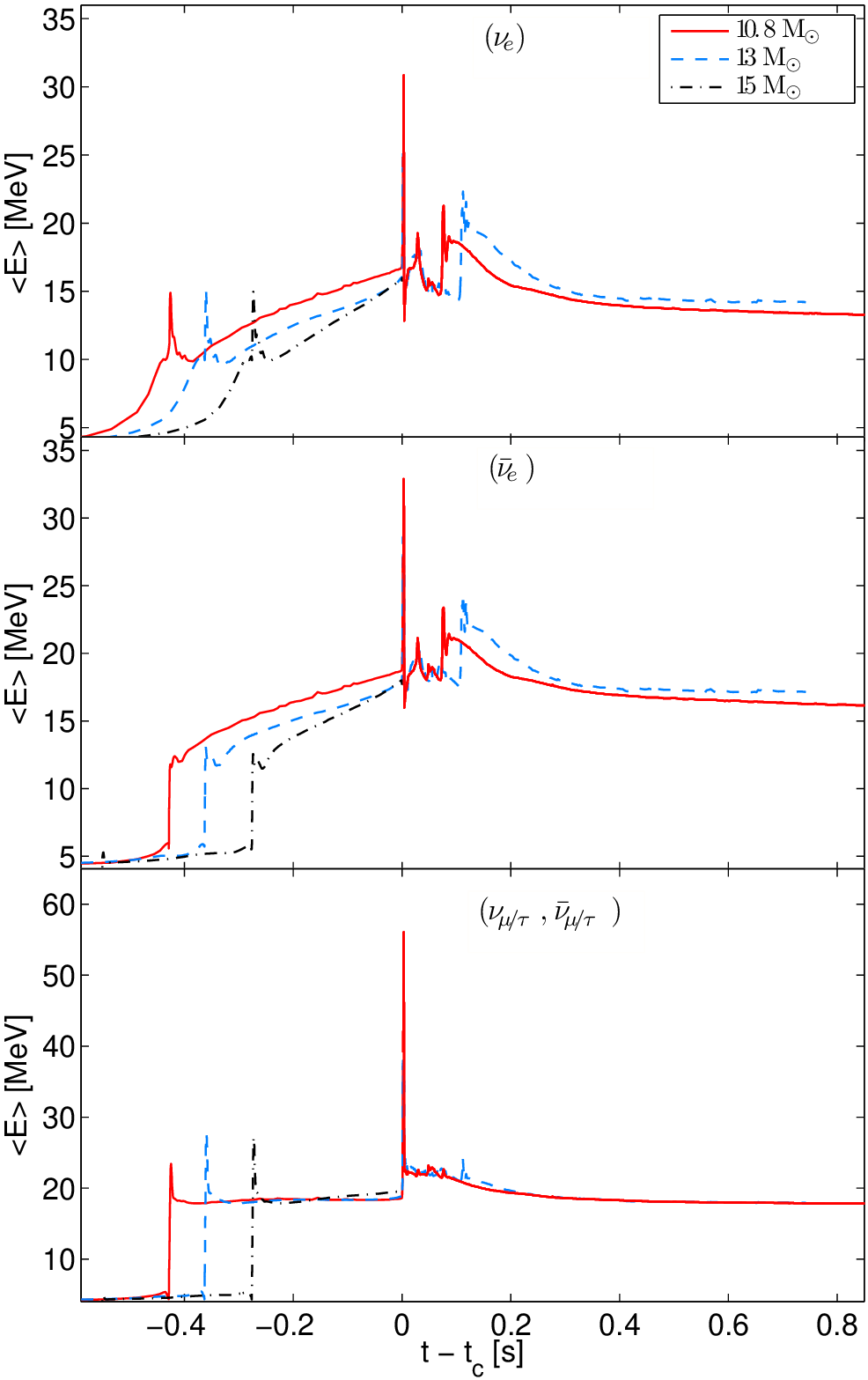}}
\caption{Mean neutrino energies for the different progenitor models
under investigation.
The same configuration as Fig.~\ref{fig-lumin-comp}.}
\label{fig-rms-comp}
\end{figure*}

\subsection{Robustness of the observable features}

The peak-like rise of the luminosities and mean neutrino energies
is an indirect consequence of the quark-hadron phase transition
as discussed for the reference case in \S~3.
It was found to occur for all models under investigation.
The results are shown in Figs.~\ref{fig-lumin-comp} and
\ref{fig-rms-comp} for the models using the hybrid EOS1
(in graphs~(a)) and EOS2 (in graph~ (b)).
The different post bounce times $t_c$ (as listed in Table~\ref{tab-summary2}
together with the corresponding central maximum densities $\rho_\text{max}$)
for the appearance of the second peak and the magnitude of the peaks
depend on the microphysics and the progenitor model.
The quantity $\rho_\text{max}$ relates to the maximum
central density obtained at the end of the PNS collapse.

For all models under investigation (see Fig.~\ref{fig-lumin-comp}
and \ref{fig-rms-comp}), the luminosities and mean neutrino
energies follow a similar behavior.
After a sharp rise on timescales on the order of milliseconds,
the luminosities and mean neutrino energies increase and decrease
in accordance to the expanding and descending standing accretion shock
at the PNS surface, between $10$--$50$~ms after the burst.
A quasi-stationary state is obtained after about $100$~ms
after the sharp rise.
The luminosities and mean neutrino energies decrease continuously
indicating the ongoing explosions.
All models evolve in a similar fashion as discussed in \S~3,
through the PNS collapse and the subsequent evolution.
Hence, the second peaks are dominated by $\bar{\nu}_e$
and $(\nu_{\mu/\tau},\bar{\nu}_{\nu/\tau})$ for all models under investigation
(see Fig.~\ref{fig-lumin-comp}).
Furthermore, the bursts are accompanied by significant increase
of the mean neutrino energies (see Fig.~\ref{fig-rms-comp}).

The magnitude and the width of the luminosities and mean neutrino
energies of the second neutrino burst found are on the same order for all
models under investigation, independent of the progenitor model
and independent of the bag constant.
On the other hand, the onsets for the second neutrino bursts take
place at different times post bounce and depend on both the
progenitor model and the bag constant.
As shown in the Figs.~\ref{fig-lumin-comp} and \ref{fig-rms-comp},
the second neutrino bursts for the 10.8~M$_\odot$ and 13~M$_\odot$
models using EOS1 occur at similar times
because the conditions for the PNSs to become gravitationally
unstable are obtained at similar post bounce times
(see Table~\ref{tab-summary1}).
Since the structure of these two progenitor models is relatively
similar, they evolve in a similar fashion and on a similar timescale
post bounce until the PNSs collapse.
The 15~M$_\odot$ progenitor model on the other hand is
more compact and the conditions for the PNS collapse using
EOS1 are obtained slightly earlier by about 40~ms
(see Table~\ref{tab-summary1}
and compare with the post bounce time of the second neutrino burst
in Figs.~\ref{fig-lumin-comp} and \ref{fig-rms-comp} (a)).

The simulations using the hybrid EOS2 is shown in the
Figs.~\ref{fig-lumin-comp}~(b) and \ref{fig-rms-comp}~(b).
They bring the differences between the progenitor models to light, due
to the different post bounce accretion times before the PNSs become
gravitationally unstable and collapse.
The lower progenitor masses relate to longer accretion times 
and hence later second neutrino bursts.
The delay for the different models explored in this article
is on the order of about 100~ms for the 10.8 to the 13
as well as between the 13 and 15~M$_\odot$ progenitor models.
In other words, the differences between the post bounce times for the
release of the second neutrino bursts relate to the same critical
conditions for the quark-hadron phase transition but different
evolutionary scenarios due to the different progenitor models.
The future observation of such multi-peaked neutrino spectra
might allow us to extract fundamental information about the
state of matter at extreme conditions.
The magnitude of the second burst and its delay after the deleptonization
burst after core bounce contains information about the quark and hadron
EoSs.
If the progenitor model is known and the pure hadronic EoS is
fixed, it may be possible to extract the critical conditions for the
quark-hadron phase transition from the observed neutrino signal.
The post bounce times for the release of the second neutrino burst
and the corresponding central conditions for the PNS collapse
are listed in Table~\ref{tab-summary1} for all models under investigation.
First results have been analyzed with respect to the possible observation
of such a neutrino burst by \citet{Dasgupta:etal:2010} based on results
discussed in \citet{Sagert:etal:2009}, exploring the 10.8~M$_\odot$
progenitor model where the hybrid EOS1 was used.
They find that the operating neutrino detectors,
Super-Kamiokande and IceCube, could detect such a
$\bar{\nu}_e$-burst from a future Galactic event.

\begin{table}[htp]
\centering
\caption{Estimated neutron star masses $M_\text{NS}$ and
explosion energies $E_\text{expl}$ and
the maximum central densities $\rho_\text{max}$
for all models under investigation.}
\begin{tabular}{ c c c c c c }
\hline \hline
Prog. 
&  EoS
& $M_\text{NS}$
\footnote{Neutron star mass (baryon mass) and explosion energy,
estimated at several 100~ms after the onset of explosion.} 
& $E_\text{expl}$ $^a$
& $t_c$
\footnote{Post bounce times $t_c$; correspond to the moments when the
maximum central densities $\rho_\text{max}$ are reached at the end
of collapse}
& $\rho_\text{max} $
\\
$[$M$_\odot]$
&
&  $[$M$_\odot]$  
& $[10^{51}$~erg$]$
& $[$ms$]$ 
& $[10^{15}$\\
&
&
&
&
&g/cm$^3]$ 
\\
\hline 
10.8 & EOS1 & 1.431 & 0.373 & 248.78 & 1.291 \\
10.8 & EOS2 & 1.479 & 1.194 & 429.81 & 1.806 \\
13    & EOS1 & 1.465 & 0.232 & 241.20 & 1.323 \\
13    & EOS2 & 1.496 & 0.635 & 364.06 & 1.788 \\
15    & EOS1 & 1.608 & 0.420 & 175.07 & 1.487 \\
15    & EOS2 & 1.641 & unknown
\footnote{black hole formation before positive explosion energy is
achieved}
 & 277.10
\footnote{time of black hole formation}  & 15.362
\footnote{central density at the time of black hole formation}
\\
15 & EOS3 & 1.674 & 0.458 & 312.99 & 1.342 \\
\hline
\end{tabular}
\label{tab-summary2}
\end{table}
%

%.alphaS corrections.......................................................
\subsection{Corrections of the strong coupling constant}

\begin{figure}[htp]
\centering
\includegraphics[width=0.9\columnwidth]{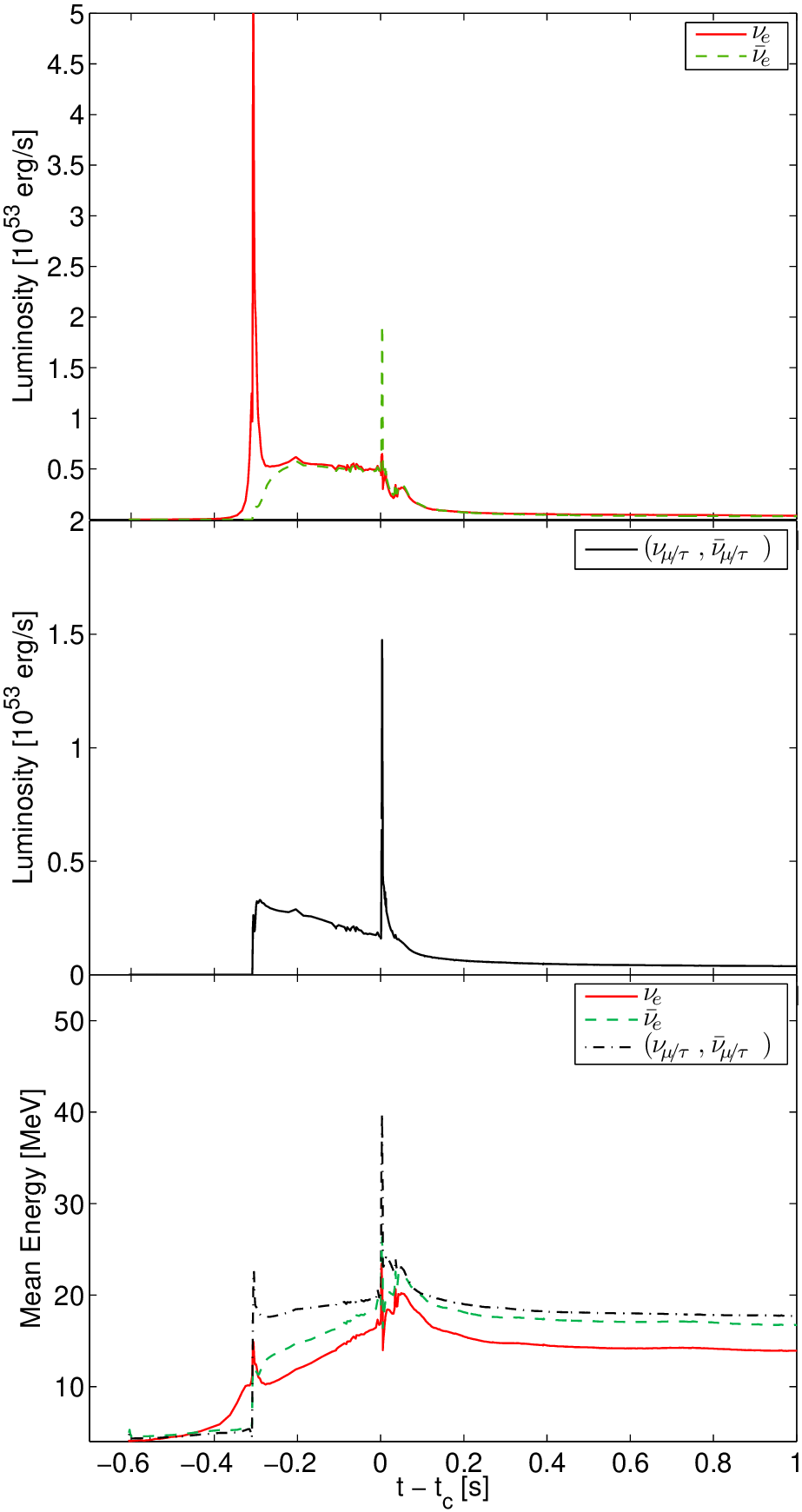}
\caption{Luminosities and mean neutrino energies for the
15~M$_\odot$ progenitor model using the hybrid EOS3.
The time gauge $t_c$ is again the moment when the maximum
central density is obtained during the PNS collapse.}
\label{fig-lumin-h15y}
\end{figure}

Massive stars of 15~M$_\odot$ are in the expected mass
range of exploding progenitors, that result in stable PNSs at least
during the time of explosion.
The evolutionary scenario of black hole formation applying the hybrid
EOS2 as discussed above in \S~4.2, indicates that important ingredients
are missing.
For that particular model, the PNS collapses to a black hole
because the maximum stable mass of the configuration,
which is given by the hybrid EoS, is reached during the collapse.
The collapse in turn is triggered due to the presence of quark
matter in the interior.
The inclusion of $\alpha_s$-corrections in the hybrid EoS as introduced
in \S.2.3 allows for more massive hybrid star configurations.
It increases the maximum stable mass significantly.
Applying the hybrid EOS3 ($B^{1/4}=155$~MeV, $\alpha_s=0.3$),
leads to similar critical conditions for the onset of quark matter
in comparison to EOS2, however with slightly higher critical density
(see Table~\ref{tab-summary1}).
Furthermore, the post bounce time for the PNS to become gravitationally
unstable and collapse (see Table~\ref{tab-summary1}) is slightly delayed
by about 50~ms (see Fig.~\ref{fig-lumin-h15y}), which results in
slightly higher temperatures (see Table~\ref{tab-summary1}).
The consequent evolutionary scenario is in qualitative agreement
with the reference model as discussed in \S.~3,
i.e. the 10.8~M$_\odot$ using EOS2.
The remaining hybrid star has a baryon mass of about 1.65~M$_\odot$
at the moment when the simulation is stopped.
Due to the larger maximum stable mass of the configuration,
the PNS is found to be stable and does not collapse to a black hole.

The evolution of the corresponding neutrino luminosities and mean
energies are shown in Fig.~\ref{fig-lumin-h15y}, which is in qualitative
agreement with the spectra discussed for example of the reference
model in \S 3.

\begin{figure}[htp]
\centering
\subfigure[With respect to time after bounce.]{
\includegraphics[width=0.4\textwidth]{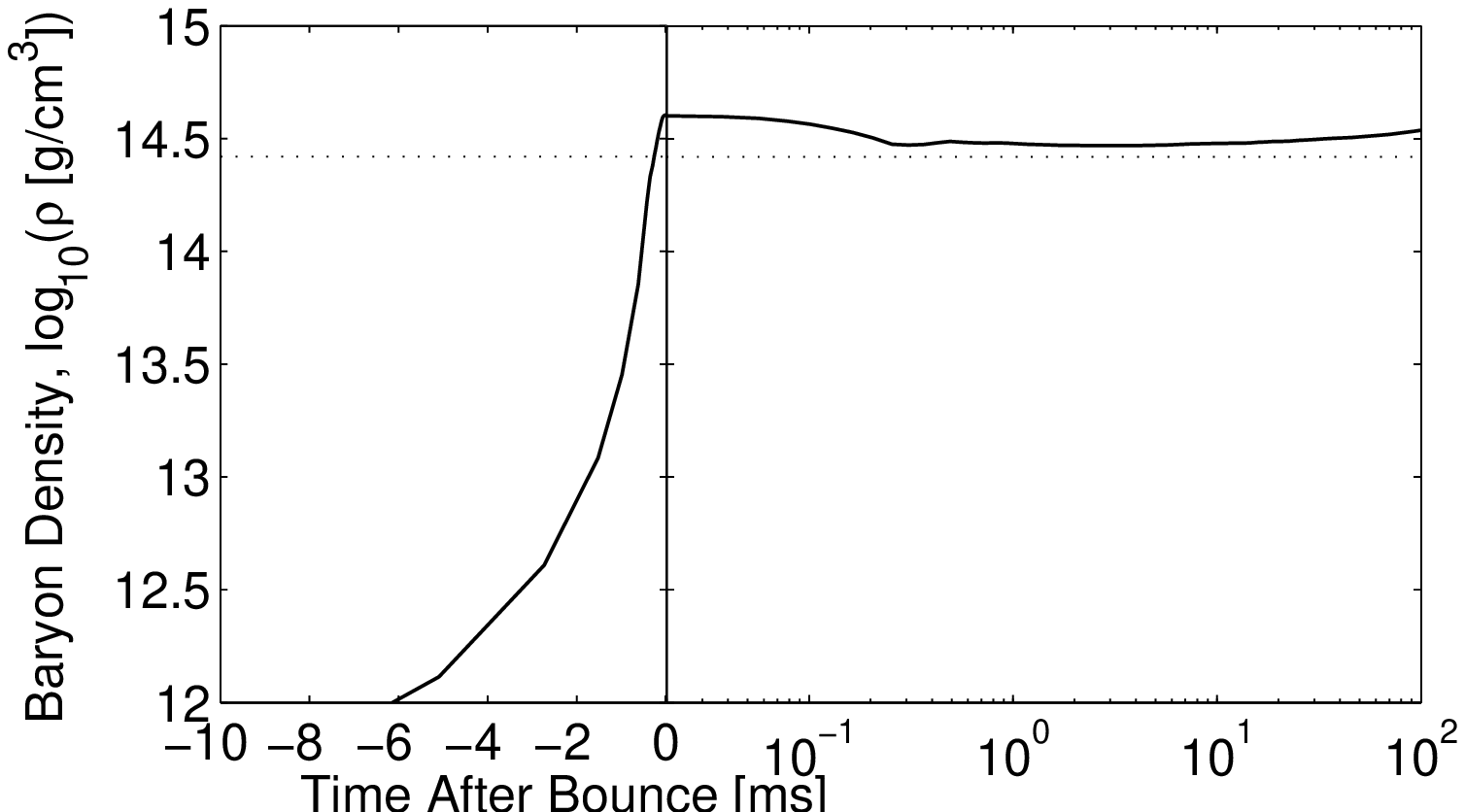}}\\
\subfigure[With respect to time after the PNS collapse
when the maximum central density is obtained, i.e.
$t_c=429.81$~ms after bounce.]{
\includegraphics[width=0.4\textwidth]{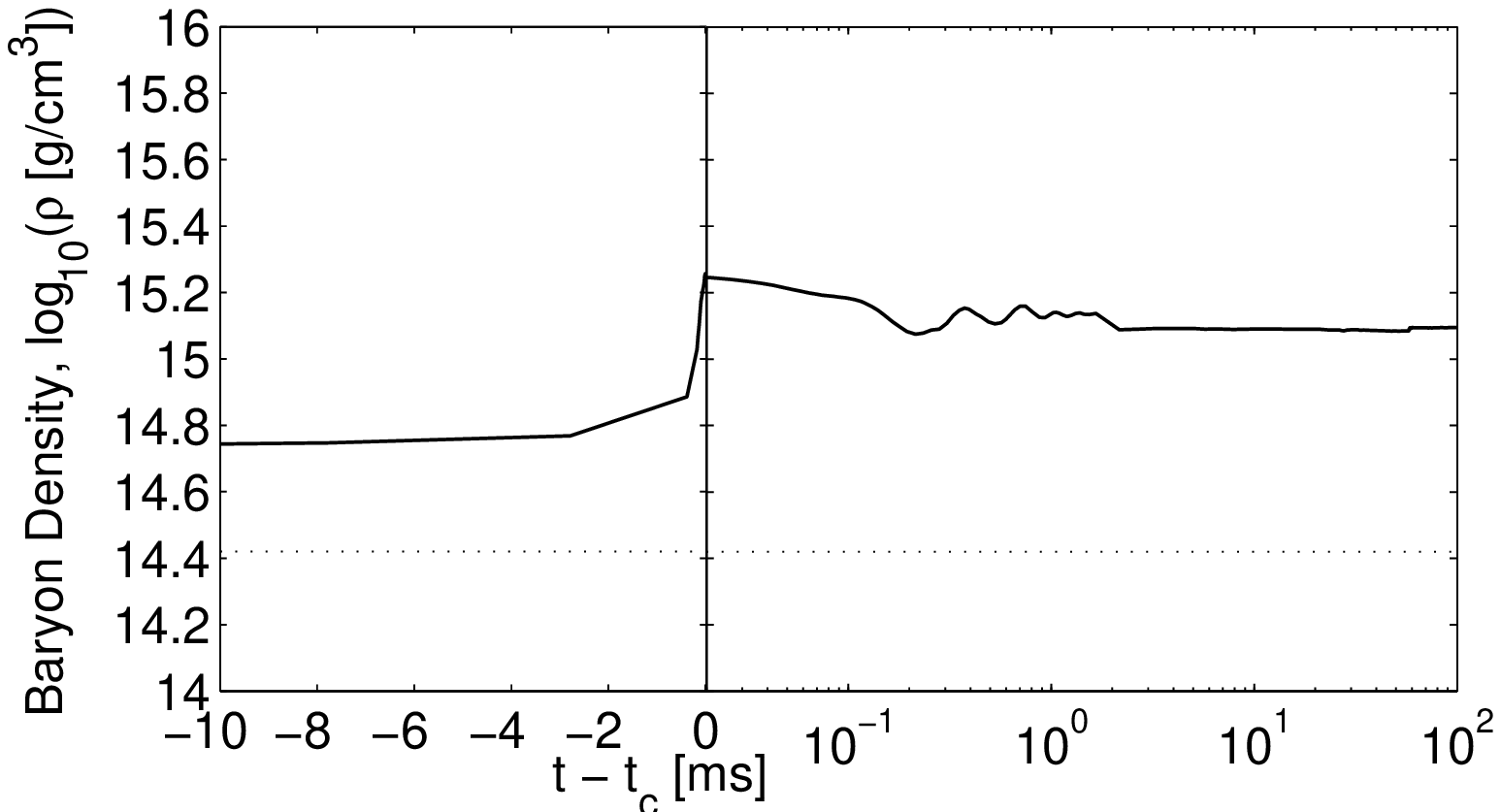}}
\caption{Evolution of the central density for the 10.8~M$_\odot$
reference model with respect to time after bounce.
The dotted line marks nuclear saturation density.}
\label{fig-density-h10n}
\end{figure}
%

%.PNS evolution and magn field..............................................
\subsection{PNS structure and the early evolution}

The rise of the central density at the final phase of the
Fe-core collapse of massive progenitor stars  is shown in
Fig.~\ref{fig-density-h10n} (a) for the 10.8~M$_\odot$
reference model.
It reaches $10^{12}$ g/cm$^3$ at about 6~ms before bounce
and increases rapidly on a timescale on the order of milliseconds
up to nuclear densities, on the order of $2-4\times10^{14}$ g/cm$^3$.
In general, the maximum values obtained for the central
density depend on the EoS and on the progenitor model.
The moment of bounce, i.e. $t=0$, is determined when the
maximum central density is obtained
(see Fig.~\ref{fig-density-h10n} (a)).
After bounce, the central density decreases slightly
on the same short timescale on the order of milliseconds,
until after about 10~ms post bounce the central density
increases back again.
It relates to the stalling of the bounce shock.
After bounce, the central density stays above nuclear saturation
density (see the horizontal dotted line in Fig.~\ref{fig-density-h10n}).
During the later post bounce evolution, which is determined by
the presence of the standing accretion shock and hence
mass accretion, the central density increases up to
$6.3\times10^{14}$ g/cm$^3$ at about $428$~ms
post bounce (see Fig.~\ref{fig-density-h10n} (b)).
At about $428.5357$~ms post bounce, the PNS contraction
proceeds into a collapse.

The density increases rapidly up to $1.29\times10^{15}$~g/cm$^3$
(see Fig.~\ref{fig-density-h10n} (b)).
The post bounce time when the maximum central density
is reached during the PNS collapse is labeled $t_c=429.3$~ms.
After $t_c$, the central density decreases again slightly,
which relates to the initial expansion of the
second shock wave formed inside the PNS.
At about $0.1$~ms after $t_c$, the central density starts
to oscillate at a short timescale on the order of milliseconds,
shown in Fig.~\ref{fig-density-h10n} (b).
This oscillation is due to the oscillating PNS interior, which is triggered
by the mass accretion onto the outwards propagating accretion shock
where nucleons are dissociated into quarks.
Hence, the quark core continues to grow in mass.
About 5~ms after $t_c$, the saturation value is obtained,
where $\rho_\text{central}\simeq1.25\times10^{15}$ g/cm$^3$
which remains constant with respect to time on the order of
100~ms.
The central density will continue to rise only at later times and
at a longer timescale, which is given by fall-back of material
enclosed inside the mass cut and due to deleptonization
(i.e. cooling).

\begin{figure}[htp]
\centering
\includegraphics[width=.45\textwidth]{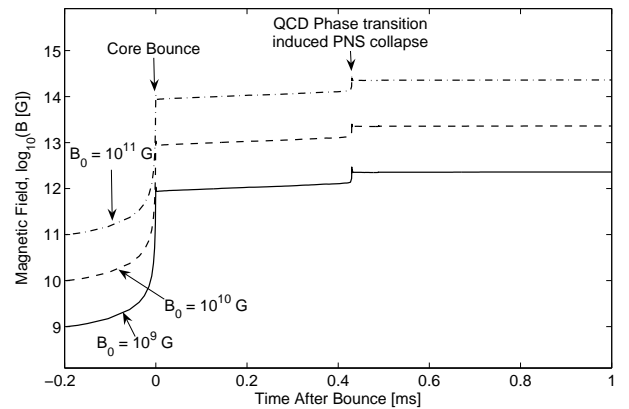}
\caption{Evolution of the central magnetic field strength for the
10.8~M$_\odot$ reference model with respect to time after
bounce for three different initial magnetic field strengths
($10^9$~G, $10^{10}$~G, $10^{11}$~G).}
\label{fig-Bfield-h10n}
\end{figure}

The scenario explored here might reveal a connection between
magnetars and neutron stars with quark matter in the interiors.
Due to flux conservation, the following relation between the
baryon density $\rho$ and the magnetic field strength $B$
applies,
\begin{equation}
\frac{B_0}{B_1} = \left(\frac{\rho_0}{\rho_1}\right)^{\frac{2}{3}}.
\label{eq-Bfield}
\end{equation}
This expression indicates a $B$-field increase during the Fe-core
collapse post bounce as well as an additional $B$-field increase
during the quark-hadron phase transition, in accordance with the
density increase.
The evolution of the central magnetic field, according to the
above relation, is shown in Fig.~\ref{fig-Bfield-h10n}
with respect to time after bounce.
It corresponds to the density evolution shown in Fig.~\ref{fig-density-h10n}.
Illustrated are three different initial magnetic field strengths,
i.e. $B_0=10^9$~G, $B_0=10^{10}$~G and $B_0 = 10^{11}$~G.
These are values which might be obtained at pre-collapse stellar
models \citep[see e.g.][]{HegerLanger:2000, Heger:etal:2005}.
At core bounce, for these initial configurations and according
to the density evolution as discussed above, the central $B$-fields
reach values of $10^{12}$, $10^{13}$ and $10^{14}$~G.
The post bounce compression increases the $B$-fields
additionally by a factor of about $2$ until the PNS collapses
at about $428.3757$~ms post bounce where, due to the density jump,
the $B$-fields rise to about $2.5\times10^{12-14}$~G on a short
timescale on the order of milliseconds.
After the quark-hadron phase transition, the central density
increases on a longer timescale on the order of $100$~ms
up to seconds and hence the magnetic field strength increases
slowly on the same timescale.

In addition to the evolution of the PNS interior in
Fig.~\ref{fig-Bfield-h10n} for the central density evolution,
close to the PNS surface the density decreases over several
orders of magnitude.
Hence we expect the magnetic field strengths at the PNS
surface to be smaller in comparison to the
evolution at the PNS center.
Surprisingly, the $B$-fields reach values only
slightly below $10^{12-14}$~G close to the PNS surface.
During the later PNS evolution after the explosion has been
launched, the PNS continues to contract.
The subsequent density increase due to deleptonization
on a longer timescale on the order of seconds,
may cause an additional increase of the $B$-field
where based on the three initial choices used
values of $10^{15}$~G might be obtained
at approximate simulation times on the order of $10$
seconds post bounce.

In this ad-hoc approach, important effects such as the
presence of rotation and fluid instabilities have been
neglected. Due to angular momentum conservation, the PNS
spins up during the (second) collapse. Even if at the
onset of the (second) collapse the PNS is in solid-body
rotation, the fact that the collapse proceeds non-homologously
leads to differential rotation.
Since a rotating body has the least rotational energy when it
rotates solidly, there is some free energy stored in differential rotation.
The free energy in differential rotation may be converted
into magnetic energy by winding up poloidal field into toroidal
field. Further, depending on the angular velocity profile, the
PNS may also be subject to the magneto-rotational instability (MRI)
\citep[see][]{BalbusHawley:1998}.
While field winding leads to a linear field strength growth, the
MRI may lead to exponential field growth.
Also convective motion from negative entropy and/or lepton gradients
generated by the propagation of the second shock and the second
neutrino burst may lead to dynamo action
\citep[see e.g.][]{ThompsonDuncan:1993}.
Therefore our estimated values for the evolution of the
magnetic field strength represent a lower limit.
In order to obtain a more precise prediction of the magnetic field
strength, detailed multidimensional investigations are necessary.

\begin{figure}[htp]
\centering
\includegraphics[width=.45\textwidth]{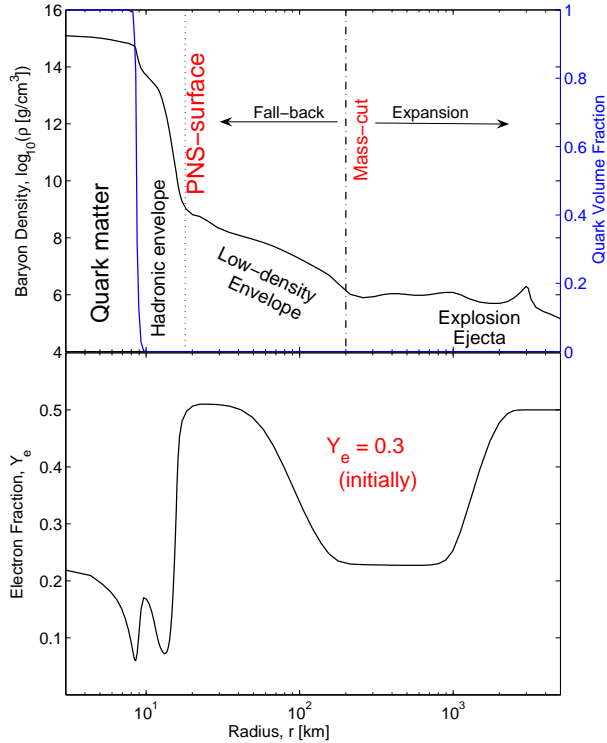}
\caption{Schematic composition, illustrating the radial profile of the
remaining hybrid PNS configuration for the reference 10.8
M$_\odot$ progenitor model at late post bounce time.}
\label{fig-density-final}
\end{figure}

Finally, Fig.~\ref{fig-density-final} schematically shows the
resulting PNS structure, with quark matter in the interior.
More precisely, the PNS is composed of an extended quark core
surrounded by a high-density (including a reasonable amount
of matter close to nuclear saturation density) hadronic mantle
as well as a low-density envelope.
The surface of the PNS can be defined via the contracting
neutrinospheres.
The evolution of the PNS is determined by fall-back of very
low-density hadronic material enclosed inside the mass cut.
The mass cut estimate is a dynamical quantity which changes
during the evolution.
The mass enclosed inside the mass cut will form the remnant
hybrid star.
The remaining hybrid star cools after mass accretion vanishes,
initially via deleptonization up to about one minute and later
via photons up to millions of years.
Matter outside the mass cut has become gravitationally unbound
during the ongoing explosion and mass ejection from the PNS
surface via neutrino heating (i.e. the neutrino-driven wind).
This mass will be ejected into the interstellar medium.
As can be seen from Fig.~\ref{fig-density-final}, a reasonable
amount of mass (about $7\times10^{-3}$~M$_\odot$) which corresponds
to the ejected matter, has become neutron-rich with an electron
fraction of $Y_e\simeq0.3$.
During the later expansion, the electron fraction increases
and values between $Y_e\simeq0.35-0.5$ are obtained.
Detailed explosive nucleosynthesis investigations are required
in order to confirm or exclude the scenario as a possible site
for the synthesis of heavy elements via the $r$-process.

\subsection{The explosion ejecta}

The mass elements which are ejected directly (ballistically),
are located at several $10^4$~km during the Fe-core collapse,
bounce and until the quark-hadron phase transition.
They have densities and temperatures of several $10^6$~g/cm$^3$
and about 0.2~MeV, which remain constant with respect to time.
These mass elements are relatively far away from the
center and belong to the surrounding part of the Fe-core.
The composition is dominated by $^{28}$Si and
$^{32}$S, where $Y_e\simeq 0.5$.
The structure of these layers is given by the progenitor model
and changes only insignificantly during the collapse,
bounce and early post bounce phases.

The situation changes when the expanding explosion shock
crosses these mass elements.
This happens at a post bounce time of about 450~ms
for the 10.8~M$_\odot$ reference model.
The mass elements contract rapidly towards the PNS
surface to about 10~km, during the PNS collapse.
Density and temperature increase to several $10^{12}$~g/cm$^3$
and several tens of~MeV, on a short timescale on the order of
milliseconds.
Furthermore, the entropy per baryon increases from values below
10~k$_B$ to about 60~k$_B$ and the electron fraction
decreases from $Y_e\simeq0.5$ to $Y_e\simeq0.1$.
During the later expansion, density and temperature decrease
initially on the same short timescale, where a larger value for
the electron fraction is obtained, between $Y_e\simeq0.33-0.48$,
and the entropy reduces to values close to 40~k$_B$ per baryon.
After the initial rapid expansion, these direct explosion ejecta
expand on a longer timescale on the order of 100~ms up to seconds.
Density and temperature decrease on the same timescale and reach
values between $10^{3-4}$~g/cm$^3$ and below 0.1~MeV
at about 2~seconds post bounce.
For these obtained conditions, the reaction rates become much smaller
and hence the electron fraction changes on a longer timescale
on the order of seconds.
In addition, since the later expansion proceeds adiabatically,
the entropy per baryon remains constant with respect
to time as well, where values of $\simeq 40$~k$_B$ per baryon
are obtained.

Mass elements which are not ejected directly in the initial shock
expansion fall back onto the PNS surface at the center.
During the short infall timescale on the order of milliseconds,
density and temperature increase up to $10^{13}$~g/cm$^3$
and 5--10~MeV.
During the mass accretion onto the PNS surface,
weak-equilibrium establishes at a low value of the electron
fraction of $Y_e\simeq0.05$ and the entropy per baryon
decreases to values below 5~k$_B$.
However, after about 1.5~seconds post bounce
the neutrino-driven wind appears as discussed in
e.g. \citet{Arcones:etal:2007, Fischer:etal:2010b}.
One of the most important result of the appearance of the
neutrino-driven wind relates to the high entropy
conditions obtained for the accelerated matter.
It results in generally proton-rich material, where $Y_e\simeq0.55$
(first seen in \citet{Liebendoerfer:etal:2003}
and explained in \citet{Froehlich:etal:2006}
and \citet{Pruet:etal:2006}).
%

%......................................................................................................................
\section{Summary}

We performed core-collapse supernova simulations of massive stars
in the mass range of 10.8 to 15~M$_\odot$, based on
general relativistic radiation hydrodynamics employing
a sophisticated EoS and three flavor Boltzmann neutrino transport.
A description for three flavor quark matter is implemented
based on the bag model with the different parameter choices,
($B^{1/4}=162$~MeV, $\alpha_s=0$~MeV),
($B^{1/4}=165$~MeV, $\alpha_s=0$~MeV) and
($B^{1/4}=155$~MeV, $\alpha_s=3$~MeV).
The resulting quark EoSs, EOS1, EOS2 and EOS3, are coupled
to the hadron EoS from \citet{Shen:etal:1998}, where Gibbs conditions
are applied for the transition between hadron and quark matter.
It results in an extended co-existence region in the phase diagram
where nucleons and quarks are present, the mixed phase.
The thermodynamic conditions for the appearance of quark
matter relate to critical densities close to nuclear matter density,
for temperatures of tens of~MeV and a low proton-to-baryon ratio.

During the evolution of massive stars, we find that quark matter
appears already at the Fe-core bounce for the chosen parameters.
However, the quark matter volume fraction is found to be rather
small with $\chi\leq0.1$.
The quark matter volume fraction increases slowly over several
100~ms during the early post bounce mass accretion phase,
during which central density and temperature increase while the
central electron fraction decreases.
The EoS in the mixed phase is significantly softer compared
to the pure hadronic and the pure quark phases.
Hence, the PNS contraction accelerates during the post bounce
evolution as more and more matter of the PNS interior is
converted into the mixed phase.
Finally, the central PNS configuration becomes gravitationally
unstable and the contraction proceeds into an adiabatic collapse.
Density and temperature increase on a short timescale on the
order of milliseconds, which in turn favors quark matter over hadronic matter.
The stiffening of the EoS in the pure quark phase halts the
collapse and a strong hydrodynamic shock wave forms.
The shock appears initially as a pure accretion front.
It propagates outward along the decreasing density of the PNS.
The propagation is determined via the balance of ram pressure
from the supersonically infalling material ahead of the shock
and the thermal pressure of the quarks as well as neutrino heating
behind the shock.
At the PNS surface where the density decreases over several
orders of magnitude, the accretion shock accelerates and positive
matter velocities are obtained.
The accretion front turns into a dynamic shock wave,
which continues to expand.
This moment establishes the onset of explosion.
It finally merges with the standing accretion shock from the Fe-core bounce,
which remained unaffected from the happenings inside the PNS.
All models under investigation follow the same evolutionary behavior
and lead to explosions, if the PNS mass does not exceed the 
maximum stable mass of the configuration given by the hybrid EoS.

Investigating the large uncertainty in core-collapse supernova input
physics, i.e. the state of matter at high densities and temperatures
and with a low proton-to-baryon ratio, we were able to explore
a new explosion mechanism in simulations of massive stars.
It serves as an addition to the well explored neutrino-driven,
the magnetically-driven and the acoustic mechanisms.
Note that these appear to be working only in multiple spatial
dimensions (except for the low mass O-Ne-Mg-core), due to the
nature of the physics directly involved in these explosion mechanisms.
The explosions obtained by taking QCD degrees of freedom into
account are presumably not restricted to dimensional limitations.
However, moderate explosion energies on the order of $10^{51}$~erg
could only be obtained for the 10.8~M$_\odot$ progenitor model
and for a particular choice of parameters.
Taking multi-dimensional phenomena into account,
such as rotation and the development of fluid instabilities,
may shift the explosion energies to robust values.

The sudden density rise during the seconds collapse
indicates an additional increase of the magnetic field strength.
It enables us to give a lower estimate of the expected magnetic field
strength for protoneutron stars with quark matter cores.
However, with initial magnetic fields that are confirmed by stellar evolution
calculations, a connection between hybrid stars and the special neutron
star class {\em Magnetars} with magnetic fields up to several $10^{15}$~G
could not be established.
Important aspects of the magneto-hydrodynamics evolution have not been
taken into account due to the current restriction to spherical symmetry.
They will be explored in a detailed study in an upcoming article.

Direct observables which allow us to identify the appearance of quark
matter in the PNS interior cannot be expected in the neutrino signal,
because matter is opaque for neutrinos.
Neutrinos can only diffuse out of the PNS interior on timescales
on the order of seconds.
However, the presence of a strong hydrodynamic shock wave changes
the situation.
It releases an additional outburst of neutrinos when crossing
the neutrinospheres.
This millisecond neutrino burst is dominated by $\bar{\nu}_e$,
because matter is neutron-rich where positron captures dominate
over electron captures, and by similar fluxes of
$\nu_e$ and $(\nu_{\mu/\tau},\bar{\nu}_{\mu/\tau})$
that are produced via pair process.
The delay of the second neutrino burst to the deleptonization
burst from the Fe-core bounce and its magnitude,
contain information about the progenitor model,
the hadronic and quark EoSs, including the critical
conditions for the onset of quark matter as well as the nature
of the phase transition from hadronic matter to quark matter.
It will become observable for future Galactic events, as
confirmed recently by \citet{Dasgupta:etal:2010} for the
operating neutrino detectors Super-Kamiokande and IceCube.
The knowledge of the progenitor model and the hadronic EoS
may allow us to decode hidden information about the state
of matter at high densities and temperatures, conditions which
will likely be reached in future heavy-ion collision experiments
at FAIR/GSI (Germany) and NICA/Dubna (Russia).
Furthermore, direct observables from the quark-hadron phase
transition can be expected in the emission of gravitational waves.
These might become observable in the near future if the sensitivity
improvements of the gravitational wave detectors continue.
They will serve as important addition to the neutrino spectra.

The composition of the ejecta can also be used to probe the explosion scenario.
Although core-collapse supernovae have been considered as the favored site for
possible $r$-process nucleosynthesis for a long time, the required conditions
have not been found in simulations so far.
Therefore it would be interesting to explore whether an $r$-process
can ensue in the layers being ejected here.
We defer, however, a thorough investigation to a separate paper.
The reduced nuclear reaction network included here to account for energy
generation is not suited to follow the $Y_e$ evolution and nucleosynthesis
at late times.
Nevertheless, there is only a small range of layers to be considered.
Neutron-rich matter from the deep layers close to the PNS surface attains
$Y_e\simeq$0.33--0.48 after shock heating and is ejected ballistically.
Layers in a second class are not directly ejected, rather they fall back onto
the PNS surface, where they experience continued neutrino heating and can
subsequently be ejected.
Initially, these layers can be very neutron-rich but turn proton-rich
as they expand in the neutrino-wind developing at later times.
Therefore, only the directly ejected zones may contain viable conditions
for neutron-rich nucleosynthesis, provided they are located not too far
above the PNS surface.

With this study we address a deeper understanding of the
inclusion of QCD degrees of freedom in radiation hydrodynamics studies
of astrophysical scenarios, explored at the example of core-collapse
supernova simulations of massive stars.
The quark matter description applied here leads to an
early onset of the chiral phase transition.
This is an active subject of research where several aspects,
such as the chiral phase transition in general, the hadronic freeze out,
the existence of a possible quarkyonic phase
\citep[][]{McLerranPisarski:2007}
and the nature of the transition between hadronic and quark
matter, are investigated.
Further developments of quark matter EoSs where different conditions
with respect to the critical conditions for the onset of deconfinement may
favor different evolutionary scenarios, will be studied in the future.
Furthermore, simulations in multiple spatial dimensions are
required in order to support the findings of the present article.
%

%..............................................................................
\section*{Acknowledgments}

The project was funded by the Swiss National Science Foundation
grant.~no.~PP00P2-124879/1 and 200020-122287 and
the Helmholtz Research School for Quark Matter Studies,
and the Helmholtz International Center (HIC) for FAIR.
T.F. is supported by HIC for FAIR project~no.~62800075
and G.M.P is partly supported by the Sonderforschungsbereich 634,
the ExtreMe Matter Institute EMMI and HIC for FAIR.
I.S. is supported by the Alexander von Humboldt foundation via the
Feodor-Lynen fellowship and the work of G.P. is supported by the
Deutsche Forschungsgemeinschaft (DFG) under Grant No. PA~1780/2-1.
J.S.-B. is supported by the DFG through the Heidelberg Graduate
School of Fundamental Physics.
M.H. acknowledges support from the High Performance and High
Productivity Computing (HP2C) project.
The authors are additionally supported by CompStar, a research
networking program of the European Science Foundation, and the Scopes
project funded by the Swiss National Science Foundation
grant.~no.~IB7320-110996/1.
The authors would also like to thank
Dr. Mary Beard from the University of Notre Dame
for proofreading the manuscript with focus on the
English language.\\

%.........................................................................
%\bibliography{references}

\end{document}